%% file: main.tex
\documentclass[draftcls,12pt,onecolumn]{IEEEtran}

\usepackage{authblk}
\usepackage{mathtools}
\usepackage{amsmath}
\usepackage{amssymb}
\usepackage{graphicx}
\usepackage{array}
\usepackage{makecell}
\usepackage{tikz}
\usepackage{pgfplots}	
\usepackage{booktabs}
\usepackage{floatflt}
\usepackage{enumerate}
\usepackage{psfrag}	
\usepackage{array}
\usepackage{multirow,hhline}
\usepackage{exscale}
\usepackage{color}
\usepackage{colortbl}
\usepackage{epsfig,subfigure}
\usepackage{cite}
\usepackage{amsthm}
\usepackage{xfrac}
\usepackage{algorithm}
\usepackage{algpseudocode}
\usepackage[font=small]{caption}
\usepackage[latin1]{inputenc} 
\usepackage[T1]{fontenc} 
\usepackage{enumerate}
\usepackage{siunitx}
\usepackage{nopageno}
\usepackage[left=0.58in, right=0.72in, top=0.7in]{geometry}
\usepackage{braket}

\newtheorem{theorem}{Theorem}
\newtheorem{definition}{Definition}
\newtheorem{lemma}{Lemma}

\newtheorem{remark}{Remark}

\usepackage{todonotes}

\usetikzlibrary{shapes,snakes}
\usetikzlibrary{calc}
\usetikzlibrary{patterns}
\usetikzlibrary{decorations.pathmorphing}

\newcommand{\Ali}{\textcolor{black}}
\newcommand{\Hossein}{\textcolor{black}}

\newcommand{\TxAntenna}[3]{
	\coordinate (a) at (#1,#2);
	\draw[line width=0.25pt,scale=(#3)] (a)--($(a)+(0.2,0)$)--($(a)+(0.2,0.7)$)--
	($(a)+(0.1,0.8)$)--($(a)+(0.3,0.8)$)--($(a)+(0.2,0.7)$);
}

\newcommand{\RxAntenna}[4]{
	\coordinate (a) at (#1,#2);
	\draw[line width=0.25pt,scale=(#3)] (a)--($(a)+(-0.2,0)$)--($(a)+(-0.2,0.7)$)--
	($(a)+(-0.1,0.8)$)--($(a)+(-0.3,0.8)$)--($(a)+(-0.2,0.7)$);
}

\allowdisplaybreaks
\definecolor{almond}{rgb}{0.94, 0.87, 0.8}
\definecolor{amber}{rgb}{1.0, 0.75, 0.0}
\definecolor{applegreen}{rgb}{0.55, 0.71, 0.0}
\definecolor{aqua}{rgb}{0.0, 1.0, 1.0}
\begin{document}
\IEEEoverridecommandlockouts
\title{Robust Transceiver Design for IRS-Assisted Cascaded MIMO Systems}
\author{
\IEEEauthorblockN{Hossein Esmaeili, Ali Kariminezhad, and Aydin Sezgin}\\
\thanks{
H. Esmaeili, A. Kariminezhad and A. Sezgin are with the Institute of Digital Communication Systems, Ruhr-Universit\"at Bochum (RUB), Germany (emails: \{hossein.esmaeili, ali.kariminezhad, aydin.sezgin\}@rub.de). This work is supported in part by the German Research Foundation, Deutsche Forschungsgemeinschaft (DFG), Germany, under the project grant number SE1967/16.
}}

\maketitle
\thispagestyle{empty}
\begin{abstract}
Robust transceiver design against unresolvable system uncertainties is of crucial importance for reliable communication.
We consider a MIMO multi-hop system, where the source, the relay, and the destination are equipped with multiple antennas. Further, an intelligent reconfigurable surface (IRS) is established to cancel the RSI as much as possible. The considered decode-and-forward (DF) hybrid relay can operate in either half-duplex or full-duplex mode, and the mode changes adaptively depending on the RSI strength. We investigate a robust transceiver design problem, which maximizes the throughput rate corresponding to the worst-case RSI under a self-interference channel uncertainty bound constraint. To the best of our knowledge, this is the first work that uses the IRS for RSI cancellation in MIMO full-duplex DF relay systems. The yielded problem turns out to be a non-convex optimization problem, where the non-convex objective is optimized over the cone of semidefinite matrices. We propose a closed-from lower bound for the IRS worst case RSI cancellation. Eventually, we show an important result that, for the worst case scenario, IRS can be helpful only if the number of IRS elements are at least as large as the size of the interference channel. Moreover, a novel method based on majorization theory is proposed to find the best response of the transmitters and relay against worst case RSI. Furthermore, we propose a multi-level water-filling algorithm to obtain a locally optimal solution iteratively.  Finally, we obtain insights on the optimal antenna allocation at the relay input-frontend and output-frontend, for relay reception and transmission, respectively. 

\begin{IEEEkeywords}
Intelligent reconfigurable surface (IRS), robust design,  relay systems, MIMO.
\end{IEEEkeywords}
\end{abstract}
\section{Introduction}
 Reliability and throughput are two of the most crucial requirements for the next generation of wireless networks.  Optimally relaying  the  signal  from  a  source  to  a  destination  can  help enhance  reliability  and  capacity  of  networks  and  is  currently an  active  research  area~\cite{Kariminezhad2017}.

 {Another emerging candidate for relaying signals are reconfigurable intelligent surfaces (IRSs) \cite{wu2019towards}. IRS is a device equipped with multiple passive reconfigurable reflectors that can reflect the colliding waves with an adjustable phase. One of the biggest advantages of IRSs is that they work in a real time manner without consuming noticeable amount of power~\cite{di2020smart}. However, the characteristics of the IRS (e. g. the lack of signal amplification and decode-and-forward processes) can potentially limit its functionality. As a result, in cases where reliability and throughput are of greater importance than the power consumption, conventional relays might still be a better option than IRSs. For instance, authors in~\cite{bazrafkan2021simple} show that a simple full-duplex relay can outperform an IRS in terms of throughput under certain conditions. In this paper, we exploit a full-duplex (FD) relay as a primary means of communication, and utilize an IRS to help it through self-interference (SI) cancelation.}

IRSs can be utilized in various ways to help the direct links enhance the performance of the system. In~\cite{guo2020weighted}, authors have utilized an IRS to maximize the weighted sum rate MISO system. Authors in~\cite{wu2019intelligent} proposed a method to minimize the power consumption in a MISO system equipped with an IRS and authors in~\cite{huang2019reconfigurable} investigate the problem of energy efficiency in an IRS-assisted MISO downlink system. The problem of rate maximization in a MIMO system has been presented in~\cite{zhang2020capacity}, where the authors propose an iterative algorithm to find the best IRS pattern assuming the perfect channel state information (CSI) is given. While the performance of IRS communication systems has been extensively studied, there are not much researches that are considering robust design when the perfect CSI is not available~\cite{omid2021low}. 

One of the earliest studies of the robust transmission designs of IRS-assisted systems was done in~\cite{zhou2020robust}, where a bounded CSI error model is applied to a problem of power miminization in a MISO transmission system. There, by the virtue of semidefinite progtramming (SDP), the authors turn the original problem into a sequence of convex sub-problems. The robust power minimization subject to the outage probability constraints under statistical cascaded channel error model is considered in~\cite{zhou2020framework}, where the aims is to optimize the system under worst-case rate constraint. Authors in~\cite{zhang2020robust} have proposed a robust algorithm for mean squared error (MSE) minimization for a single user MISO system equipped with an IRS. Their method provides a closed form solution for each iteration. However, it can be used only for the case of single user system and cannot be extended to more general cases where there are multiple users. Recently, a rubust algorithm based on penalty dual decomposition (PDD) technique is proposed in \cite{omid2021low} for sum-rate maximization where they assumed that the channel estimation error follows a complex normal distribution.

 Exploiting a relay to improve communication throughput rate is a classic alternative for IRS in communication systems. However, utilizing a relay in a network raises some important questions to be answered. For instance, how should the relay process the received signal before dispatching it to the destination? Now, the relay can receive a signal from the source, process it and transmit it towards the destination in a successive manner. This type of relaying technique is known as half-duplex relaying. Alternatively, while receiving a signal at a certain time instant, the relay can simultaneously transmit the previously received signals. This technique is known as full-duplex relaying~\cite{Bliss2007}. 
 
As a consequence of transmitting and receiving at a common resource unit, the relay is confronted with SI. Note that, full-duplex relaying potentially increases the total throughput rate of the communication compared to the half-duplex counterpart, only if the SI is handled properly at the relay input. By physically isolating the transmitter and receiver frontends of the relay, a significant portion of SI can be reduced~\cite{Sabharwal2014}. Moreover, analog and/or digital signal processing at the relay input can be utilized to cancel a portion of SI~\cite{ Eltawil2015,Vogt2018,Lee2014,Irigaray2018}. This can be realized if the estimate of the SI can be obtained at the relay. These SI cancellation procedures can effectively mitigate the destructive impact of SI up to a certain level. Hence, the remaining portion, the so-called residual self-interference (RSI), is still present at the relay input. The distribution of the RSI is investigated in~\cite{Irio2018,Irio2019}. 
This RSI is mainly due to the channel estimation uncertainties and transmitter noise. Therefore, the quality of channel estimation plays an important role for limiting RSI if the conventional modulation techniques are utilized. Interestingly, the authors in~\cite{Koohian2017} employ a superimposed signaling procedure (asymmetric modulation constellation) in the basic point-to-point FD communication for cancelling the SI and further retrieving the desired information contents without requiring channel estimates. They show that for the same average energy per transmission block, the bit error rate of their proposed method is better than that of conventional ones. The RSI degrades the performance of the communication quality evidently. To this end, the authors in~\cite{Chae2017} study the degrees-of-freedom (DoF), i.e., the slope of the rate curve at asymptotically high SNR, and its relation to the performance of FD cellular network in the presence of RSI. Moreover, the authors in~\cite{KariminezhadTWC} and~\cite{Kariminezhad2017B} investigate the joint rate-energy and delivery time optimization of FD communication, respectively, when RSI is still present. Furthermore, the authors in~\cite{Herath2013} study the sum rate capacity of the FD channel with and without such degradation. In the presence of RSI, the authors in~\cite{Zlatanov2017} study the capacity of Gaussian two-hop FD relay.

Robust transceiver design against the worst-case RSI channel helps find the threshold for switching between HD and FD operating modes. This set up is commonly known as hybrid relay systems \cite{riihonen2011hybrid}. The authors in~\cite{Taghizadeh2014} investigate a robust design for multi-user full-duplex relaying with multi-antenna DF relay. In that work, the sources and destinations are equipped with single antennas. Moreover, the authors in~\cite{Cirik2016} investigate a robust transceiver design for FD multi-user MIMO systems for maximizing the weighted sum-rate of the network.

\textit{Contribution:} Motivated by above, in this work, we consider a DF multi-hop system with multiple antennas at the source, relay and destination along with an IRS to provide additional links. Then we try to maximize the sum rate throughput for the worst case RSI scenario. To the best of our knowledge, this is the first time that the throughput rate maximization against the worst case RSI is evaluated for IRS-assisted DF full-duplex relay in MIMO systems. First, we simplify the problem by finding an analytical lower bound for the performance of the IRS. Then, the optimization of maximum achievable rate of the DF full-duplex relaying is cast as a non-convex optimization problem. \Ali{Thereafter, we propose a low complexity method to find the solution using majorization theory.} We propose an efficient algorithm to solve this problem in polynomial time. Finally, the transmit signal covariances at the source and the relay are designed efficiently to improve robustness against worst-case RSI channel in a given uncertainty bound. \Hossein{Notice that once the covariances are known, one can easily find the precoders using conventional methods such as SVD decomposition etc.}

\section{System Model}
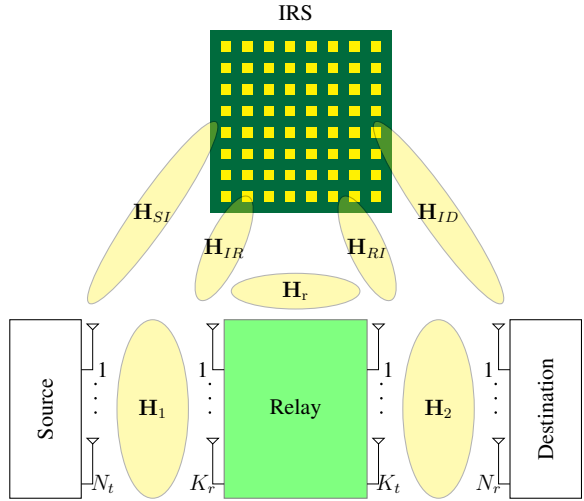
\begin{figure}
\centering
\tikzset{every picture/.style={scale=.95}, every node/.style={scale=0.7}}%
\input{SystemModel}
\vspace*{0.4cm}
\caption{System model of a IRS assisted full-duplex relay. In our model both source and destination are equipped with $N_t$ and $N_r$ antennas respectively. Also relay is equipped with $K_\mathrm{t}$ transmitting and $K_\mathrm{t}$ receiving antennas and IRS has $M$ passive elements.}
\label{fig:SystemModel}
\end{figure}

We consider the communication from a source equipped with $N_t$ antennas to a destination with $N_r$ antennas. The reliable communication from the transmitter to the destination is assumed to be only feasible by means of a relay with $K_\mathrm{t}$ transmitter and $K_{r}$ receiver antennas at the output and input frontends, respectively. This means that the direct link from the transmitter to the destination and the link from the transmitter to the IRS and to the destination has a negligible impact on the throughput. This assumption is realistic for the scenarios where the path loss is high due to the high frequency ranges such as mmWave and Terahertz or due to far distances \cite{han2016distance}. An IRS consisting of $M$ elements is established to either cancel the RSI or help enhance one of the transmitter-relay/relay-destination links. \Hossein{The system model is given in~\figurename{ \ref{fig:SystemModel}}.  As depicted, there are three different scenarios at which the IRS can perform. First, IRS can be used to eliminate the RSI. In this case, the received signals at the relay and destination are given by
\begin{align}
\mathbf{y}_{r}&= \mathbf{H}_1\mathbf{x}_{s}+\kappa \left(\mathbf{H}_{r}+\mathbf{H}_{Ir}\mathbf{\Theta}\mathbf{H}_{rI}\right)\mathbf{x}_{r}+\mathbf{n}_{t},\\
\mathbf{y}_{d}&= \mathbf{H}_2\mathbf{x}_{r}+\mathbf{n}_\mathrm{d},
\end{align}
respectively, where $\kappa\in\{0,1\}$. Notice that, $\kappa=0$ coincides with HD relaying and $\kappa=1$ denotes FD relaying. The transmit signal of the source is denoted by $\mathbf{x}_{s}\in\mathbb{C}^{N_t}$ with the covariance matrix $\mathbf{Q}_{s}=\mathbb{E}[\mathbf{x}_{s}\mathbf{x}^H_{s}]$, and the transmit signal of the relay is represented by $\mathbf{x}_{r}\in\mathbb{C}^{K_{\mathrm{t}}}$, with the covariance matrix $\mathbf{Q}_{r}=\mathbb{E}[\mathbf{x}_{r}\mathbf{x}^H_{r}]$. The additive noise vectors at the relay and destination are denoted by $\mathbf{n}_{r}\in\mathbb{C}^{K_{r}}$ and $\mathbf{n}_\mathrm{d}\in\mathbb{C}^{N_r}$, respectively, which are assumed to follow zero-mean Gaussian distributions with covariance matrices $\sigma^2_t\mathbf{I}$ and $\sigma_d^2\mathbf{I}$ respectively, where $\mathbf{I}$ is identity matrix. The source-relay channel is represented by $\mathbf{H}_1\in\mathbb{C}^{K_{r}\times N_t}$ and the relay-destination channel is denoted by $\mathbf{H}_2\in\mathbb{C}^{N_r\times K_{t}}$. These channels are assumed to be partially known. {Similarly, the partially known channels from the relay's transmitter to the IRS and from the IRS to the relay's receiver are denoted by $\mathbf{H}_{rI}\in\mathbb{C}^{M \times K_{t}}$ and $\mathbf{H}_{Ir}\in\mathbb{C}^{K_{r}\times M}$ respectively, see~\figurename{ \ref{fig:SystemModel}}}. $\mathbf{\Theta} \in \mathbb{C}^{M \times M}$ is a diagonal matrix representing the IRS phase profile. Throughout this paper, we assume that the IRS is passive and therefore for each element we have $|\theta_m| \leq 1$ \cite{bafghi2022degrees}. Furthermore, the channels from the source to the IRS and from the IRS to the destination are also assumed to be partially known and denoted as $\mathbf{H}_\mathrm{SI}$ and $\mathbf{H}_\mathrm{ID}$ respectively. Finally, the SI channel at the relay is represented by $\mathbf{H}_{r}$, which is assumed to be known only imperfectly. In what follows, we present the achievable throughput rates for the HD and FD relaying. We start with the HD relay, in which $\kappa=0$.
In the second case, IRS can be exploited to enhance the quality of the channel between the source and the relay. In such a case, the received signals at the relay and destination can be expressed as
\begin{align}
\mathbf{y}_{r}&= \left(\mathbf{H}_{1}+\mathbf{H}_{IR}\mathbf{\Theta}\mathbf{H}_{SI}\right)\mathbf{x}_{s}+\kappa\mathbf{H}_{r}\mathbf{x}_{r}+\mathbf{n}_{t},\\
\mathbf{y}_\mathrm{d}&= \mathbf{H}_2\mathbf{x}_{r}+\mathbf{n}_\mathrm{d},
\end{align}
where $\mathbf{H}_{SI}\in\mathbb{C}^{N_{\mathrm{t}} M}$ is the channel from the source to the IRS. Finally, IRS can be used to help the channel from the relay to the destination. In this case, the received signals are going to be
\begin{align}
\mathbf{y}_{r}&= \mathbf{H}_{1}\mathbf{x}_{s}+\kappa\mathbf{H}_{r}\mathbf{x}_{r}+\mathbf{n}_{t},\\
\mathbf{y}_\mathrm{d}&= \left(\mathbf{H}_{2}+\mathbf{H}_{ID}\mathbf{\Theta}\mathbf{H}_{RI}\right)\mathbf{x}_{r}+\mathbf{n}_\mathrm{d},
\end{align}
where $\mathbf{H}_{ID}\in\mathbb{C}^{M \times N_{\mathrm{r}}}$ is the channel from the source to the IRS.}

\Hossein{In this paper, we consider the channel uncertainty model used for FD communication in \cite{ng2012dynamic,kim2013effects,riihonen2011mitigation} in which some information about the channel state is provided to the receiver. Therefore, the receiver can use the estimation of the channel, denoted as $\hat{\mathbf{H}}_{x}$ where $x \in \left\{1,2,SI,IR,RI,ID\right\}$. The channel estimation error is then expressed as $\bar{\mathbf{H}}_{x} = {\mathbf{H}}_{x} - \hat{\mathbf{H}}_{x}$, where $\hat{\mathbf{H}}_{x}$ and $\bar{\mathbf{H}}_{x}$ are uncorrelated. Further we assume that $\text{Tr}(\bar{\mathbf{H}}^H_{x}\bar{\mathbf{H}}_{x}) \leq T_{x}$. Notice that this assumption is more genereal than zero mean circularly symmetric complex Gaussian entries or non-i.i.d channel estimation errors applied in \cite{cirik2014achievable} and \cite{day2012full,day2012full2} respectively.
In what follows, we find the achievable rate for three aforementioned cases and compare them to see under what conditions each of them should be applied.}

\section{Achievable Rate (Full-Duplex Relay)}
\subsection{Overview}
Suppose that the relay employs a DF strategy. In the full-duplex scenario both source-relay and relay-destination links are active at the same time. As a result, the signals from the relay transmitter interfere with the receiving signal at the relay receiver. We assume that an estimate of the SI channel $\mathbf{H}_{r}$ is available at the relay denoted by $\hat{\mathbf{H}}_{r}$. Hence, the RSI is represented by $\bar{\mathbf{H}}_{r}$ is given as
\begin{align}
\bar{\mathbf{H}}_{r}=\mathbf{H}_{r}-\hat{\mathbf{H}}_{r}.
\end{align}
\subsection{Mathematical Preliminaries}
Considering a FD DF relay, the following rates are achievable \cite{wang2005capacity},
\begin{align}
R^{\mathrm{FD}}=\min(R^{\mathrm{FD}}_\mathrm{sr},R^{\mathrm{FD}}_\mathrm{rd}),
\end{align}
in which depending on how the IRS is applied to the system, the three following sets of rates are possible. First,
\begin{align}
R^{\mathrm{FD}}_\mathrm{sr}&=\log_2\frac{\big|\sigma_t^2\mathbf{I}_{K_{r}}+\hat{\mathbf{H}}_1\mathbf{Q}_{s}\hat{\mathbf{H}}_1^H+\bar{\mathbf{H}}_1\mathbf{Q}_{s}\bar{\mathbf{H}}_1^H+{\mathbf{H}}_{tot_{1}}\mathbf{Q}_{r}{\mathbf{H}}^H_{tot_{1}}\big|}{\big|\sigma_t^2\mathbf{I}_{K_{r}}+\bar{\mathbf{H}}_1\mathbf{Q}_{s}\bar{\mathbf{H}}_1^H+{\mathbf{H}}_{tot_{1}}\mathbf{Q}_{r}{\mathbf{H}}^H_{tot_{1}}\big|},\label{eq:FD_srA}\\
R^{\mathrm{FD}}_\mathrm{rd}&=\log_2\frac{\big|\sigma_d^2\mathbf{I}_{N}+\hat{\mathbf{H}}_2\mathbf{Q}_{r}\hat{\mathbf{H}}^H_2+\bar{\mathbf{H}}_2\mathbf{Q}_{r}\bar{\mathbf{H}}^H_2\big|}{\big|\sigma_d^2\mathbf{I}_{N}+\bar{\mathbf{H}}_2\mathbf{Q}_{r}\bar{\mathbf{H}}^H_2\big|},
\end{align}
where ${\mathbf{H}}_{tot_{1}} = \left(\bar{\mathbf{H}}_{r}+\mathbf{H}_{rI}\mathbf{\Theta}\mathbf{H}_{Ir}\right)$ when the IRS is used to cancel the self interference. Second,
\begin{align}
R^{\mathrm{FD}}_\mathrm{sr}&=\log_2\frac{\big|\sigma_t^2\mathbf{I}_{K_{r}}+\hat{\mathbf{H}}_{tot_{2}}\mathbf{Q}_{s}\hat{\mathbf{H}}^H_{tot_{2}}\!+\!\bar{\mathbf{H}}_{tot_{2}}\mathbf{Q}_{s}\bar{\mathbf{H}}^H_{tot_{2}}+\bar{\mathbf{H}}_{r}\mathbf{Q}_{r}\bar{\mathbf{H}}^H_{r}\big|}{\big|\sigma_t^2\mathbf{I}_{K_{r}}\!+\!\bar{\mathbf{H}}_{tot_{2}}\mathbf{Q}_{s}\bar{\mathbf{H}}^H_{tot_{2}}+\bar{\mathbf{H}}_{r}\mathbf{Q}_{r}\bar{\mathbf{H}}_{r}^H\big|},\label{eq:FD_srA}\\
R^{\mathrm{FD}}_\mathrm{rd}&=\log_2\frac{\big|\sigma_d^2\mathbf{I}_{N}+\hat{\mathbf{H}}_2\mathbf{Q}_{r}\hat{\mathbf{H}}^H_2+\bar{\mathbf{H}}_2\mathbf{Q}_{r}\bar{\mathbf{H}}^H_2\big|}{\big|\sigma_d^2\mathbf{I}_{N}+\bar{\mathbf{H}}_2\mathbf{Q}_{r}\bar{\mathbf{H}}^H_2\big|},
\end{align}
where $\hat{\mathbf{H}}_{tot_{2}} = \left(\hat{\mathbf{H}}_{1}+\mathbf{H}_{rI}\mathbf{\Theta}\hat{\mathbf{H}}_{Ir}\right)$ and $\bar{\mathbf{H}}_{tot_{2}} = \left(\bar{\mathbf{H}}_{1}+\bar{\mathbf{H}}_{rI}\mathbf{\Theta}\bar{\mathbf{H}}_{Ir}\right)$ if the IRS is established to help the source-relay channel and finally
\begin{align}
R^{\mathrm{FD}}_\mathrm{sr}&=\log_2\frac{\big|\sigma_t^2\mathbf{I}_{K_{r}}+\hat{\mathbf{H}}_{1}\mathbf{Q}_{s}\hat{\mathbf{H}}^H_{1}+\bar{\mathbf{H}}_1\mathbf{Q}_{s}\bar{\mathbf{H}}_1^H+\bar{\mathbf{H}}_{r}\mathbf{Q}_{r}\bar{\mathbf{H}}_{r}^H\big|}{\big|\sigma_t^2\mathbf{I}_{K_{r}}+\bar{\mathbf{H}}_1\mathbf{Q}_{s}\bar{\mathbf{H}}_1^H+\bar{\mathbf{H}}_{r}\mathbf{Q}_{r}\bar{\mathbf{H}}_{r}^H\big|},\label{eq:FD_srA}\\
R^{\mathrm{FD}}_\mathrm{rd}&=\log_2\frac{\big|\sigma_d^2\mathbf{I}_{N}+\hat{\mathbf{H}}_{tot_{3}}\mathbf{Q}_{r}\hat{\mathbf{H}}^H_{tot_{3}}+{\bar{\mathbf{H}}}_{tot_{3}}\mathbf{Q}_{r}{\bar{\mathbf{H}}}^H_{tot_{3}}\big|}{\big|\sigma_d^2\mathbf{I}_{N}+{\bar{\mathbf{H}}}_{tot_{3}}\mathbf{Q}_{r}{\bar{\mathbf{H}}}^H_{tot_{3}}\big|},
\end{align}
where $\hat{\mathbf{H}}_{tot_{3}} = \left(\hat{\mathbf{H}}_{2}+\hat{\mathbf{H}}_{rI}\mathbf{\Theta}\hat{\mathbf{H}}_{Ir}\right)$ and $\bar{\mathbf{H}}_{tot_{3}} = \left(\bar{\mathbf{H}}_{2}+\bar{\mathbf{H}}_{rI}\mathbf{\Theta}\bar{\mathbf{H}}_{Ir}\right)$ if the IRS is utilized to enhance the rate of relay-destination channel. Notice that, assuming that the RSI remains uncanceled, a robust transceiver against the worst-case RSI channel is required which is formulated as an optimization problem as follows
\begin{subequations}\label{P:FDa}
\begin{align}
 \max_{\mathbf{Q}_{s},\mathbf{Q}_{r},\mathbf{\Theta}}\  \min_{\bar{\mathbf{H}}_{r}}\quad & \min\bigg( R^{\mathrm{FD}}_\mathrm{sr}, R^{\mathrm{FD}}_\mathrm{rd} \bigg) \tag{\ref{P:FDa}}\\
\text{subject to}\quad\quad & \mathrm{Tr}(\mathbf{Q}_{s})\leq P_{s},\label{P:FDa:ConsA}\\ 
&\mathrm{Tr}(\mathbf{Q}_{r})\leq P_{r},\label{P:FDa:ConsB}\\
&\mathrm{Tr}(\bar{\mathbf{H}}_{x}\bar{\mathbf{H}}^H_{x})\leq T_{x},\nonumber\\ 
&x \in \left\{1,2,r,RI,IR,ID,SI\right\}
\label{P:FDa:ConsCF}\\
&|\theta_m| \leq 1, \forall m \label{P:FDa:ConsDF}
\end{align}
\end{subequations}
in which the throughput rate with respect to the worst-case RSI channel is maximized. Two constraints $P_s$ and $P_r$ represent the transmit power budgets at the source and the relay respectively. In constraint~\eqref{P:FDa:ConsCF}, $T_x$ represents the RSI or the channel estimation error bound corresponding to $\mathbf{H}_x$. Notice that, $\mathrm{Tr}(\bar{\mathbf{H}}_{x}\bar{\mathbf{H}}^H_{x})$ represents the sum of the squared singular values of $\mathbf{H}_{x}$. It should be noted that, using a bounded matrix norm is the most common way for modeling the uncertainty of a matrix~\cite{wang2013robust,shen2013robust}. In practice, $T_x$ can be found using stochastic methods when the distribution of the channel error is known. Otherwise, one may find it using sample average approximation method. {Finally, constraints~\eqref{P:FDa:ConsDF} are due to the unit modulus limitation of the IRS elements}. 

The problem~\eqref{P:FDa} is non-convex and hard to solve. As a result, for each of the above-mentioned scenarios, we propose a simplified version of the optimization problem and try to solve it instead. Note that as we are interested in finding the throughput corresponding to the worst case RSI, any simplification in the optimization problem should be in favor of the RSI and interference. In the next, we first analyse the performance of the system when the IRS is helping the relay to cancel the RSI. Consequently, we show that the problem~\eqref{P:FDa} can be simplified to the following optimization problem

\begin{subequations}\label{PP:FDaa}
\begin{align}
 \max_{\mathbf{Q}_{s},\mathbf{Q}_{r}}\  \min_{{\mathbf{H}}_{tot}}\quad & \min\bigg( R^{\mathrm{FD}}_\mathrm{sr}, R^{\mathrm{FD}}_\mathrm{rd} \bigg) \tag{\ref{PP:FDaa}}\\
\text{subject to}\quad\quad & \mathrm{Tr}(\mathbf{Q}_{s})\leq P_{s},\nonumber\\ 
&\mathrm{Tr}(\mathbf{Q}_{r})\leq P_{r},\nonumber\\
&\mathrm{Tr}({\mathbf{H}}_{tot}{\mathbf{H}}^H_{tot})\leq T'(T_r,\mathbf{\Theta}).\nonumber\\
&\mathrm{Tr}(\bar{\mathbf{H}}_{x}\bar{\mathbf{H}}^H_{x})\leq T_{x},\quad x \in \left\{1,2\right\}\nonumber
\label{P:FDa:ConsCF}\\
\end{align}
\end{subequations}
where
\begin{subequations}\label{OPP:T}
\begin{align}
T'(T_r,\mathbf{\Theta}) = \min_{\mathbf{\Theta}}\  \max_{\bar{\mathbf{H}}_{r}}\quad & ||\bar{\mathbf{H}}_{r}+\mathbf{H}_{rI}\mathbf{\Theta}\mathbf{H}_{Ir}||_F^2\\
\text{subject to}\quad\quad 
&\mathrm{Tr}(\bar{\mathbf{H}}_{r}\bar{\mathbf{H}}^H_{r})\leq T_r,\label{OPP:FDa:ConsC}\\
&||Vec(\mathbf{\Theta})||_2^2 \leq 1,
\end{align}
\end{subequations}
and where $Vec(\cdot)$ denotes the vector of all non-zero elements of its input matrix.
We can equivalently write $T'$ as

\begin{subequations}\label{OPP:TT}
\begin{align}
T'(T_r,\mathbf{\Theta}) = \min_{\mathbf{\Theta}}\  \max_{\bar{\mathbf{H}}_{r}}\quad & ||Vec(\bar{\mathbf{H}}_{r})+(\mathbf{H}_{Ir}*\mathbf{H}_{rI}^T)Vec(\mathbf{\Theta})||_2^2\\
\text{subject to}\quad\quad 
&\mathrm{Tr}(\bar{\mathbf{H}}_{r}\bar{\mathbf{H}}^H_{r})\leq T_r,\label{P:FDa:ConsC}\\
&||Vec(\mathbf{\Theta})||_2^2 \leq 1,
\end{align}
\end{subequations}
where $*$ denotes column-wise Khatri-Rao product defined as below
\begin{equation}
    A*B = \left[A_1\otimes B_1|A_2\otimes B_2|\cdots|A_n\otimes B_n\right],
\end{equation}
and where $A_i$ is the $i$'th column of $A$ and $\otimes$ denotes the Kronecker product. See Appendix I for proof.

One can show that $T'\leq (\sqrt{T_r}-\sigma_{\min}(\mathbf{H}_{Ir}*\mathbf{H}_{rI}^T))^2$. As mentioned before, problem~\eqref{PP:FDaa} is a simplification of the problem~\eqref{P:FDa}. This means every achievable rate which is inside the feasible set of~\eqref{PP:FDaa} is also inside the feasible set of~\eqref{P:FDa}\footnote{Notice that the reverse is not necessarily true .i .e every achievable rate which is a feasible solution of~\eqref{P:FDa} is not necessarily a feasible solution for~\eqref{PP:FDaa} as well. However, as we look for achievable rates, we can still use this method.}. The intuition behind that is, in problem~\eqref{P:FDa} the minimization over RSI happens only one time and the worst case RSI simultaneously tries to cancel the effect of best configuration of IRS and the best covariance matrices. Whereas in~\eqref{PP:FDaa}, first, the RSI does its worst damage on the performance of the best IRS configuration, and after that, performs another optimization to bring the worst power allocation against the best covariance matrices\footnote{This will be more clear later on when the geometrical representation of the problem is given.}. In what follows, we provide our proposed ways to deal with optimization problems~\eqref{OPP:T} and~\eqref{PP:FDaa} respectively.

\begin{theorem}\label{the:1}
 For the optimization problem~\eqref{PP:FDaa}, one can show that $T'(T,\Theta)\leq (\sqrt{T_r}-\sigma_{\min}(\mathbf{H}_{Ir}*\mathbf{H}_{rI}^T))^2$.

\begin{proof}
We begin the proof with an intuitive example and then extend it to the more general case. Assume that $K_t = 1,K_r = 2$ and $M = 3$. Then we have

\begin{subequations}\label{max-min}
\begin{align}
T' = \max_{\mathbf{\Theta}}\  \min_{\bar{\mathbf{h}}_{r}}\quad & ||\bar{\mathbf{h}}_{r}+\mathbf{H}_{Ir}diag(\mathbf{H}_{rI}^T)Vec(\mathbf{\Theta})||_2^2\label{mesalsade}\\
\text{subject to}\quad\quad 
&\bar{{h}}_{11}^2 + \bar{{h}}_{21}^2 \leq T_r,\label{P:georepa:ConsC}\\
&\theta_1^2 \leq 1, \ \theta_2^2 \leq 1, \ \theta_3^2 \leq 1.\label{P:georepa:cube}
\end{align}
\end{subequations} 
Also, consider the following optimization problem
\begin{subequations}\label{max-min2}
\begin{align}
T'' = \max_{\mathbf{\Theta}}\  \min_{\bar{\mathbf{h}}_{r}}\quad & ||\bar{\mathbf{h}}_{r}+\mathbf{H}_{Ir}diag(\mathbf{H}_{rI}^T)Vec(\mathbf{\Theta})||_2^2\label{mesalsade}\\
\text{subject to}\quad\quad 
&\bar{{h}}_{11}^2 + \bar{{h}}_{21}^2 \leq T_r,\label{P:georepa:ConsC}\\
&\theta_1^2+\theta_2^2+\theta_3^2 \leq 1,\label{P:georepa:sphere}
\end{align}
\end{subequations}


Here, notice that $\mathbf{H}_{Ir}diag(\mathbf{H}_{rI}^T)$ is a linear map from a three-dimensional into a two-dimensional space. One simple example of such a mapping can be found in Fig~\ref{fig:threetotwo}. Here, an example of mapping from three-dimensional to two-dimensional space is shown. The left shape shows the feasible set for the IRS with three elements in a real valued space. The cube belongs to the case of $T'$, i. e. constraints  $-1\leq\theta_m\leq1,\forall m$, while the sphere shows the constraint $\theta_1^2+\theta_2^2+\theta_3^2 \leq 1$ which belongs to $T''$. In the right, the feasible sets belonging to two aforementioned regions after performing mapping $f$ are presented as an example. It can be seen that the mapping of the first set of constraints (the hexagon) covers the whole area of that of the second one (the ellipse). One important key is, as mapping is a linear function, we have $\mathcal{A} \subset \mathcal{B} \xrightarrow[]{} f(\mathcal{A}) \subset f(\mathcal{B})$, where $\mathcal{A}$ and $\mathcal{B}$ are two arbitrary sets and $f$ is the mapping.

\Hossein{With the above considerations in mind, now we are ready to deal with the optimization problem. Fig~\ref{fig:georep} shows the overview of the optimization problem~\eqref{PP:FDaa}. Due to the constraint~\eqref{P:georepa:ConsC}, the feasible set of all possible choices of $\bar{\mathbf{h}}_r$ creates a circle with radius $\sqrt{T}$ centered at $(0,0)$. The parallelogram and the ellipse show the areas corresponding to the mapping of the cubic and the spherical feasible sets (12c) and (13c) imposed by the IRS constraints respectively. Notice that, the latter is always a subset of the former, i.e, ${S}_{ellipse} \subset {S}_{parallelogram}$, where $S$ represents the area of the set. Now, considering the max-min problem~\eqref{max-min}, we are able to have a geometrical interpretation of the solution. Note that ~\eqref{max-min} indicates the distance between $\bar{\mathbf{h}}_r$ and $-\mathbf{H}_{Ir}diag(\mathbf{H}_{rI}^T)\mathbf{\Theta}$. While $\bar{\mathbf{h}}_r$ wants to maximize this distance, the job of $\mathbf{\Theta}$ is to minimize it as much as possible. It can be seen in the figure that the best choice of RSI for $\bar{\mathbf{h}}_r$ is the point that is farthest from the parallelogram. It is evident that this point is somewhere on the circumference of the outer circle. On the other hand, the best choice for the ${\Theta}$ is to be chosen in a way to be as close as possible to $\bar{\mathbf{h}}_r$. These points are indicated respectively as $G$ and $F$ on the figure. In general, as the number of IRS elements or the dimensions of $\bar{\mathbf{h}}_r$ increase, the mapping of the hypercube becomes more and more complicated and finding the optimal distance becomes more difficult. However, there is an upper bound for this distance. As shown in the figure, if instead of the cube, we limit the feasible set of IRS elements to the sphere inside the cube, i.e. replacing (12c) with (13c), the solution to the problem becomes $GE \geq GF$. It turns out that finding $GE$ is very simple as by the definition we have $\sigma_{\min}(\mathbf{H}_{Ir}diag(\mathbf{H}_{rI}^T))=OE$, and also we know that $\sqrt{T_r}=GO$. Therefore, we can conclude $GE = \sqrt{T_r}-\sigma_{\min}(\mathbf{H}_{Ir}diag(\mathbf{H}_{rI}^T))$.
Finally, we use one last upper bound to make the original problem even easier to solve. Note that, if instead of the ellipse, we consider the circle inscribed in it, we will have $  \max_{\bar{\mathbf{h}}_r}\ \min_{\mathbf{\Theta}}||\bar{\mathbf{h}}_{r}+(\mathbf{H}_{Ir}*\mathbf{H}_{rI}^T){\theta}||_2 = \sqrt{T_r}-\sigma_{\min}(\mathbf{H}_{Ir}*\mathbf{H}_{rI}^T), \forall \bar{\mathbf{h}}_r$. As a result, we have 
\begin{equation}
||\bar{\mathbf{h}}_{tot}\bar{\mathbf{h}}_{tot}^H||_2 \leq T' ,\quad \forall \bar{\mathbf{h}}_r,  
\end{equation}
where $T'=\left(\sqrt{T_r}-\sigma_{\min}(\mathbf{H}_{Ir}*\mathbf{H}_{rI}^T)\right)^2$. It is worth mentioning that the geometrical representation for the optimization problem~\eqref{P:FDa} is different, because there, considering that the RSI wants to bring the worst representation against the IRS configuration and covariance matrices simultaneously, the RSI cannot freely span the whole circle. This is due to the fact that some regions in the cicle might not be a good choice when it comes to RSI design against covariance matrices. However, if the best representation of RSI against the covariance matrices also provides the best RSI against the IRS configuration, the solution to~\eqref{PP:FDaa} and~\eqref{P:FDa} will be the same. Eventually, instead of optimizaton problem~\eqref{P:FDa} one can solve optimization probem~\eqref{PP:FDaa}. The solution to the new problem is guaranteed to be achievable by the original problem as well.}

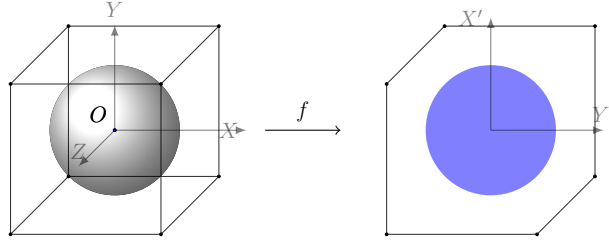
\begin{figure}
\centering
\tikzset{every picture/.style={scale=.5}, every node/.style={scale=0.7}}%
\input{threetotwo}
\vspace*{0.4cm}
\caption{An example of mapping from three-dimensional to two-dimensional space.}
\label{fig:threetotwo}
\end{figure}

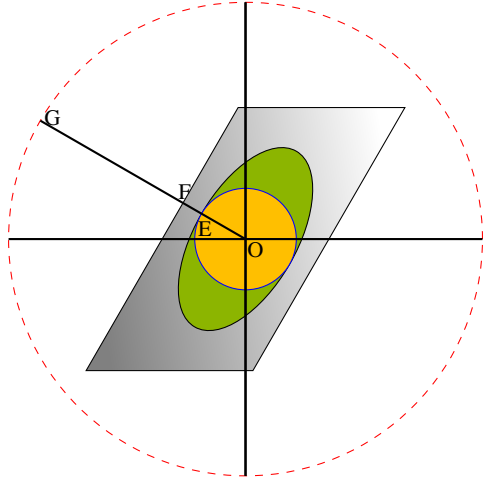
\begin{figure}
\centering
\tikzset{every picture/.style={scale=.45}, every node/.style={scale=0.7}}%
\input{georep}
\vspace*{0.4cm}
\caption{Geometrical representation of optimization problem~\eqref{max-min}.}
\label{fig:georep}
\end{figure}

\Hossein{Notice that one can readily extend this interpretation into the complex domain, as the constraint~\eqref{P:georepa:sphere} will still be a subset of constraints~\eqref{P:georepa:cube}. It remains to show one can generalize the geometrical proof for arbitrary large dimensions. This means that the channel dimensions and the number of IRS elements can be any natural numbers. Interestingly, it is enough to show that the geometrical proof based on $\ell_2$ norms and Euclidean distance exists for higher dimensions. This proof is given in Appendix I where it is shown $||\bar{\mathbf{H}}_{r}+\mathbf{H}_{rI}\mathbf{\Theta}\mathbf{H}_{Ir}||_F^2 = ||Vec(\bar{\mathbf{H}}_{r})+(\mathbf{H}_{Ir}*\mathbf{H}_{rI}^T)Vec(\mathbf{\Theta})||_2^2$.}
\end{proof}
\end{theorem}



The next is to solve problem~\eqref{PP:FDaa}. Solving this problem is hard in general as it is non-convex. Hence, we first use the following lemma and theorem to solve it. There, it is shown that for every possible choice of ${\mathbf{H}}_\mathrm{1}$ and ${\mathbf{H}}_\mathrm{2}$, there exists at least one set of simultaneously diagonalizable matrices ${\mathbf{H}}_{tot}$, ${\mathbf{Q}}_{s}$ and ${\mathbf{Q}}_{r}$ that are the solutions to the problem~\eqref{PP:FDaa}.
%
%
\begin{lemma}\label{lem:1}
For two positive semi-definite and positive definite matrices $\mathbf{A}$ and $\mathbf{B}$ with eigenvalues $\lambda_1\left(\mathbf{A}\right)\geq \lambda_2\left(\mathbf{A}\right) \geq ... \geq \lambda_N\left(\mathbf{A}\right) $ and $\lambda_1 \left(\mathbf{B}\right) \geq \lambda_2\left(\mathbf{B}\right) \geq ... \geq \lambda_N\left(\mathbf{B}\right)$ respectively, the following inequalities hold,
\begin{equation}
    \prod_{i = 1}^{N} \left( 1 + \frac{\lambda_i\left(\mathbf{A}\right)}{\lambda_i\left(\mathbf{B}\right)} \right) \leq \Big| \mathbf{I}+\mathbf{A}\mathbf{B}^{-1} \Big| \leq \prod_{i = 1}^{N} \left( 1 + \frac{\lambda_i\left(\mathbf{A}\right)}{\lambda_{N+1-i}\left(\mathbf{B}\right)} \right).\label{eq:Fiedler0}
\end{equation}

\begin{proof}
Consider Fiedler's inequality given by~\cite{fiedler1971bounds},
\begin{equation}
\prod_{i = 1}^{N} \left( \lambda_i\left(\mathbf{A}\right) \!+\! \lambda_i\left(\mathbf{B}\right) \right) \!\leq \!\Big| \mathbf{B}+\mathbf{A} \Big| \!\leq \!\prod_{i = 1}^{N} \left( \lambda_i\left(\mathbf{A}\right) \!+\! \lambda_{N+1-i}\left(\mathbf{B}\right) \right).\label{eq:Fiedler}
\end{equation}
Furthermore, given $\mathbf{B}$ as a positive definite matrix, the followings hold,
\begin{align}
\big| \mathbf{B}^{} \big| &> 0,\\
\big| \mathbf{B}^{-1} \big| &= \prod_{i = 1}^{N}\frac{1}{\lambda_i\left(\mathbf{B}\right)}.
\end{align}
Now, dividing the sides of~\eqref{eq:Fiedler} by $\big|\mathbf{B}\big|$, one can readily obtain~\eqref{eq:Fiedler0}.
\end{proof}
\end{lemma}

Note that, in~\eqref{eq:Fiedler0} the inequalities hold with equality if and only if $\mathbf{A}$ and $\mathbf{B}$ are diagonalizable over a common basis. Using the result of Lemma~\ref{lem:1}, $R_{\mathrm{sr}}^{\mathrm{FD}}$ can be lowerbounded as
\begin{align}
       &\log_2\frac{\big|\sigma_t^2\mathbf{I}_{K_{r}}+\hat{\mathbf{H}}_1\mathbf{Q}_{s}\hat{\mathbf{H}}_1^H+\bar{\mathbf{H}}_1\mathbf{Q}_{s}\bar{\mathbf{H}}_1^H+{\mathbf{H}}_{tot_{1}}\mathbf{Q}_{r}{\mathbf{H}}^H_{tot_{1}}\big|}{\big|\sigma_t^2\mathbf{I}_{K_{r}}+\bar{\mathbf{H}}_1\mathbf{Q}_{s}\bar{\mathbf{H}}_1^H+{\mathbf{H}}_{tot_{1}}\mathbf{Q}_{r}{\mathbf{H}}^H_{tot_{1}}\big|}
        \nonumber\geq \\ 
        &\sum_{i = 1}^{\min{(M,K_r)}}\! \log_2\! \left(\! 1+\! \frac{\lambda_i\left(\mathbf{H}_1\mathbf{Q}_{s}{\mathbf{H}_1}^H\right)}{\lambda_i\!\left(\!\sigma_t^2\mathbf{I}_{K_{r}}\!+\!\bar{\mathbf{H}}_1\!\mathbf{Q}_{s}\!\bar{\mathbf{H}}_1^H\!+\!{\mathbf{H}}_{tot_{1}}\!\mathbf{Q}_{r}\!{\mathbf{H}}^H_{tot_{1}}\!\right)} \right). \label{eq:InequalityA}
\end{align}
Also it holds that $\lambda_i\!\left(\!\sigma_t^2\mathbf{I}_{K_{r}}\!+\!\bar{\mathbf{H}}_1\!\mathbf{Q}_{s}\!\bar{\mathbf{H}}_1^H\!+\!{\mathbf{H}}_{tot_{1}}\!\mathbf{Q}_{r}\!{\mathbf{H}}^H_{tot_{1}}\!\right)= \!\sigma_t^2\! + \lambda_i\!\left(\!\bar{\mathbf{H}}_1\!\mathbf{Q}_{s}\!\bar{\mathbf{H}}_1^H\!+\!{\mathbf{H}}_{tot_{1}}\!\mathbf{Q}_{r}\!{\mathbf{H}}^H_{tot_{1}}\!\right)$. Hence, we obtain
\begin{align}
        &\log_2\frac{\big|\sigma_t^2\mathbf{I}_{K_{r}}+\hat{\mathbf{H}}_1\mathbf{Q}_{s}\hat{\mathbf{H}}_1^H+\bar{\mathbf{H}}_1\mathbf{Q}_{s}\bar{\mathbf{H}}_1^H+{\mathbf{H}}_{tot_{1}}\mathbf{Q}_{r}{\mathbf{H}}^H_{tot_{1}}\big|}{\big|\sigma_t^2\mathbf{I}_{K_{r}}+\bar{\mathbf{H}}_1\mathbf{Q}_{s}\bar{\mathbf{H}}_1^H+{\mathbf{H}}_{tot_{1}}\mathbf{Q}_{r}{\mathbf{H}}^H_{tot_{1}}\big|}
        \nonumber\geq \\
        &\sum_{i = 1}^{\min{(M,K_r)}}\! \log_2\! \left(\! 1+\! \frac{\lambda_i\left(\mathbf{H}_1\mathbf{Q}_{s}{\mathbf{H}_1}^H\right)}{\!\sigma_t^2\!+\!\lambda_i\!\left(\!\bar{\mathbf{H}}_1\!\mathbf{Q}_{s}\!\bar{\mathbf{H}}_1^H\!+\!{\mathbf{H}}_{tot_{1}}\!\mathbf{Q}_{r}\!{\mathbf{H}}^H_{tot_{1}}\!\right)} \right). \label{eq:InequalityAA}
\end{align}
Note that, the inequality holds with equality whenever $\mathbf{H}_1\mathbf{Q}_{s}{\mathbf{H}_1}^H$ and $\bar{\mathbf{H}}_1\!\mathbf{Q}_{s}\!\bar{\mathbf{H}}_1^H\!+\!{\mathbf{H}}_{tot_{1}}\!\mathbf{Q}_{r}\!{\mathbf{H}}^H_{tot_{1}}$ share a common basis. 
 Next, we use the following inequality
 \begin{align}
        &\sum_{i = 1}^{\min{(M,K_r)}}\! \log_2\! \left(\! 1+\! \frac{\lambda_i\left(\mathbf{H}_1\mathbf{Q}_{s}{\mathbf{H}_1}^H\right)}{\!\sigma_t^2\!+\!\lambda_i\!\left(\!\bar{\mathbf{H}}_1\!\mathbf{Q}_{s}\!\bar{\mathbf{H}}_1^H\!+\!{\mathbf{H}}_{tot_{1}}\!\mathbf{Q}_{r}\!{\mathbf{H}}^H_{tot_{1}}\!\right)} \right)
        \nonumber\geq \\
        &\sum_{i = 1}^{\min{(M,K_r)}}\! \log_2\! \left(\! 1+\! \frac{\lambda_i\left(\mathbf{H}_1\mathbf{Q}_{s}{\mathbf{H}_1}^H\right)}{\!\sigma_t^2\!+T_1 P_s +  \lambda_i\!\left(\!{\mathbf{H}}_{tot_{1}}\!\mathbf{Q}_{r}\!{\mathbf{H}}^H_{tot_{1}}\!\right)} \right). \label{eq:InequalityAB}
\end{align}
The above inequality holds true since $\lambda_i(\bar{\mathbf{H}}_1Q_s\bar{\mathbf{H}}_1^H) \leq T_1 P_s$.

Now, instead of doing the minimization over the left-hand side (LHS) of equation~\eqref{eq:InequalityAA}, we can first minimize the right-hand side (RHS) of equation~\eqref{eq:InequalityAB}  to find an achievable rate. 
Similarly, for $R_{\mathrm{rd}}^{\mathrm{FD}}$ we have
\begin{align}
        &\log_2\frac{\big|\sigma_d^2\mathbf{I}_{N}+\hat{\mathbf{H}}_2\mathbf{Q}_{r}\hat{\mathbf{H}}^H_2+\bar{\mathbf{H}}_2\mathbf{Q}_{r}\bar{\mathbf{H}}^H_2\big|}{\big|\sigma_d^2\mathbf{I}_{N}+\bar{\mathbf{H}}_2\mathbf{Q}_{r}\bar{\mathbf{H}}^H_2\big|}
        \geq \sum_{i = 1}^{\min{(M,K_r)}}\! \log_2\! \left(\! 1+\! \frac{\lambda_i\left(\mathbf{H}_2\mathbf{Q}_{r}{\mathbf{H}_2}^H\right)}{\!\sigma_d^2+T_2 P_r} \right). \label{eq:InequalityAAAA}
\end{align}
\begin{remark}\label{Remark1}
Having the equality $\mathbf{C}=\bar{\mathbf{H}}_{r}\mathbf{Q}_{r}\bar{\mathbf{H}}^H_{r}$, one can generally conclude the rule of multiplication is determinant i. e. $\det \left(\mathbf{C}\right)=\det\left(\bar{\mathbf{H}}^H_{r}\bar{\mathbf{H}}_{r}\right)\det\left(\mathbf{Q}_{r}\right)$. Further, using the properties of determinants we can also conclude $\prod_{i=1}^N\lambda_i\left(\mathbf{C}\right)=\prod_{i=1}^N\left(\lambda_{\rho\left(i\right)}\left(\bar{\mathbf{H}}^H_{r}\bar{\mathbf{H}}_{r}\right)\lambda_i\left(\mathbf{Q}_{r}\right)\right)$ where ${\rho\left(i\right)}$ is a random permutation of $i$ and indicates that there is no need for $\lambda_{\rho\left(i\right)}\left(\bar{\mathbf{H}}^H_{r}\bar{\mathbf{H}}_{r}\right)$ to be in decreasing order. However, one cannot generally conclude $\lambda_i\left(\mathbf{C}\right)=\lambda_{\rho\left(i\right)}\left(\bar{\mathbf{H}}^H_{r}\bar{\mathbf{H}}_{r}\right)\lambda_i\left(\mathbf{Q}_{r}\right), \forall i$, unless $\bar{\mathbf{H}}^H_{r}\bar{\mathbf{H}}_{r}$ and $\mathbf{Q}_{r}$ share common basis.
\end{remark}
As a result of Remark~\ref{Remark1}, in general, we cannot rewrite~\eqref{eq:InequalityAA} in terms of $\lambda_i\left(\bar{\mathbf{H}}^H_{r}\bar{\mathbf{H}}_{r}\right)$, $\lambda_i\left(\mathbf{Q}_{r}\right)$, $\lambda_i\left({\mathbf{H}}^H_\mathrm{1}{\mathbf{H}}_\mathrm{1}\right)$ and $\lambda_i\left(\mathbf{Q}_{s}\right)$. However, if we show that for every choice of $\mathbf{Q}_{s}$, there exists a matrix $\mathbf{Q}'_{s}$ with properties: 1) $\lambda_i({\mathbf{H}}_\mathrm{1}\mathbf{Q}_{s}{\mathbf{H}}^H_\mathrm{1})=\lambda_i({\mathbf{H}}_\mathrm{1}\mathbf{Q}'_{s}{\mathbf{H}}^H_\mathrm{1})$; 2) $\lambda_i({\mathbf{H}}_\mathrm{1}\mathbf{Q}'_{s}{\mathbf{H}}^H_\mathrm{1}) = \lambda_i(\mathbf{Q}'_{s})\lambda_i({\mathbf{H}}^H_\mathrm{1}{\mathbf{H}}_\mathrm{1})$ and 3) $\text{Tr}(\mathbf{Q}'_{s}) \leq \text{Tr}(\mathbf{Q}_{s})$; then we can use $\mathbf{Q}'_{s}$ instead and rewrite ~\eqref{eq:InequalityAA} in terms of $\lambda_i\left({\mathbf{H}}^H_\mathrm{1}{\mathbf{H}}_\mathrm{1}\right)$ and $\lambda_i\left(\mathbf{Q}'_{s}\right)$ to simplify the problem. The first property implies that both $\mathbf{Q}_{s}$ and $\mathbf{Q}'_{s}$ have the exact same impact on the capacity. Hence, if we find  a $\mathbf{Q}_{s}$ which is the solution to the problem~\eqref{PP:FDaa}, its corresponding $\mathbf{Q}'_{s}$ will also be a solution. The second property means, unlike $\mathbf{Q}_{s}$, $\mathbf{Q}'_{s}$ actually shares the common basis with ${\mathbf{H}}^H_\mathrm{1}{\mathbf{H}}_\mathrm{1}$. The last property implies that $\mathbf{Q}'_{s}$ is at least as good as $\mathbf{Q}_{s}$ in terms of power consumption. Observe that if we show for every feasible $\mathbf{Q}_{s}$ there exists at least one such $\mathbf{Q}'_{s}$, then we can solve the problem~\eqref{PP:FDaa} in a much easier way. The reason is, in such a case, instead of searching for optimal $\mathbf{Q}_{s}$ over the whole feasible set, we can search for the optimal $\mathbf{Q}'_{s}$. Unlike $\mathbf{Q}_{s}$, finding $\mathbf{Q}'_{s}$ does not need a complete search over the whole feasible set since $\mathbf{Q}'_{s}$ shares a common basis with ${\mathbf{H}}^H_\mathrm{1}{\mathbf{H}}_\mathrm{1}$. Therefore, we can limit our search only to the portion of the feasible set in which the matrices have eigendirections identical to those of ${\mathbf{H}}^H_\mathrm{1}{\mathbf{H}}_\mathrm{1}$. Similarly, if we show for every choice of $\bar{\mathbf{H}}_{r}$, there exist at least one $\bar{\mathbf{H}}'_{r}$ for which we have three conditions $\lambda_i(\bar{\mathbf{H}}_{r}\mathbf{Q}_{r}\bar{\mathbf{H}}^H_{r})=\lambda_i(\bar{\mathbf{H}}'_{r}\mathbf{Q}_{r}\bar{\mathbf{H}}'^H_{r})$, $\lambda_i(\bar{\mathbf{H}}_{r}\mathbf{Q}_{r}\bar{\mathbf{H}}'^H_{r}) = \lambda_i(\mathbf{Q}_{r})\lambda_i(\bar{\mathbf{H}}'^H_\mathrm{1}\bar{\mathbf{H}}'_\mathrm{1})$ and $\text{Tr}(\bar{\mathbf{H}}'^H_\mathrm{1}\bar{\mathbf{H}}'_\mathrm{1}) \leq \text{Tr}(\bar{\mathbf{H}}^H_\mathrm{1}\bar{\mathbf{H}}_\mathrm{1})$, we can simplify our search to finding $\bar{\mathbf{H}}'_{r}$ instead of $\bar{\mathbf{H}}_{r}$. In the next theorem, we show that such $\mathbf{Q}'_{s}$ and $\bar{\mathbf{H}}'_{r}$ exist.


\begin{theorem}\label{Theorem:1}
For every matrices $\mathbf{Q}_{s}$ and ${\mathbf{H}}_\mathrm{1}$, there exists at least one matrix $\mathbf{Q}'_{s}$ that satisfies the following conditions, 
\begin{align}
    \lambda_i \big( \mathbf{H}_1\mathbf{Q}_{s}\mathbf{H}^H_1 \big) &= \lambda_i \big( \mathbf{H}_1\mathbf{Q}'_{s}\mathbf{H}^H_1 \big),\label{P:newo1a}\\ 
    \lambda_i({\mathbf{H}}_\mathrm{1}\mathbf{Q}'_{s}{\mathbf{H}}^H_\mathrm{1}) &= \lambda_{\rho\left(i\right)}(\mathbf{Q}'_{s})\lambda_i({\mathbf{H}}^H_\mathrm{1}{\mathbf{H}}_\mathrm{1}),\label{P:newo1b}\\
    \text{Tr}(\mathbf{Q}'_{s}) &\leq \text{Tr}(\mathbf{Q}_{s}),\label{P:newo1c}
\end{align} \label{P:newo1}
where ${\rho\left(i\right)}$ is a random permutation of $i$ and indicates that there is no need for $\lambda_{\rho\left(i\right)}(\mathbf{Q}'_{s})$ to be in decreasing order.
\begin{proof}
The proof is given in Appendix II.
\end{proof}
\end{theorem}
For the sake of simplicity, we use following notions for the rest of the paper,
\begin{align}
& \gamma_{s_i} = \lambda_i(\mathbf{Q}_{s}),\\
&\gamma_{r_i} = \lambda_i(\mathbf{Q}_{r}),\\
& \sigma^2_{1_i} = \lambda_i({\mathbf{H}}_\mathrm{1}{\mathbf{H}}^H_\mathrm{1}),\\
& \sigma^2_{r_i} = \lambda_i(\bar{\mathbf{H}}_{tot_1}\bar{\mathbf{H}}^H_{tot_1}),\\
& \sigma^2_{2_i} = \lambda_i({\mathbf{H}}_\mathrm{2}{\mathbf{H}}^H_\mathrm{2}).
\end{align}
Now, using Theorem~\ref{Theorem:1} alongside Lemma~\ref{lem:1}, we infer that with no loss of generality, instead of optimising over matrices, one can do the optimization over eigenvalues to find the optimal value for RSH of~\eqref{eq:InequalityAB}. Then we have
\begin{subequations}\label{P:FDbbb}
\begin{align}
\max_{\boldsymbol{\gamma_{s}},\boldsymbol{\gamma_{r}}}\  \min_{\boldsymbol{\sigma_{r}}}\ & \min\!\Bigg(\! \sum_{i=1}^{\min(M,K_{r})} \!\log{\!\left(\!1+\!
\frac{\sigma_{1_i}^2\gamma_{{s}_{{\rho{\left(i\right)}}}}}{\sigma_t^2\!+T_1 P_s +\gamma_{{r}_i}\sigma_{{r}_{{\rho{\left(i\right)}}}}^2}\right)},\nonumber\\ &{\hspace*{1cm}\sum_{i=1}^{{\min(K_{\mathrm{t}},N)}} \!\log{\!\Big(1+\frac{\sigma^2_{2_i} \gamma_{r_i}}{\sigma_d^2+T_2 P_r }\Big)} \Bigg)}\tag{\ref{P:FDbbb}}\\
\text{subject to}\quad & \|\boldsymbol{\gamma}_{s}\|_1\leq P_{s},\label{34A}\\
&\|\boldsymbol{\gamma}_{r}\|_1 \leq P_{r},\label{34B}\\
&\|\boldsymbol{\sigma}^2_{r}\|_1\leq T',\label{34C}\\
& \sigma_{1_i}^2\gamma_{{s}_{{\rho{\left(i\right)}}}} \geq \sigma_{1_{i+1}}^2\gamma_{{s}_{{\rho{\left(i+1\right)}}}},\ \forall i \leq \min(M,K_{{r}}),\label{P:FDb.ConsA}\\
& \gamma_{{r}_i}\sigma_{{r}_{{\rho{\left(i\right)}}}}^2 \geq \gamma_{{r}_{i+1}}\sigma_{{r}_{{\rho{\left(i+1\right)}}}}^2,\ {\forall i \leq \min(K_{\mathrm{t}},N).}\label{P:FDb.ConsB}
\end{align}
\end{subequations}
Note that, the two additional constraints~\eqref{P:FDb.ConsA} and~\eqref{P:FDb.ConsB} need to be satisfied due to the conditions of Lemma~\ref{lem:1} (i.e. eigenvalues have to be in decreasing order). 
Interestingly, these two additional constraints are affine. The above optimization problem can further be simplified using the following lemma,
\begin{lemma}\label{lemm2}
The objective function of the optimization problem~\eqref{P:FDbbb} is optimized when the constraints\eqref{34A} and~\eqref{34C} are satisfied with equality.
\end{lemma}
\begin{proof}
 Intuitively, as the objective function is an increasing and decreasing function of each element of $\mathbf{\gamma}_s$ and $\mathbf{\sigma}^2_r$ respectively, at convergence, the constraints are met with equality. See Appendix III for the proof.
\end{proof}
%
%
%
%
%
\begin{algorithm}
\caption{Robust Transceiver Design}
\begin{algorithmic}[1]

\State Define $U = P_{r}$, $L = 0$, $\bar{P}_{r}^{(1)}=\frac{P_{r}}{2}$
\While {$|U-L|$ is large}
\State Determine $\boldsymbol\gamma^{}_{r}=[\tau^{}_{r}-\frac{1}{\boldsymbol{\sigma}^2_2}]^{+}$, s.t. $\|\boldsymbol\gamma^{}_{r}\|_1=\bar{P}_{r}^{}$
\State Define $\boldsymbol{\sigma}_{r}^{(0)^2} = \boldsymbol{0}$ and $\boldsymbol{\sigma}_{r}^{(1)^2} = \boldsymbol{1}$ and $q=0$
\While {$\|\boldsymbol{\sigma}^{(q)^2}_{r} - \boldsymbol{\sigma}^{(q-1)^2}_{r}\|_1$ is large}

\State Obtain $\boldsymbol{\sigma}^{(q)^2}_{r}$, using equation $38$
\State Obtain $\boldsymbol\gamma^{(q)}_{s}$, using Algorithm 2
\State $q=q+1$
\EndWhile
\State Calculate $R^{}_\mathrm{sr}$ and $R^{}_\mathrm{rd}$
\If{$R^{}_\mathrm{sr}>R^{}_\mathrm{rd}$}
\State $U = \bar{P}_{r}$
\ElsIf{$R^{}_\mathrm{sr}<R^{}_\mathrm{rd}$}
\State $L = \bar{P}_{r}$
\EndIf
\State $\bar{P}_{r} = \frac{U+L}{2}$
\EndWhile 
\end{algorithmic}
\label{alg:MaterialCharacterization}
\end{algorithm}

\color{black}
\begin{algorithm}
\caption{The optimal $\boldsymbol{\gamma}_{s}$}
\begin{algorithmic}[1]
\State Find power allocation $\boldsymbol{P}^0$ using water-filling algorithm 
\While {$|\boldsymbol{P}^{(q)}-\boldsymbol{P}^{(q-1)}|$ is large}
\State Define $\text{temp}=0$
\For{$i$}
\State Calculate $\text{cap}_i = \min_{1\leq i' \leq i-1}\{\sigma_{1_{i'}}^2\gamma_{{s}_{\rho{\left(i'\right)}}}\}/\sigma_{1_{i}}^2$
\If{$P_i>\text{cap}_i$}
\State $P_i=\text{cap}_i$
\State $\text{temp}=\text{temp}+P_i-\text{cap}_i$
\EndIf
\EndFor
\State $\boldsymbol{P}=\boldsymbol{P}+\frac{\text{temp}}{\text{number of channels}}$
\EndWhile 
\end{algorithmic}
\label{alg:MaterialCharacterization}
\end{algorithm}
\subsection{Algorithm Description}
Now, we need to solve the optimization problem~\eqref{P:FDbbb}. It can be readily shown that $R_{\text{rd}}^{\text{FD}}$ is a monotonically increasing function of $P_r$. Furthermore, one can show that $R_{\text{sr}}^{\text{FD}}$ is an increasing function w.r.t. $P_s$ and decreasing function w.r.t. $T'$ and $P_r$ (See Appendix IV). Consequently, the worst-case RSI chooses a strategy to reduce the spectral efficiency, while the relay and the source cope with such strategy for improving the system robustness. That means, on one hand the RSI hurts the stronger eigendirections of the received signal space more than the weaker ones. However, on the other hand the source tries to cope with this strategy adaptively by smart eigen selection.
This process clearly makes the optimization problem complicated at the source-relay hop. Unlike the source-relay hop, the resource allocation problem at the relay-receiver hop is rather easy. Since at the relay-receiver hop there is only one maximization and we can find the sum capacity simply by using the well-known water-filling algorithm.

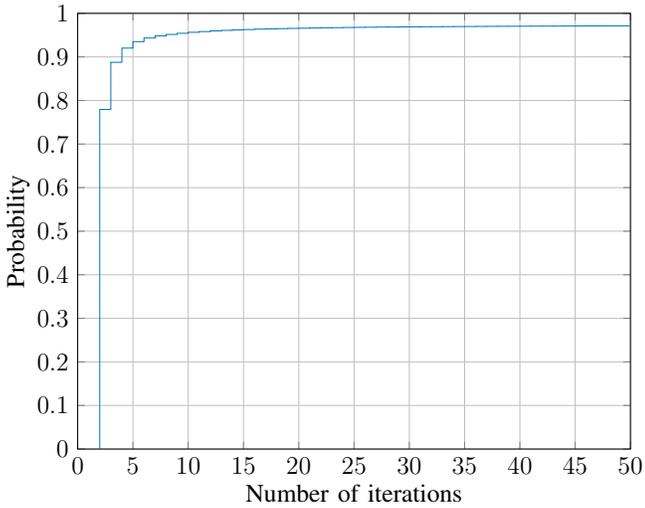
\begin{figure}
\hspace*{-.6cm}
\tikzset{every picture/.style={scale=.8}, every node/.style={scale=1}}%
\input{histog}

\caption{Cumulative distribution function (cdf) of iterations when $P_{s}=5$ and $P_\mathrm{r}=1$ and $M = K_\mathrm{t} = K_\mathrm{r} = N = 10$. The maximum number of iterations is set to be $50$. Cases that took more than $50$ iterations to converge are considered to be divergent.}
\label{fig:histo}
\end{figure}

Observe that although finding each of $R_{\text{sr}}^{\text{FD}}$ and $R_{\text{rd}}^{\text{FD}}$ separately is a convex problem, the problem~\eqref{P:FDbbb} as a whole is not convex. Therefore in this paper, we find the optimal $R_{\text{sr}}^{\text{FD}}$ by keeping $R_{\text{rd}}^{\text{FD}}$ fixed. Then we use the resulted $R_{\text{sr}}^{\text{FD}}$ to find optimal $R_{\text{rd}}^{\text{FD}}$ and again, using the new resulted $R_{\text{rd}}^{\text{FD}}$ to find optimal $R_{\text{sr}}^{\text{FD}}$. This iterative process repeats until the convergence. Our simulation showed that the algorithm has a very fast convergence and only in rare cases it takes more than 20 iterations for the algorithm to converge. This is mainly due to the fact that inequalities \eqref{P:FDb.ConsA} and \eqref{P:FDb.ConsB} restrict the eigenvalues to vary up to a certain limit, which in turn, makes the whole outputs more stable. Fig~\ref{fig:histo} depicts a typical histogram of iterations. As it can be seen, only less than $3\%$ of cases did not converge until $50$ iterations. 

Notice that the optimum values for the transmission power on relay hop may not sum to ${P}_{r}$. The reason is that $R^{\mathrm{FD}}_\mathrm{sr}$ is a monotonically decreasing function of ${P}_{r}$ and as we are interested in the $\min (R^{\mathrm{FD}}_\mathrm{sr},R^{\mathrm{FD}}_\mathrm{rd})$, with $R^{\mathrm{FD}}_\mathrm{sr}<R^{\mathrm{FD}}_\mathrm{rd}$ we will have $\min (R^{\mathrm{FD}}_\mathrm{sr},R^{\mathrm{FD}}_\mathrm{rd}) = R^{\mathrm{FD}}_\mathrm{sr}$. Therefore, it is in our interest to keep ${P}_{r}$ as low as possible to increase $R^{\mathrm{FD}}_\mathrm{sr}$ as much as possible. Analogously, in the case of $R^{\mathrm{FD}}_\mathrm{sr}>R^{\mathrm{FD}}_\mathrm{rd}$ we have $\min (R^{\mathrm{FD}}_\mathrm{sr},R^{\mathrm{FD}}_\mathrm{rd}) = R^{\mathrm{FD}}_\mathrm{rd}$ which can be increased by increasing the total power usage of relay's transmitter. As a result the well-known bisection method can be used to find the optimal rate where we have $R^{\mathrm{FD}}_\mathrm{sr}=R^{\mathrm{FD}}_\mathrm{rd}$, unless the case $R^{\mathrm{FD}}_\mathrm{sr}\geq R^{\mathrm{FD}}_\mathrm{rd}$ happens even if the maximum allowed power is used at the relay transmitter. In such a case, the relay-destination link becomes the bottleneck.

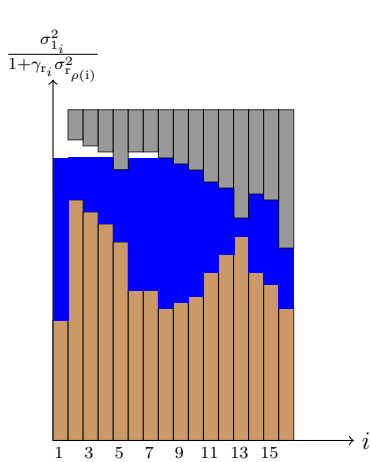
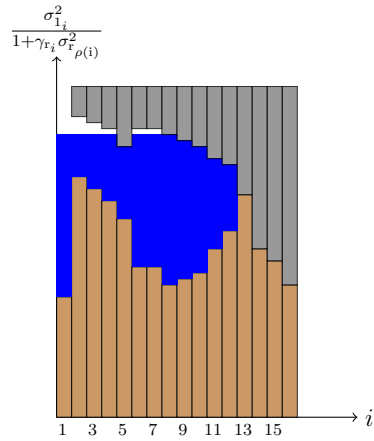
\begin{figure*}
\centering
\begin{minipage}{0.4\textwidth}
\subfigure[No subchannel with optimum power equal to zero. Note that due to the power cap constraint~\eqref{P:FDb.ConsA}, water has not the same level for all subchannels.]{
\tikzset{every picture/.style={scale=.8}, every node/.style={scale=0.8}}%
\input{WaterLevel}
}
\end{minipage}\quad\quad\quad\quad\quad\quad
\begin{minipage}{0.4\textwidth}
\subfigure[Subchannel $i=13$ gets the optimum of zero as its input power. Notice that in this case, due to the additional power cap constraint~\eqref{P:FDb.ConsA} all the remaining subchannels $i>13$ also get zero power.]{
\tikzset{every picture/.style={scale=.8}, every node/.style={scale=0.8}}%
\input{WaterLevelB}
}
\end{minipage}
\caption{Examples of multilevel water-filling for two different cases}
\label{fig:WaterLevel}
\end{figure*}

Now we focus on how to find $R^\mathrm{FD}_\mathrm{sr}$. In order to find the sum rate for the source-relay hop, we assume that we are already given $\boldsymbol{\gamma}^\star_{r}$ which is the vector of relay input powers that maximizes the sum rate at relay-destination hop. The next step is to do the minimization over $\boldsymbol{\sigma}_{r}$ and the maximization over $\boldsymbol{\gamma}_{s}$. One approach to solve this problem is to solve it iteratively. With this method, first one finds the optimal $\boldsymbol{\gamma}_{s}$ by solving the maximization part of \eqref{P:FDbbb} under the assumption that the optimal $\boldsymbol{\sigma}_{r}$ is given, and then, having the optimal $\boldsymbol{\gamma}_{s}$ the minimization part of \eqref{P:FDbbb} can be solved efficiently. This process goes on until the convergence of $\boldsymbol{\gamma}_{s}$ and/or $\boldsymbol{\sigma}_{r}$. The maximization part is performed using water-filling method. However, the additional conditions $\forall \ i \leq \min(M,K_{{r}}), \sigma_{1_i}^2\gamma_{{s}_{\rho{\left(i\right)}}} \geq \sigma_{1_{i+1}}^2\gamma_{{s}_{\rho(i+1)}}$ should be taken into account. For instance, if the optimal value for $\gamma_{s_i}$ turns out to be equal to zero, then we should have $\gamma_{s_j}=0$ for all $j>i$ irrespective of their SNR. Fig. \ref{fig:WaterLevel} depicts two different examples of multi-level water-filling algorithm. As it can be seen, first, regular water-filling algorithm is considered where for each subchannel we have $\frac{\sigma^2_{1_i}}{1+\gamma_{r_i}\sigma^2_{r_{\rho(i)}}}$ as its channel gain. After finding the water-level in this way, we need to impose $\gamma_{{s}_{\rho(i+1)}} \leq \frac{\min_{1\leq i' \leq i}\{\sigma_{1_{i'}}^2\gamma_{{s}_{\rho{\left(i'\right)}}}\}}{\sigma_{1_{i+1}}^2}$. These additional restrictions act like caps on top of the water and create multilevel water-filling which can be interpreted as a cave. Fig. \ref{fig:WaterLevel} (a) shows the case where these caps do not make any subchannel to have zero power. However, Fig. \ref{fig:WaterLevel} (b) shows the case where subchannel $i=13$ has to be zero as a result of the cap imposed by the additional constraints \eqref{P:FDb.ConsA}. In this case we have $\gamma_{{s}_{\rho(13)}}=0$ and as a result $\min_{1\leq i' \leq 13}\{\sigma_{1_{i'}}^2\gamma_{{s}_{\rho{\left(i'\right)}}}\}=0$. Thus, this condition forces all other subchannels (i.e. $i > 13$) to get no power. Algorithm 2 provides the detail of multilevel water-filling. For the minimization part, Lagrangian multiplier is used. We have
\begin{align}
    L = &\sum_{i=1}^{\min(M,K_{r})} \!\log_2{\!\left(1+
\frac{\sigma_{1_i}^2\gamma_{{s}_{\rho{\left(i\right)}}}}{\sigma^2_{t}+T_1P_s+\gamma_{{r}_i}\sigma_{{r}_{\rho{\left(i\right)}}}^2}\right)} + \lambda \left( \sum_{i=0}^{N}\sigma^2_{r_i} - T_r\right).
\end{align}
Calculating $\frac{\partial L}{\partial \sigma^2_{r_i}}=0$ we arrive at
{\begin{align}
\sigma^2_{r_i}\! = \! \left[\frac{\sqrt{\left(\sigma^2_{1_i}\gamma_{s_i}\right)^2\!+\!\frac{4\sigma^2_{1_i}\gamma_{s_i}\gamma{r_i}}{\lambda}}\!-\!\sigma^2_{1_i}\gamma_{s_i}\!-\!2(\sigma^2_{t}\!+\!T_1P_s)}{2\gamma_{r_i}}\right]^+,\label{eq:condition}
\end{align}}
where $\lambda$ is the water level. Similarly to the maximization case, there are additional constraints $\gamma_{{r}_i}\sigma_{{s}_{\rho{\left(i\right)}}}^2 \geq \gamma_{{r}_{i+1}}\sigma_{{r}_{i+1}}^2$ that must be considered during the minimization process. However, it can be shown that if the constraints $\gamma_{r_{i}} \geq \gamma_{r_{i+1}}$ and $\sigma_{1_i}^2\gamma_{{s}_{\rho{\left(i\right)}}} \geq \sigma_{1_{i+1}}^2\gamma_{{s}_{i+1}}$ are met, then the constraint $\gamma_{{r}_i}\sigma_{{s}_{\rho{\left(i\right)}}}^2 \geq \gamma_{{r}_{i+1}}\sigma_{{r}_{i+1}}^2$ becomes redundant. Please refer to Appendix V for proof. 

Next we deal with the optimization for the cases where IRS is utilized to help either the source-relay or relay-destination channels. In such cases, the optimization part over the covariance matrices remains the same as the above-mentioned case. Also, the optimization of the IRS elements can be done using eigenvalue decomposition and the algorithm introduced in \cite{zhang2020capacity}. Notice that for the case in which IRS is assisting the source relay link, the term $T_1 P_s$ in \eqref{eq:InequalityAB} should be replaced with $(T_1+T_{SI}T_{IR})P_s$, and for the case where IRS helps the relay-destination link, the term $T_2 P_s$ in \eqref{eq:InequalityAAAA} should be replaced with $(T_2+T_{RI} T_{ID} )P_r$.

\section{Achievable Rate (Half-Duplex Relay)}
\Hossein{We consider a simple HD relay, where the source and the relay transmit in two subsequent time instances. Notice that for the case of HD, IRS can be used to assist both the source-relay and relay-destination channels as the singal is being sent over each of these channels in a different time slot. therefore, the received signals at the relay and the destination can respectively be expressd as}
\begin{align}
\mathbf{y}_\mathrm{r}&= \!\left(\hat{\mathbf{H}}_{1}\!+\!\hat{\mathbf{H}}_{IR}\mathbf{\Theta}\hat{\mathbf{H}}_\mathrm{SI}\right)\mathbf{x}_\mathrm{r}\!+\!\left(\bar{\mathbf{H}}_{1}\!+\!\bar{\mathbf{H}}_{IR}\mathbf{\Theta}\bar{\mathbf{H}}_\mathrm{SI}\right)\mathbf{x}_\mathrm{r}+\mathbf{n}_\mathrm{t},\\
\mathbf{y}_\mathrm{d}&= \!\left(\hat{\mathbf{H}}_{2}\!+\!\hat{\mathbf{H}}_{ID}\mathbf{\Theta}\hat{\mathbf{H}}_{RI}\right)\mathbf{x}_{r}\!+\!\left(\bar{\mathbf{H}}_{2}\!+\!\bar{\mathbf{H}}_{ID}\mathbf{\Theta}\bar{\mathbf{H}}_{RI}\right)\mathbf{x}_{r}+\mathbf{n}_\mathrm{d}.
\end{align}
Consequently, the achievable rates for the transmitter-relay and relay-destination links can be expressed as bellow 
\newpage
\begin{align}
R&^{\mathrm{HD}}_\mathrm{sr}\!=\!\log\Big|\mathbf{I}_{K_{r}}\!+\!\left(\hat{\mathbf{H}}_{1}\!+\!\hat{\mathbf{H}}_{IR}\mathbf{\Theta}\hat{\mathbf{H}}_\mathrm{SI}\right)\mathbf{Q}_{s}\left(\hat{\mathbf{H}}_{1}\!+\!\hat{\mathbf{H}}_{IR}\mathbf{\Theta}\hat{\mathbf{H}}_\mathrm{SI}\right)^H\nonumber\\
&\left(\sigma^2_t\mathbf{I}_{K_r}\!\!+\!\left(\bar{\mathbf{H}}_{1}\!+\!\bar{\mathbf{H}}_{IR}\mathbf{\Theta}\bar{\mathbf{H}}_\mathrm{SI}\right)\mathbf{Q}_{s}\left(\bar{\mathbf{H}}_{1}\!+\!\bar{\mathbf{H}}_{IR}\mathbf{\Theta}\bar{\mathbf{H}}_\mathrm{SI}\right)\!^H\!\right)^{-1}\!\Big|\\
&\geq \sum_{i = 1}^{\min{(M,K_r)}}\! \log_2\! \left(\! 1+\! \frac{\lambda_i\left(\mathbf{H}'_1\mathbf{Q}_{s}{\mathbf{H}'_1}^H\right)}{\!\sigma_t^2+T_1 P_s} \right),\\
R&^{\mathrm{HD}}_\mathrm{rd}\!=\!\log\Big|\mathbf{I}_{N}\!+\!\left(\hat{\mathbf{H}}_{2}\!+\!\hat{\mathbf{H}}_{ID}\mathbf{\Theta}\hat{\mathbf{H}}_\mathrm{RI}\right)\mathbf{Q}_{r}\left(\hat{\mathbf{H}}_{2}\!+\!\hat{\mathbf{H}}_{ID}\mathbf{\Theta}\hat{\mathbf{H}}_\mathrm{RI}\right)^H\nonumber\\
&\left(\sigma^2_d\mathbf{I}_{N}\!\!+\!\left(\bar{\mathbf{H}}_{2}\!+\!\bar{\mathbf{H}}_{ID}\mathbf{\Theta}\bar{\mathbf{H}}_\mathrm{RI}\right)\mathbf{Q}_{r}\left(\bar{\mathbf{H}}_{2}\!+\!\bar{\mathbf{H}}_{ID}\mathbf{\Theta}\bar{\mathbf{H}}_\mathrm{RI}\right)^H\right)^{-1}\Big|\\
&\geq \sum_{i = 1}^{\min{(M,K_r)}}\! \log_2\! \left(\! 1+\! \frac{\lambda_i\left(\mathbf{H}'_2\mathbf{Q}_{r}{\mathbf{H}'_2}^H\right)}{\!\sigma_d^2+T_2 P_r} \right). \label{eq:InequalityAAAA}
\end{align}
where $\mathbf{H}'_1=\hat{\mathbf{H}}_{1}\!+\!\hat{\mathbf{H}}_{IR}\mathbf{\Theta}\hat{\mathbf{H}}_\mathrm{SI}$ and $\mathbf{H}'_2=\bar{\mathbf{H}}_{2}\!+\!\bar{\mathbf{H}}_{ID}\mathbf{\Theta}\bar{\mathbf{H}}_\mathrm{RI}$. Also, $R^{\mathrm{HD}}_\mathrm{sr}$ and $R^{\mathrm{HD}}_\mathrm{rd}$ are the achievable rates on the source-relay and relay-destination links, respectively  Using time sharing, the achievable rate between the source and destination nodes is given by
\begin{align}
R^{\mathrm{HD}}=\min(\alpha R^{\mathrm{HD}}_\mathrm{sr},(1-\alpha)R^{\mathrm{HD}}_\mathrm{rd}),
\end{align}
where $\alpha$ is the time-sharing parameter. Note that, in half-duplex relaying the source and relay transmissions are conducted in separate channel uses. Hence, the transmit covariance matrices $\mathbf{Q}_{s}\in\mathbb{H}^{N_{t}\times N_{t}}$ and $\mathbf{Q}_{r}\in\mathbb{H}^{K_\mathrm{t}\times K_\mathrm{t}}$ are optimized by maximizing the achievable rate from the source to the destination. Here, the convex cone of Hermitian positive semidefinite matrices of dimensions $N_{t}\times N_{t}$ and $K_\mathrm{t}\times K_\mathrm{t}$ are represented by $\mathbb{H}^{N_{t}\times N_{t}}$ and $\mathbb{H}^{K_\mathrm{t}\times K_\mathrm{t}}$, respectively. Importantly, for maximizing this achievable rate, the time-sharing parameter, i.e., $\alpha$ needs to be optimized alongside the system parameters, e.g., power allocation. Readily, optimal $\alpha$ occurs at $\alpha R^{\mathrm{HD}}_\mathrm{sr}=(1-\alpha)R^{\mathrm{HD}}_\mathrm{rd}$. Therefore, the achievable rate becomes as follows
\begin{align}
R^{\mathrm{HD}}=\frac{R^{\mathrm{HD}}_\mathrm{sr}R^{\mathrm{HD}}_\mathrm{rd}}{R^{\mathrm{HD}}_\mathrm{sr}+R^{\mathrm{HD}}_\mathrm{rd}}.
\end{align}
\Hossein{Notice that as  the objective function of the above optimization problem is a monotonically increasing function of both $R_\mathrm{sr}^\mathrm{HD}$ and $R_\mathrm{rd}^\mathrm{HD}$, the problem can be simplified to maximizing each of $R^{\mathrm{HD}}_\mathrm{sr}$ and $R^{\mathrm{HD}}_\mathrm{rd}$ separately}.
\Hossein{Next, we provide the solution to the rate optimization problem when IRS is assisting the source-relay link. We have}
\begin{subequations}\label{P:HDa}
\begin{align}
\max_{\mathbf{Q}_{s}}\quad & \min_{\bar{\mathbf{H}}_{SI}, \bar{\mathbf{H}}_{1}, \bar{\mathbf{H}}_\mathrm{IR}}\quad R^{\mathrm{HD}}_\mathrm{sr}\\
\text{subject to}\quad\quad & \mathrm{Tr}(\mathbf{Q}_{s})\leq P_{s},\label{P:HDa:ConsA}\\ 
&\mathrm{Tr}(\bar{\mathbf{H}}_{1}\bar{\mathbf{H}}_{1}^H)\leq T_{1},\label{P:HDa:ConsB}\\ 
&\mathrm{Tr}(\bar{\mathbf{H}}_{SI}\bar{\mathbf{H}}_{SI}^H)\leq T_{SI},\label{P:HDa:ConsC}\\ 
&\mathrm{Tr}(\bar{\mathbf{H}}_{IR}\bar{\mathbf{H}}_{IR}^H)\leq T_{IR},\label{P:HDa:ConsD}
\end{align}
\end{subequations}
The above optimization problem follows the same approach applied for the optimization of relay-destination link in FD scenario. As a result, the same method could be applied to find it. In other words, the well-known water-filling algorithm can be used to find the optimal covariance matrices along with the algorithm introduced in \cite{zhang2020capacity} to find the best IRS pattern. This process continues iteratively until it finally converges. The solution to $R_{rd}^{HD}$ is the same as well and the same procedure can be applied to find $\mathbf{Q}_r$.
\section{Numerical Results}
We assume the transmit power budgets at the source and at the relay are $P_{s}=5$ and $P_{r}=1$ respectively. Moreover, the AWGN spectral density is assumed to be $-175$dBm and the bandwidth is $BW=180$ MHz. In this section, we investigate the performance of IRS-assisted full-duplex relaying with RSI channel uncertainty bound $T_r$, i.e.,  $\mathrm{Tr}(\bar{\mathbf{H}}_{r}\bar{\mathbf{H}}^H_{r})\leq T_r$. We consider all the channels to follow the Rician distribution with the factor $\epsilon = 0.1$ and the specificaiton given in Table~\ref{tab}. 
\begin{table}[H]
\centering
\caption{Simulation Parameters\label{tab}}
\begin{tabular}{c|c}
\hline Parameters & Values \\
\hline Transmitter location & $(0 \mathrm{~m}, 0 \mathrm{~m})$ \\
\hline IRS location & $(4000 \mathrm{~m}, 20 \mathrm{~m})$ \\
\hline Relay location & $(4000 \mathrm{~m}, 0 \mathrm{~m})$ \\
\hline Receiver location & $(8000 \mathrm{~m}, 0 \mathrm{~m})$ \\
\hline Path-loss & $32.6+36.7 \mathrm{log} (d)$ \\
\hline Transmission bandwidth $B$ & $180 \mathrm{Mb}$ \\
\hline
\end{tabular}
\end{table}
We also assume $T_x = 0.001, x \in \left\{1,2,SI,IR,RI,ID\right\}$. We perform Monte-Carlo simulations with $L=10^3$ realizations from random channels and noise vectors. Hence, the average worst-case throughput rate is defined as the average of worst-case rates for $L$ randomizations, i.e., 
$
R_{\mathrm{av}}=\frac{1}{L}\sum_{l=1}^{L}R_l.
$
Notice that, for each set of realizations, we solve the robust transceiver design as is elaborated in Algorithm 1. We run different sets of simulations as described in the following subsections.
\begin{figure*}
\centering
\begin{minipage}{0.5\textwidth}
\centering
\subfigure[$\{N_t,K_{r}+K_\mathrm{t},N_r\}=\{4,10,4\}$]{
\includegraphics[scale=.5]{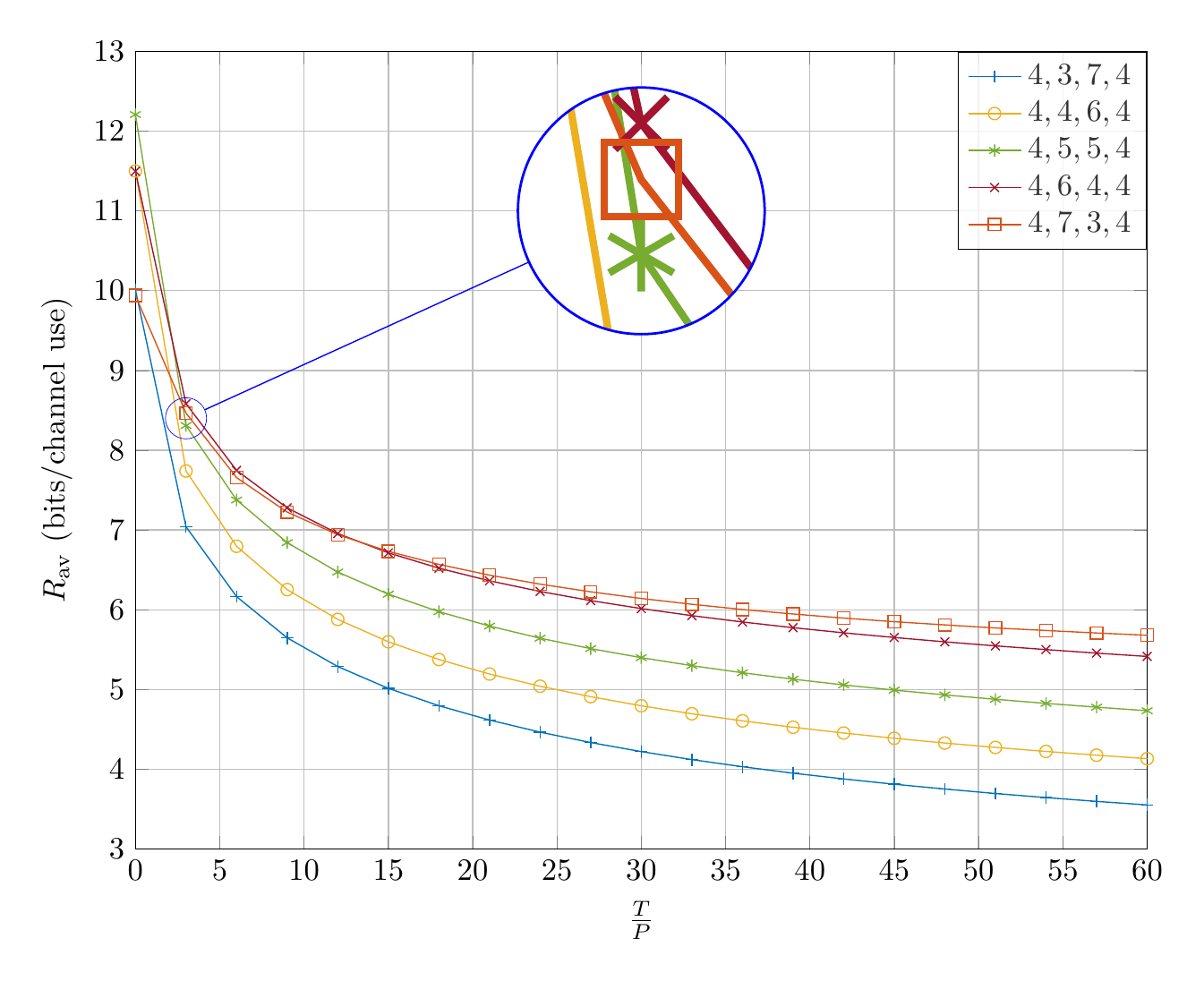}
\label{fig:HdVsFdE}
}
\end{minipage}%
\begin{minipage}{0.5\textwidth}
\centering
\subfigure[$\{N_t,K_{r}+K_\mathrm{t},N_r\}=\{10,24,10\}$]{
\includegraphics[scale=.5]{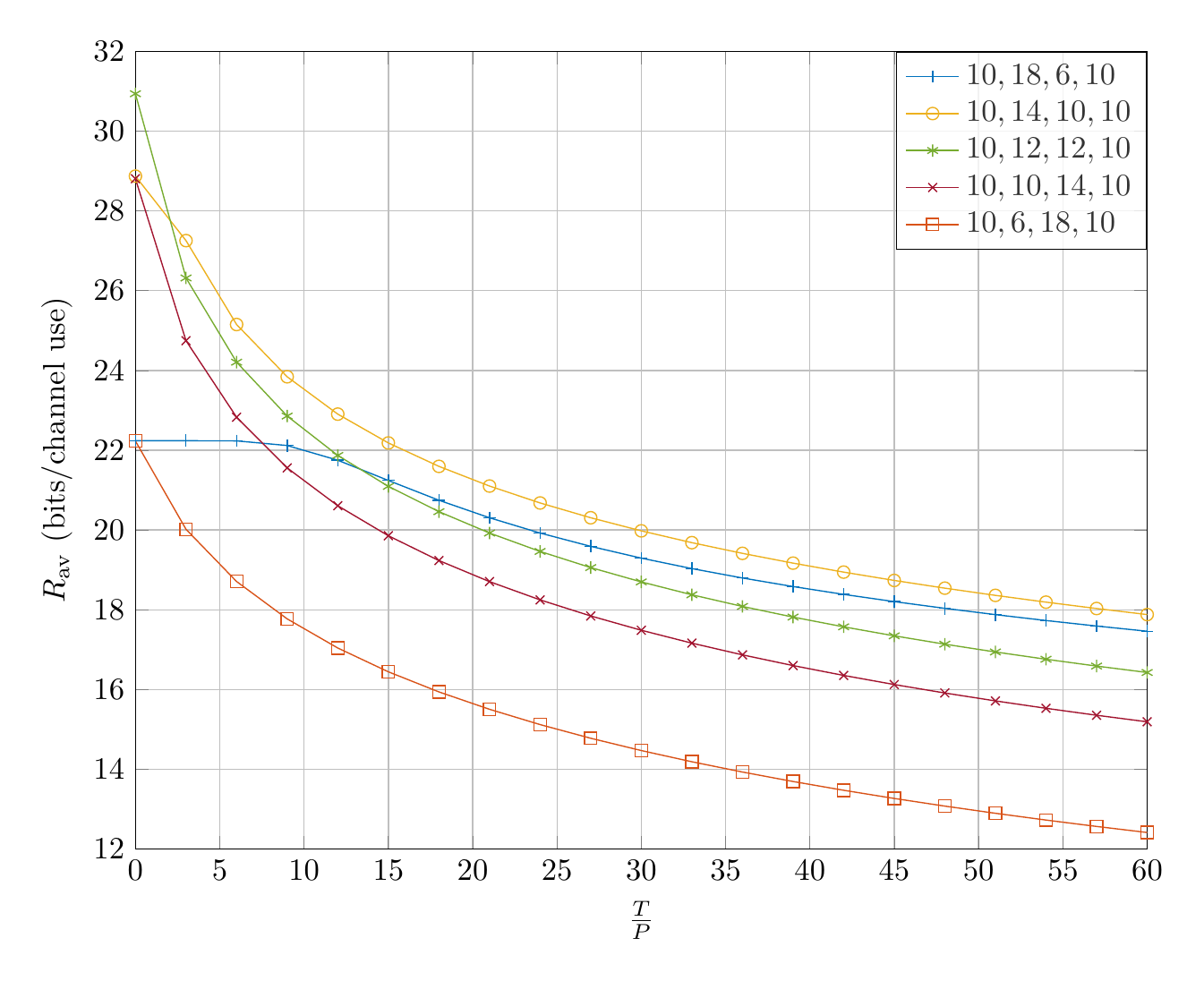}
\label{fig:HdVsFdE}
}
\end{minipage}
\caption{Sum rate throughput of the FD mode as a function of normalized interference $\frac{T}{P}$ with different number of antennas at the source, relay and destination. The transmit power budget at the source and the relay are assumed to be equal, i.e., $P_{s}=5$ and $P_\mathrm{r}=1$.}
\label{fig:HdVsFdBD1}
\end{figure*}
\subsection{Antenna Array Increment with no IRS}
\Hossein{In this part, first we assume that there is no IRS installed. Then we evaluate the performance of the system usign different strategies. Thereafter, we examine how installing an IRS can help increase the throughput. We consider two cases, where the source, relay and destination are equipped with (a)- small antenna array and (b)- large antenna arrays. In order to see the impact of IRS, we first assume that there is no IRS installed. For these cases, we have
\begin{enumerate}[(a)-]
\item $N_t=4,K_{r}+K_\mathrm{t}=10,N_r=4$,
\item $N_t=10,K_{r}+K_\mathrm{t}=24,N_r=10$.
\end{enumerate}
These cases are considered to highlight the performance of full-duplex DF relaying as a function of the number of antennas with the worst-case RSI. Interestingly, as the number of antennas at the source, relay and destination increase, full-duplex relaying achieves a higher throughput rate even with strong RSI. This can be seen by comparing rates from~\figurename{~\ref{fig:HdVsFdBD1}}(a) to  those from ~\figurename{~\ref{fig:HdVsFdBD1}}(b). Furthermore, notice that the worst-case RSI casts strong interference on the strong streams from the source to the destination. With very low RSI power $T_r\rightarrow 0$, full-duplex almost doubles the throughput rate compared to the half-duplex counterpart. This can be seen in~\figurename{~\ref{fig:HdVsFdBD1}}, where the curves have their intercept point with the vertical axis. However, as $T_r$ increases, the efficiency of full-duplex operation drops. It is worth noting that at low RSI power the DoF plays the most important role to have higher sum rate. For instance, consider ~\figurename{~\ref{fig:HdVsFdBD1}}(a), in which the cases $\text{FD=}{\{\text{4,5,5,4}\}}$, $\text{FD=}{\{\text{4,4,6,4}\}}$ and $\text{FD=}{\{\text{4,6,4,4}\}}$ have $\text{DoF}_{\text{total}}=4$  -- $\text{DoF}_{\text{total}}$ is the minimum of the DoF of source-relay and relay-destination channels, i.e., $\text{DoF}_{\text{total}}=\min{\left(\text{DoF}_{\text{sr}},\text{DoF}_{\text{rd}}\right)}$ -- while the cases $\text{FD=}{\{\text{4,7,3,4}\}}$ and $\text{FD=}{\{\text{4,3,7,4}\}}$ have $\text{DoF}_{\text{total}}=3$. At $T_r = 0$, there is a noticable gap between the first three cases and the last two ones, while the difference of the first three cases from each other is small. The big gap is due to the difference in $\text{DoF}_{\text{total}}$ and the small one is due to the difference in SNR. Similarly, in~\figurename{~\ref{fig:HdVsFdBD1}}(b) the three cases $\text{FD=}{\{\text{10,12,12,10}\}}$, $\text{FD=}{\{\text{10,10,14,10}\}}$ and $\text{FD=}{\{\text{10,10,14,10}\}}$ with $\text{DoF}_{\text{total}}=10$ have higher rates than the two cases $\text{FD=}{\{\text{10,6,18,10}\}}$ and $\text{FD=}{\{\text{10,18,6,10}\}}$ with $\text{DoF}_{\text{total}}=6$. Finally, it can be seen in both ~\figurename{~\ref{fig:HdVsFdBD1}}(a) and ~\figurename{~\ref{fig:HdVsFdBD1}}(b) that at $T_r = 0$ there is no difference between cases that have the same $\text{DoF}_{\text{total}}$ but different $\text{DoF}_{\text{sr}}$ and $\text{DoF}_{\text{rd}}$. 
As it can be seen in both~\figurename{~\ref{fig:HdVsFdBD1}}(a) and (b), for cases with $K_{t}>K_{r}$ the sum rate drops quickly as RSI increases. In fact, the more relative antennas at the relay transmitter compared to its receiver, the faster the sum rate drops with the rise in RSI. 
To understand this behaviour of the system better, again, consider case $\{\text{10,18,6,10}\}$ and also, suppose $T_r \rightarrow \infty$. As discussed before, we have $\text{DoF}_{\text{sr}}=10$ and $\text{DoF}_{\text{rd}}=6$. Moreover, we have $\text{DoF}_{\text{I}}=6$ for the interference channel ($\bar{\mathbf{H}}_\mathrm{r}$). Unlike the case with no interference, in this case the bottleneck is no longer the relay-destination link. This is due to the fact that interference can act to the detriment of some six of the source-relay subchannels. As we have $\text{DoF}_{\text{I}}=6$, interference can choose at most six independent subchannels and as we assumed $T_r \rightarrow \infty$, for those subchannels we get $\text{SINR} \rightarrow 0$. Therefore no information can be conveyed from those links and the bottleneck becomes the source-relay link with $4$ usable subchannels. It can be seen in ~\figurename{~\ref{fig:HdVsFdBD1}}(a) and (b) that  as $T_r$ increases, the cases with the same sum-rate at $T_r=0$ start to diverge because of the different characteristics of the interference they experience. We explain the effect of interference in the following subsection in more detail. }
\subsection{The Impact of IRS}
\begin{figure*}
\centering
\includegraphics[scale=1]{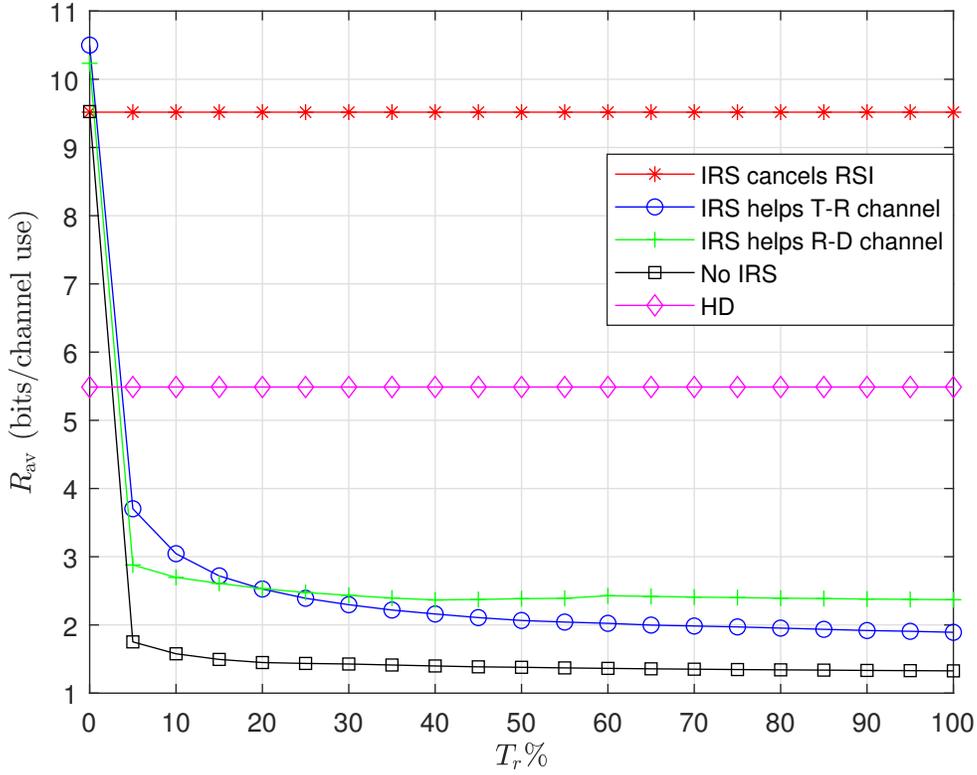}
\caption{Comparison of five different scenatios: HD with IRS, FD with no IRS, FD with IRS as RSI cancelator; FD with IRS to help transmitter-relay link; FD with IRS to help relay-destination link. We considered the case where $\{N_t,K_\mathrm{t},K_{r},N_r,M\}=\{4,5,5,4,100\}$.}
\label{lastfig}
\end{figure*}

In this part, we evaluate the impact of IRS on the throughput rate when it is used to performe different tasks.~\figurename{~\ref{lastfig}} shows the throughput for three different scenarios, namely, when IRS is used to help the transmitter-relay link, when it is applied to cancel RSI and when the IRS job is to help the realy-destination link and then the results are compared with two cases where the system is working in HD wiht IRS and the case where the system is working in FD with no IRS. It is also assumed that $\frac{T_r}{({\mathbf{H}}_{r}{\mathbf{H}}^H_{r})}= 75\%$ i. e. the system works at high RSI range. As it can be seen, the highst performance is achieved when the IRS is utilized to deal with the RSI. As a result, for the rest of the paper we use the IRS for this purpose. Also, it can be seen that when the number of

\begin{figure*}
\centering
\begin{minipage}{.5\textwidth}
\centering
\hspace{-.8cm}
\subfigure[$\{N_t,K_\mathrm{t},K_{r},N_r\}=\{4,5,5,4\}$]{
\tikzset{every picture/.style={scale=.7}, every node/.style={scale=0.8}}%
\input{xIRSvTsmall}
\label{fig:xIRSvTsmall}
}
\end{minipage}%
\begin{minipage}{.5\textwidth}
\centering
\hspace{-.8cm}
\subfigure[$\{N_t,K_\mathrm{t},K_{r},N_r\}=\{10,12,12,10\}$]{
\tikzset{every picture/.style={scale=.7}, every node/.style={scale=0.8}}%
\input{xIRSvTlarge}
\label{fig:xIRSvTlarge}
}
\end{minipage}
\caption{Sum rate throughput as a function of IRS elements. The transmit power budget at the source and the relay are assumed to be equal, i.e., $P_{s}=5$ and $P_\mathrm{r}=1$.}
\label{fig:xIRSvTsmallandlarge}
\end{figure*}
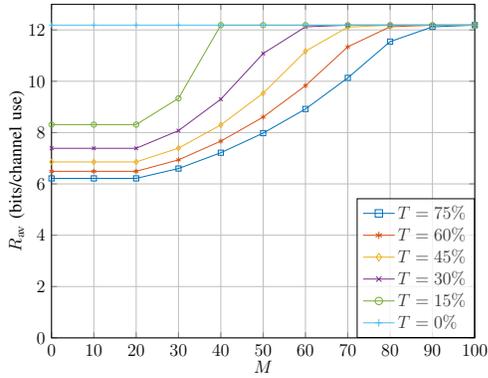
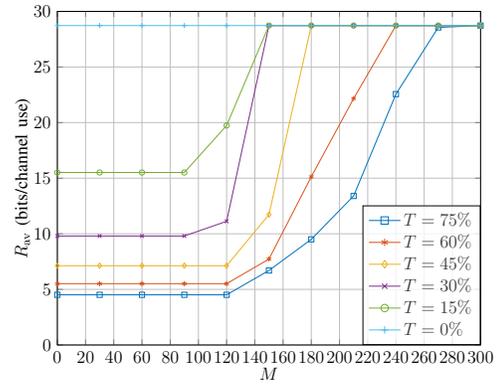
\Hossein{~\figurename{~\ref{fig:xIRSvTsmallandlarge}} shows the impact of IRS on the throughput. For~\figurename{~\ref{fig:xIRSvTsmall}} we considered the case $\{N_t,K_\mathrm{t},K_{r},N_r\}=\{4,5,5,4\}$ and for~\figurename{~\ref{fig:xIRSvTlarge}}, we considered $\{N_t,K_\mathrm{t},K_{r},N_r\}=\{10,12,12,10\}$. As shown in the figure, the number of IRS elements has a geat impact on RSI cancellation to the extend that having an IRS with $M=100$ and $M=300$ can cancel interference of $\frac{T_r}{\mathrm{Tr}({\mathbf{H}}_{r}{\mathbf{H}}^H_{r})}=0.75$ for $\{N_t,K_\mathrm{t},K_{r},N_r\}=\{4,5,5,4\}$, and $\{N_t,K_\mathrm{t},K_{r},N_r\}=\{10,12,12,10\}$ respectively. Further, it can also be seen in the figure that having IRS with 20 and 100 elements for the small and large antenna array cases respectively, is not helpful at all. This is mainly due to the fact that, unlike the average case, for the case of worst case scenario, the number of IRS elements should be at least as large as the dimension of $\bar{\mathbf{H}}_r$, otherwise, the the IRS feasible set cannot span into all dimensions of $\bar{\mathbf{H}}_r$. Therefore, there is always least one representation for $\bar{\mathbf{H}}_r$ in which IRS cannot do any RSI cancellation. Also, comparing two figures~\figurename{~\ref{fig:xIRSvTsmall}} and ~\figurename{~\ref{fig:xIRSvTlarge}}, one can conclude that, when the dimension of $\bar{\mathbf{H}}_r$ increases, the effort that IRS has to make in order to cancel RSI remarkably increases which is consistent with the previous statement.}


\begin{figure*}
\centering
\begin{minipage}{.5\textwidth}
\centering
\hspace{-.8cm}
\subfigure[$\{N_t,K_\mathrm{t}+K_{r},N_r\}=\{4,10,4\}$ with no IRS.]{
\tikzset{every picture/.style={scale=.7}, every node/.style={scale=0.8}}%
\input{CwrtT2}
\label{fig:CwrtT}
}
\end{minipage}%
\begin{minipage}{.5\textwidth}
\centering
\hspace{-.8cm}
\subfigure[$\{N_t,K_\mathrm{t}+K_{r},N_r\}=\{4,10,4\}$ with $M = 60$.]{
\tikzset{every picture/.style={scale=.7}, every node/.style={scale=0.8}}%
\input{CwrtT}
\label{fig:CwrtT1}
}
\end{minipage}
\caption{Sum rate throughput as a function of relay receiver antennas $K_r$ with and without RSI. The transmit power budget at the source and the relay are assumed to be equal, i.e., $P_{s}=5$ and $P_\mathrm
r=1$.}
\label{fig:HdVsFdBD}
\end{figure*}
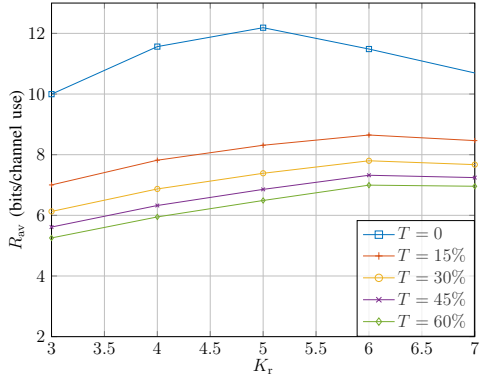
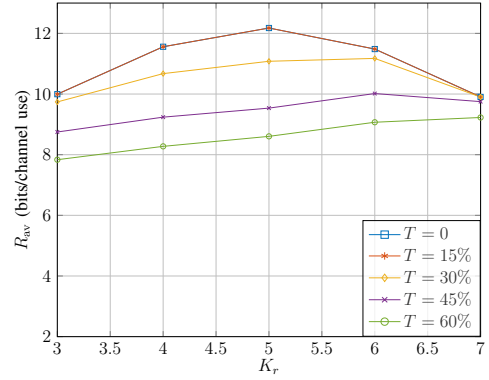
\subsection{Relay Tx/Rx Antenna allocation}
\Hossein{Suppose that the relay has $K_\mathrm{t}+K_{r}=8$ in total. Furthermore, following cases in which the number of antenna at the source and destination are $\{N_t,N_r\}=\{4,4\}$. The question is, from $8$ antennas at the relay, how many should be used for reception for the robust design?}
\Hossein{~\figurename{~\ref{fig:HdVsFdBD}} shows the sum rate as a function of $K_{r}$ for different values of $T$ where there is no IRS and there is an IRS with $M=60$ elements respectively. As it can be seen, by using more antennas for reception than for transmission, i.e., $K_{r}>K_\mathrm{t}$, at the relay, the throughput rate is maximized. This is due to the fact that, increasing the signal-to-noise ratio (SNR) of the source-relay streams enhances the overall throughput rate more than increasing the number of antennas for transmission in order to enhance the $\mathrm{DoF}$ of the relay-destination link. Furthermore, notice that in this setup the overall $\mathrm{DoF}$ from the source to destination is limited by the $\mathrm{DoF}$ of the source-relay link, i.e., the bottleneck is in the first hop.}

\Hossein{By comparing two scenarios we see that having an IRS not only improves the rates in all cases, but also it mat change the best antenna allocation. For instance, for the case of $T = 15\%$, it is best to have 6 antennas at the relay receiver and 4 antennas at the relay transmitter. However, after establishing the IRS, the best antenna allocation changes to 5 antennas at each end.
For instance, the results show that although the $\text{DoF}_\text{total}$ for both $\{N_t,K_\mathrm{t},K_{r},N_r\}=\{4,3,5,4\}$ and $\{N_t,K_\mathrm{t},K_{r},N_r\}=\{4,5,3,4\}$ is 3, the sum rate capacity of the latter is much better than that of the former at high interference. This is because of the fact when $\text{DoF}_\mathrm{sr}>\text{DoF}_\mathrm{rd}$, the source relay link enjoys $\text{DoF}_\mathrm{sr}-\text{DoF}_\mathrm{rd}$ subchannels with no interference. Therefore, the source can manage to obtain higher sum rate by choosing its power allocation wisely. However, in the case of $\mathrm{DoF}_\mathrm{sr}\leq\mathrm{DoF}_\mathrm{rd}$, no matter how well the power allocation is done, all sub channels suffer from interference at the source-relay end.}
\subsection{Full-Duplex vs Half-Duplex}
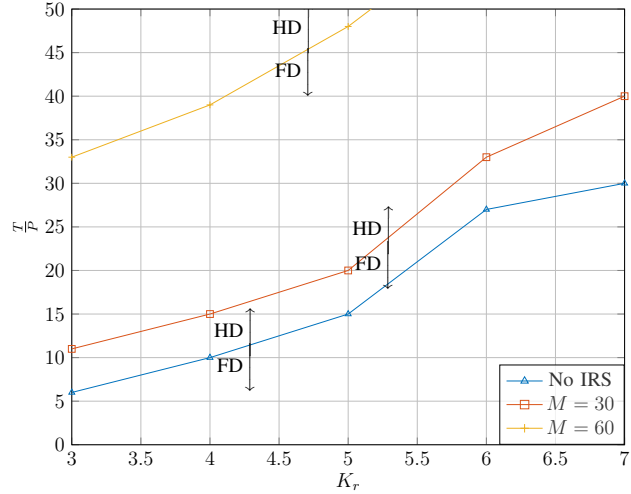
\begin{figure}
\centering
\hspace*{-.8cm}
\tikzset{every picture/.style={scale=.8}, every node/.style={scale=0.8}}%
\input{criticalPoint1.tex}
\vspace*{-.4cm}
\caption{Thresholds for different $K_r$ and $M$. The region above each curve indicates values of $\frac{T}{P}$ for which HD outperforms FD. In contrast, points below the curve belong to cases where FD performs better than HD.}
\label{fig:FDvsHD1}
\end{figure}

In this subsection, we determine the thresholds where the HD relaying outperforms the FD relaying. This thresholds provide a mode-switching threshold in hybrid HD/FD relay systems. As it can be seen in ~\figurename{~\ref{fig:FDvsHD1}}, for each case of $K_r$, there is a maximum value of $\frac{T}{P}$ above which the HD mode outperforms the FD mode in terms of sum rate maximization. Furthermore, ~\figurename{~\ref{fig:FDvsHD1}} shows the threshold for different different IRS configurations. For this part we continued with the case of $N_t=4, K_{r}=5, K_\mathrm{t}=5, N_r=4$. As it can be seen, by increasing the number of antennas, the threshold occurs at higher RSI. This is in fact a direct result of getting better performance by having more antennas at the relay's receiver. It is worth noting that the IRS has a great impact on the performance of FD relaying. For instance, by having an IRS consists of only 60 elements, FD mode outperforms HD mode in almost all cases.

\section{Conclusion}
In this paper, we investigated a multi-antenna source communicating with a multi-antenna destination through a multi-antenna relay. The relay is assumed to exploit a decode-and-forward (DF) strategy. An IRS is installed to hep the relay cope with the RSI. The transceivers are designed in order to be robust against the worst-case residual self-interference (RSI). To this end, the worst-case achievable throughput rate is maximized. This optimization problem turns out to be a non-convex problem. Assuming that the degrees-of-freedom (DoF) of the source-relay link is less than the DoF of the relay-destination link, we determined the left and right matrices of the singular vectors of the worst-case RSI channel. Then, the problem is simplified to the optimal power allocation at the transmitters, which guarantees robustness against the worst-case RSI singular values. This simplified problem is still non-convex. Based on the intuitions for optimal power allocation at the source and relay, we proposed an efficient algorithm to capture a stationary point. Hence, in a DF relay with multi-stream beamforming, we determine the threshold where the half-duplex relaying outperforms the full-duplex relaying. This threshold provides a mode-switching threshold in hybrid half-duplex full-duplex relay systems. 

\section{Appendix I\\ proof of equation~\eqref{OPP:TT}}
To show the proof, we use the Bra-Ket notation. We have
\begin{subequations}
\begin{align}
\mathbf{H}_{rI}\mathbf{\Theta}\mathbf{H}_{Ir} &= \!\left(\sum^{K_r}_{r_1=1}\!\sum^{M}_{r_2=1}{h_{rI}}_{r_1,r_2} \ket{r_1}\!\!\bra{r_2}\right)\!\left(\sum^{M}_{r_3=1}{\theta}_{r_3,r_3} \ket{r_3}\!\!\bra{r_3}\right)\! \left(\sum^{M}_{r_4=1}\! \sum^{K_t}_{r_5=1}{h_{Ir}}_{r_4,r_5}\ket{r_4}\!\!\bra{r_5}\right)\\
&=\sum^{K_r}_{r_1=1}\!\!\sum^{M}_{r_2=1}\!\!\sum^{M}_{r_3=1}\!\!\sum^{M}_{r_4=1} \!\!\sum^{K_t}_{r_5=1}{h_{rI}}_{r_1,r_2} {\theta}_{r_3,r_3}{h_{Ir}}_{r_4,r_5}\!\!\ket{r_1}\!\!\braket{r_2|r_3}\!\!\braket{r_3|r_4}\!\!\bra{r_5}\\
&=\sum^{K_r}_{r_1=1}\sum^{M}_{r_3=1} \sum^{K_t}_{r_5=1}{h_{rI}}_{r_1,r_3} {\theta}_{r_3,r_3}{h_{Ir}}_{r_3,r_5}\!\!\ket{r_1}\!\!\bra{r_5}.
\end{align}
\end{subequations}
Now, after turning the matrix representation into the vector representation we get
\begin{subequations}
\begin{align}
    Vec\left(\mathbf{H}_{rI}\mathbf{\Theta}\mathbf{H}_{Ir}\right)&=\!\!\sum^{K_r}_{r_1=1}\sum^{M}_{r_3=1} \sum^{K_t}_{r_5=1}{h_{rI}}_{r_1,r_3} {\theta}_{r_3,r_3}{h_{Ir}}_{r_3,r_5}\!\!\ket{r_1,r_5}\\
    &= \!\!\sum^{K_r}_{r_1=1}\sum^{M}_{r_3=1}\sum^{M}_{r_4=1} \sum^{K_t}_{r_5=1}{h_{rI}}_{r_1,r_4} {\theta}_{r_3,r_3}{h_{Ir}}_{r_4,r_5}\!\!\ket{r_1,r_5}\braket{r_4|r_3}\\
    &= \!\!\sum^{K_r}_{r_1=1}\sum^{M}_{r_4=1} \sum^{K_t}_{r_5=1}{h_{rI}}_{r_1,r_4} {h_{Ir}}_{r_4,r_5}\!\!\ket{r_1,r_5}\bra{r_4}\sum^{M}_{r_3=1}{\theta}_{r_3,r_3}\ket{r_3}\\
    &= (\mathbf{H}_{Ir}*\mathbf{H}_{rI}^T)Vec(\mathbf{\Theta}).
\end{align}
\end{subequations}
As a result, we have, $||\bar{\mathbf{H}}_{r}+\mathbf{H}_{rI}\mathbf{\Theta}\mathbf{H}_{Ir}||_F^2=||Vec(\bar{\mathbf{H}}_{r})+(\mathbf{H}_{Ir}*\mathbf{H}_{rI}^T)Vec(\mathbf{\Theta})||_2^2$ and the proof is complete.

\section{Appendix II\\ proof of Theorem 1}\label{App:Ia}
Before stating the proof, first we introduce the following definitions.
\begin{definition}
For a vector $\boldsymbol{a}$, we denote vector $\boldsymbol{a}^{\downarrow}$ which has the same components as $\boldsymbol{a}$ except that they are sorted in a decreasing order.
\end{definition}
\begin{definition}
The vector $\boldsymbol{a}$ is said to be majorized by vector $\boldsymbol{b}$ and denoted by $\boldsymbol{a} \prec \boldsymbol{b}$ if:
\begin{align}
    \sum_{i = 1}^{K} a^{\downarrow}_{i} \leq \sum_{i = 1}^{K} b^{\downarrow}_{i},\\
    \sum_{i = 1}^{N} a^{\downarrow}_{i} = \sum_{i = 1}^{N} b^{\downarrow}_{i},
\end{align}
where $a^{\downarrow}_i$ is the $i$'th component of $\boldsymbol{a}^{\downarrow}$, $N_r$ is the number of vector components and $K \leq N$. If the last equality does not hold, $\boldsymbol{a}$ is said to be weakly majorized by $\boldsymbol{b}$ and denoted by $\boldsymbol{a} \prec_{w} \boldsymbol{b}$.
\end{definition}
\begin{definition}
The vector $\boldsymbol{a}$ is said to be multiplicatively majorized by vector $\boldsymbol{b}$ and denoted by $\boldsymbol{a} \prec_{\times} \boldsymbol{b}$ if:
\begin{align}
    \prod_{i = 1}^{K} a^{\downarrow}_{i} \leq \prod_{i = 1}^{K} b^{\downarrow}_{i},\\
    \prod_{i = 1}^{N} a^{\downarrow}_{i} = \prod_{i = 1}^{N} b^{\downarrow}_{i}.
\end{align}
Also, it is easy to check
\begin{equation}
\boldsymbol{a} \prec_{\times} \boldsymbol{b} \quad \Leftrightarrow \quad \log(\boldsymbol{a}) \prec \log(\boldsymbol{b}).
\end{equation}
\end{definition}
To begin with, we know that for $n \times m$ matrix $\boldsymbol{A}$ and $m \times n$ matrix $\boldsymbol{B}$ we have $\lambda_i(\boldsymbol{A}\boldsymbol{B}) = \lambda_i(\boldsymbol{B}\boldsymbol{A}),\ \forall i\in\{1,\cdots,\min(m,n)\}$. Also the only difference between eigenvalues of $\boldsymbol{B}\boldsymbol{A}$ and $\boldsymbol{A}\boldsymbol{B}$ are the number of eigenvalues 0. Thus, non-zero eigenvalues of  $\bar{\mathbf{H}}_{r}\mathbf{Q}_{r}\bar{\mathbf{H}}^H_{r}$ and $\mathbf{Q}_{r}\bar{\mathbf{H}}^H_{r}\bar{\mathbf{H}}_{r}$ and also ${\mathbf{H}}_\mathrm{1}\mathbf{Q}_{s}{\mathbf{H}}^H_\mathrm{1}$ and $\mathbf{Q}_{s}{\mathbf{H}}^H_\mathrm{1}{\mathbf{H}}_\mathrm{1}$ are equal, respectively. Notice that all $\mathbf{Q}_{s}$,  ${\mathbf{H}}^H_\mathrm{1}{\mathbf{H}}_\mathrm{1}$, $\mathbf{Q}_{r}$ and $\bar{\mathbf{H}}^H_{r}\bar{\mathbf{H}}_{r}$ are square matrices. For ${\mathbf{H}}^H_\mathrm{1}{\mathbf{H}}_\mathrm{1}$ and $\bar{\mathbf{H}}^H_{r}\bar{\mathbf{H}}_{r}$ we define
$\lambda_i({\mathbf{H}}^H_\mathrm{1}{\mathbf{H}}_\mathrm{1}) = \sigma_i^{2}({\mathbf{H}}_\mathrm{1})$ and $\lambda_i(\bar{\mathbf{H}}^H_{r}\bar{\mathbf{H}}_{r}) =\sigma_i^{2}(\bar{\mathbf{H}}_{r})$ respectively.

As discussed in Remark~\ref{Remark1}, the equality $\lambda_i(\mathbf{Q}_{s}{\mathbf{H}}^H_\mathrm{1}{\mathbf{H}}_\mathrm{1}) = \lambda_{\rho\left(i\right)}(\mathbf{Q}_{s})\lambda_i({\mathbf{H}}^H_\mathrm{1}{\mathbf{H}}_\mathrm{1})$ does not hold in general. However, using the definition of determinant one can arrive at the following equality
\begin{equation}
    \prod_{i = 1}^{\text{min}(M, K_r)} \lambda_i(\mathbf{Q}_{s}{\mathbf{H}}^H_\mathrm{1}{\mathbf{H}}_\mathrm{1}) = \prod_{i=1}^{\text{min}(M, K_r)} \lambda_i(\mathbf{Q}_{s})\sigma^2_i({\mathbf{H}}_\mathrm{1}).
\end{equation}
Now, we define vector $\boldsymbol{\lambda}(\mathbf{Q}'_{s})$ and set its components to be ${\lambda_{\rho{\left(i\right)}}}(\mathbf{Q}'_{s}) = \frac{\lambda_i(\mathbf{Q}_{s}{\mathbf{H}}^H_\mathrm{1}{\mathbf{H}}_\mathrm{1})}{\sigma^2_i({\mathbf{H}}_\mathrm{1})}$. By defining $\lambda_{\rho\left(i\right)}(\mathbf{Q}'_{s})$ instead of $\lambda_{i}(\mathbf{Q}'_{s})$, we emphasize that the elements of $\boldsymbol{\lambda}(\mathbf{Q}'_{s})$ are not necessarily in decreasing order. 
Then, we construct the matrix $\mathbf{Q}'_{s}$ having the same eigenvectors as those of ${\mathbf{H}}^H_\mathrm{1}{\mathbf{H}}_\mathrm{1}$ and the eigenvalues $\lambda_{\rho\left(i\right)}(\mathbf{Q}'_{s})$. One can check that for each $i$ we have $\lambda_i(\mathbf{Q}'_{s}{\mathbf{H}}^H_\mathrm{1}{\mathbf{H}}_\mathrm{1})=\lambda_i(\mathbf{Q}_{s}{\mathbf{H}}^H_\mathrm{1}{\mathbf{H}}_\mathrm{1})$. Also, by the definition of $\lambda_{\rho\left(i\right)}(\mathbf{Q}'_{s})$ we have
\begin{align}
       &\lambda_{\rho\left(i\right)}(\mathbf{Q}'_{s}) = \frac{\lambda_i(\mathbf{Q}_{s}{\mathbf{H}}^H_\mathrm{1}{\mathbf{H}}_\mathrm{1})}{\sigma^2_i({\mathbf{H}}_\mathrm{1})} ,\\ 
       \Rightarrow &\!\log{\!\left(\lambda_{\rho\left(i\right)}(\mathbf{Q}'_{s})\!\right)}\!=\!\log\left( \lambda_i(\mathbf{Q}_{s}{\mathbf{H}}^H_\mathrm{1}{\mathbf{H}}_\mathrm{1})\right)\!-\!\log\!\left(\sigma^2_i({\mathbf{H}}_\mathrm{1})\right),\\ 
       \Rightarrow &\log{\left(\boldsymbol{\lambda}(\mathbf{Q}'_{s})\right)}=\log\left( \boldsymbol{\lambda}(\mathbf{Q}_{s}{\mathbf{H}}^H_\mathrm{1}{\mathbf{H}}_\mathrm{1})\right) - \log\left(\boldsymbol{\sigma}^2({\mathbf{H}}_\mathrm{1})\right).
\end{align}

\begin{lemma}
Let $\boldsymbol{A}$ and $\boldsymbol{B}$ be semidefinite Hermitian matrices with $\lambda_{\min(m,n)}(\boldsymbol{AB}) > 0$. Then
\begin{equation}
\log(\boldsymbol{\lambda}(\boldsymbol{AB})) -\log(\boldsymbol{\lambda}(\boldsymbol{B})) \prec \log(\boldsymbol{\lambda}(\boldsymbol{A})).
\end{equation}
\end{lemma}
\begin{proof}
The proof is given in \cite[H.1,e]{marshall1979inequalities}.
\end{proof}
\noindent Using the above lemma we can conclude
\begin{equation}
\log\left(\boldsymbol{\lambda}(\mathbf{Q}'_{s})\right) \prec \log\left(\boldsymbol{\lambda}(\mathbf{Q}_{s})\right).
\end{equation}

\noindent Then, immediately we can conclude
\begin{equation}
\boldsymbol{\lambda}(\mathbf{Q}'_{s}) \prec_{\times} \boldsymbol{\lambda}(\mathbf{Q}_{s}).
\end{equation}

\begin{remark}
It is worth mentioning that, depending on channel realizations, the optimal $\mathbf{Q}_{s}$ might contain some zero eigenvalues. In such cases, we can simply ignore the zeros and construct matrix $\mathbf{Q}'_{s}$ with dimension $(n-k) \times (n-k)$. Similarly, in the cases where ${\bar{\mathbf{H}}}^H_{r}{\bar{\mathbf{H}}_{r}}$ has some zero eigenvalues, we can do the same and proceed to constitute $\bar{\mathbf{H}}'_{s}$ using only nonzero eigenvalues of ${\bar{\mathbf{H}}}^H_{r}{\bar{\mathbf{H}}_{r}}$ and add the zeros back to the result again at the end.
\end{remark}
Finally, we use the following lemma to show that $\bar{\mathbf{H}}'_{\mathrm{r}}$ and $\mathbf{Q}'_{\mathrm{s}}$ are in the feasible set.
\begin{lemma}
For two vectors $\boldsymbol{a}$ and $\boldsymbol{b}$, if $\boldsymbol{a} \prec_{\times} \boldsymbol{b}$, then $\boldsymbol{a} \prec_{w} \boldsymbol{b}$ follows.
\end{lemma}
\begin{proof}
The proof is given in~\cite[5.A.2.b]{marshall1979inequalities}.
\end{proof}

Exploiting the above lemma one concludes

\begin{align}
\boldsymbol{\lambda}(\mathbf{Q}'_{s}) \prec_{\times} \boldsymbol{\lambda}(\mathbf{Q}_{s})\ \Rightarrow \  \boldsymbol{\lambda}(\mathbf{Q}'_{s}) \prec_{w} \boldsymbol{\lambda}(\mathbf{Q}_{s}),
\end{align}
which consequently results in
\begin{align}
\sum_{i = 1}^{N}{\lambda_i}(\mathbf{Q}'_{s}) \leq \sum_{i = 1}^{N}{\lambda_i}(\mathbf{Q}_{s}) \Rightarrow  \text{Tr} (\mathbf{Q}'_{s})\leq\text{Tr}(\mathbf{Q}_{s}).
\end{align}
Therefore, there exists $\mathbf{Q}'_{s}$ and $\bar{\mathbf{H}}'_{r}$ fulfilling \eqref{P:newo1a}-\eqref{P:newo1c}, which satisfy
\begin{align}
       \sum_{i = 1}^{\min{(M,K_r)}} \log_2 &\left( 1+ \frac{\lambda_i\left(\mathbf{H}_1\mathbf{Q}_{s}\mathbf{H}^H_1\right)}{1+\lambda_i\left(\bar{\mathbf{H}}_{r}\mathbf{Q}_{r}\bar{\mathbf{H}}^H_{r}\right)} \right) = \\ &\sum_{i = 1}^{\min{(M,K_r)}} \log_2 \left( 1+ \frac{\lambda_{\rho(i)}(\mathbf{Q}'_{s})\sigma^2_i({\mathbf{H}}_\mathrm{1})}{1+\lambda_{i}(\mathbf{Q}_{r})\sigma^2_{\rho(i)}({\bar{\mathbf{H}}'}_\mathrm{r})} \right). 
\end{align}
\allowdisplaybreaks
\definecolor{almond}{rgb}{0.94, 0.87, 0.8}
\definecolor{amber}{rgb}{1.0, 0.75, 0.0}
\definecolor{applegreen}{rgb}{0.55, 0.71, 0.0}
\definecolor{aqua}{rgb}{0.0, 1.0, 1.0}
\section{Appendix III\\ proof of proposition 2}
In this section we prove that the problem 
\begin{subequations}\label{P:FDc}
\begin{align}
\max_{\boldsymbol{\gamma}_{s},\boldsymbol{\gamma}_{r}}\quad  \min_{\boldsymbol{\sigma}_{{r}}}\quad & \sum_{i=1}^{\min(M,K_{r})} \log_2{\left(1+
\frac{\sigma_{1_i}^2\gamma_{{s}_{\rho{\left(i\right)}}}}{1+\gamma_{{r}_i}\sigma_{{s}_{\rho{\left(i\right)}}}^2}\right)} \tag{\ref{P:FDc}}\\
\text{s.t.}\quad & \|\boldsymbol{\gamma}_{s}\|_1\leq P_{s},\label{P:FDc:ConsA}\\ 
&\|\boldsymbol{\gamma}_{r}\|_1\leq P_{r},\label{P:FDc:ConsB}\\
&\|\boldsymbol{\sigma}^2_{r}\|_1\leq T,\label{P:FDc:ConsC}\\
& \sigma_{1_i}^2\gamma_{{s}_{{\rho{\left(i\right)}}}} \geq \sigma_{1_{i+1}}^2\gamma_{{s}_{{\rho{\left(i+1\right)}}}},\ \forall i \leq \min(M,K_{{r}}),\label{appP:FDb:ConsA}\\
& \gamma_{{r}_i}\sigma_{{r}_{{\rho{\left(i\right)}}}}^2 \geq \gamma_{{r}_{i+1}}\sigma_{{r}_{{\rho{\left(i+1\right)}}}}^2,\ {\forall i \leq \min(K_{\mathrm{t}},N).}\label{appP:FDb.ConsB}
\end{align}
\end{subequations} 
can be further simplified to 
\begin{subequations}\label{P:FDd}
\begin{align}
\max_{\boldsymbol{\gamma}_{s},\boldsymbol{\gamma}_{r}}\quad  \min_{\boldsymbol{\sigma}_{{r}}}\quad & \sum_{i=1}^{\min(M,K_{r})} \log_2{\left(1+
\frac{\sigma_{1_i}^2\gamma_{{s}_{\rho{\left(i\right)}}}}{1+\gamma_{{r}_i}\sigma_{{s}_{\rho{\left(i\right)}}}^2}\right)} \tag{\ref{P:FDd}}\\
\text{s.t.}\quad & \|\boldsymbol{\gamma}_{s}\|_1 = P_{s},\label{P:FDd:ConsA}\\ 
&\|\boldsymbol{\gamma}_{r}\|_1\leq P_{r},\label{P:FDd:ConsB}\\
&\|\boldsymbol{\sigma}^2_{r}\|_1 = T.\label{P:FDd:ConsC}\\
& \sigma_{1_i}^2\gamma_{{s}_{{\rho{\left(i\right)}}}} \geq \sigma_{1_{i+1}}^2\gamma_{{s}_{{\rho{\left(i+1\right)}}}},\ \forall i \leq \min(M,K_{{r}}),\label{app2P:FDb:ConsA}\\
& \gamma_{{r}_i}\sigma_{{r}_{{\rho{\left(i\right)}}}}^2 \geq \gamma_{{r}_{i+1}}\sigma_{{r}_{{\rho{\left(i+1\right)}}}}^2,\ {\forall i \leq \min(K_{\mathrm{t}},N).}\label{app2P:FDb.ConsB}
\end{align}
\end{subequations}
The proof is by contradiction. Starting with the minimization, assume that the optimal vector ${\boldsymbol{\sigma}^\star}^2_{r}$, for which we have $R^{\text{FD}}_{\text{sr}}({\boldsymbol{\sigma}^\star}^2_{r}) \leq R^{\text{FD}}_{\text{sr}}(\boldsymbol{\sigma}^2_{r})$, does not sum to $T$ and thus, we have $\|\boldsymbol{\sigma^\star}^2_{r}\|_1 < T$. Then there exists $\varepsilon > 0$ for which we have $\|\boldsymbol{\sigma^\star}^2_{r}\|_1 +\varepsilon = T$. Now define
\begin{align}
    \varepsilon_i = \frac{\varepsilon \frac{\sigma^2_{1_i}}{\gamma_{r_i}}}{\sum_{j}\frac{\sigma^2_{1_{j}}}{\gamma_{r_{j}}}}.
\end{align}
Note that we have 
\begin{align}
    \sum_{i}\varepsilon_i = \varepsilon, \quad \varepsilon_i \geq 0.
\end{align}
Also, as we have $\varepsilon > 0$, there is at least one $\varepsilon_i$ which is strictly greater than zero i.e. $\varepsilon_i > 0$. 
Now define 
\begin{align}
    \sigma'^2_{r_{\rho\left(i\right)}}= {\sigma^\star}^2_{r_{\rho\left(i\right)}}+\varepsilon_i.
\end{align}
One can check that $\sum_{i} {\sigma'}^2_{r_{\rho\left(i\right)}} = T$ and ${\forall i \leq \min(K_{\mathrm{t}},N)} \Rightarrow \gamma_{{r}_i}\sigma'^2_{{r}_{{\rho{\left(i\right)}}}} \geq \gamma_{{r}_{i+1}}\sigma'^2_{{r}_{{\rho{\left(i+1\right)}}}}$. As a result $\boldsymbol{\sigma}'^2_{r_{\rho\left(i\right)}}$ meets the constraints and could be a feasible solution. Note that, as $\boldsymbol{\gamma}^\star_{s}$ is the optimal source power allocation based on all other parameters, by changing ${\boldsymbol{\sigma}^\star}^2_{r}$ to $\boldsymbol{\sigma'}^2_{r}$, $\boldsymbol{\gamma}^\star_{s}$ might also change. However, we created each ${\sigma'}^2_{r_{\rho\left(i\right)}}$ in a special way to avoid this change. To show this, first notice that we have
\begin{align}
    \gamma^\star_{s_{\rho\left(i\right)}} &= \left[\lambda - \frac{1+\gamma_{r_i}{\sigma^\star}^2_{r_{\rho\left(i\right)}}}{\sigma^2_{1_i}}\right]^+,
\end{align}
where $\lambda$ is water level and can be found based on power constraints.
Substituting the new power allocation for interference, we get new power allocation for input power as follows
\begin{align}
    \gamma_{s_{\rho\left(i\right)}} &= \left[\lambda - \frac{1+\gamma_{r_i}\sigma'^2_{r_{\rho\left(i\right)}}}{\sigma^2_{1_i}}\right]^+ = \left[\lambda - \frac{1+\gamma_{r_i}({\sigma^\star}^2_{r_{\rho\left(i\right)}}+\varepsilon_i)}{\sigma^2_{1_i}}\right]^+\\
    & = \left[\lambda - \frac{1+\gamma_{r_i}{\sigma^\star}^2_{r_{\rho\left(i\right)}}}{\sigma^2_{1_i}}+\frac{\varepsilon}{\sum_{j=1}^{N}\frac{\sigma^2_{1_{j}}}{\gamma_{r_{j}}}}\right]^+ \stackrel{(a)}{=}\left[\lambda' - \frac{1+\gamma_{r_i}{\sigma^\star}^2_{r_{\rho\left(i\right)}}}{\sigma^2_{1_i}}\right]^+= \gamma^\star_{s_{\rho\left(i\right)}},
\end{align}
where $(a)$ comes from the fact that $\sfrac{\varepsilon}{\sum_{j=1}^{N}\frac{\sigma^2_{1_{j}}}{\gamma_{r_{j}}}}$ is a constant independent of $i$, so we can define $\lambda'= \lambda+\sfrac{\varepsilon}{\sum_{j=1}^{N}\frac{\sigma^2_{1_{j}}}{\gamma_{r_{j}}}}$. This shows, for $\boldsymbol{\sigma}'^2_{r}$, all the optimal variables and parameters remain the same as those of ${\boldsymbol{\sigma}^\star}^2_{r}$. Now we compare $R^{\text{FD}}_{\text{sr}}$ for both cases. First, notice that we have $\forall i, \varepsilon_i \geq 0$ and among them there is at least one index $i'$, for which we have $\varepsilon_{i'}>0$. This means $\forall i, \sigma'^2_{r_{\rho\left(i\right)}} \geq {\sigma^\star}^2_{r_{\rho\left(i\right)}}$ and $\sigma'^2_{r_{\rho\left(i'\right)}} > {\sigma^\star}^2_{r_{\rho\left(i'\right)}}$. Now, notice that $f_i(x) = \log_2 \left( 1 + \frac{\sigma^2_{1_i}\gamma^\star_{s_{\rho\left(i\right)}}}{1+\gamma_{r_i}x} \right)$ is a monotonically decreasing function of $x$. Thus, we have $f_i(\sigma'^2_{r_{\rho\left(i\right)}}) \leq f_i({\sigma^\star}^2_{r_{\rho\left(i\right)}})$ and $f_{i'}(\sigma'^2_{r_{\rho\left(i'\right)}}) < f_{i'}({\sigma^\star}^2_{r_{\rho\left(i'\right)}})$.
Adding all above inequalities, we get
\begin{align}
\sum_{i=1}^{\min(M,K_{r})} &\log_2{\left(1+
\frac{\sigma_{1_i}^2\gamma^\star_{{s}_{\rho\left(i\right)}}}{1+\gamma_{{r}_i}\sigma'^2_{{r}_{\rho{\left(i\right)}}}}\right)} < \sum_{i=1}^{\min(M,K_{r})} \log_2{\left(1+
\frac{\sigma_{1_i}^2\gamma^\star_{{s}_{\rho\left(i\right)}}}{1+\gamma_{{r}_i}{\sigma^\star}^2_{{r}_{\rho\left(i\right)}}}\right)}.
\end{align}
The above equation indicates $R^{\text{FD}}_{\text{sr}}({\boldsymbol{\sigma}^\star}^2_{r})>R^{\text{FD}}_{\text{sr}}(\boldsymbol{\sigma}^2_{r})$ which contradicts the first assumption $R^{\text{FD}}_{\text{sr}}({\boldsymbol{\sigma}^\star}^2_{r})\leq R^{\text{FD}}_{\text{sr}}(\boldsymbol{\sigma}^2_{r})$. This completes the proof of the minimization part.

For the maximization part, the general idea is the same. Again, the proof is by contradiction. We assume the optimal vector $\boldsymbol{\gamma}^\star_{{s}}$, for which we have $R^{\text{FD}}_{\text{sr}}(\boldsymbol{\gamma}^\star_{{s}}) \geq R^{\text{FD}}_{\text{sr}}(\boldsymbol{\gamma}_{{s}})$, does not sum to $P_{s}$. Therefore, we have $\|\boldsymbol{\gamma}_{{s}}\|_1 < P_{s}$. Then there exists $\varepsilon > 0$ for which we have $\|\boldsymbol{\gamma}_{{s}}\|_1 +\varepsilon = P_{s}$. Now we define 
\begin{align}
    \varepsilon_i = \frac{\varepsilon}{\eta}\left( \frac{1+{\sigma^\star}^2_{r_i}\gamma_{r_i}}{\sigma^2_{1_i}}+\gamma^\star_{s_i}\right),
\end{align}
where, $\eta = \sum_i \left(\frac{1+{\sigma^\star}^2_{r_i}\gamma_{r_i}}{\sigma^2_{1_i}}+\gamma^\star_{s_i}\right)$. Now we define the new source power allocation as below
\begin{align}
    \gamma'_{s_{\rho\left(i\right)}}= \gamma^\star_{s_{\rho\left(i\right)}}+\varepsilon_i.
\end{align}
One can check that $\sum_i \gamma'_{s_{\rho\left(i\right)}} = P_s$ and $\sigma^2_{1_i}\gamma'_{s_{\rho(i)}} \geq \sigma^2_{1_{i+1}}\gamma'_{s_{\rho(i+1)}}$. Thus, the new source power allocation is in the feasible set. Now the remaining is to make sure the new allocation does not change the corresponding $\boldsymbol{\sigma}^2_\mathrm{r}$. Using Lagrangian multiplier we have
\begin{align}
L &= \sum_{i} \!\log_2{\!\left(1+
\frac{\sigma_{1_i}^2\gamma'_{{s}_{\rho{\left(i\right)}}}}{1+\gamma_{{r}_i}\sigma_{{s}_{\rho{\left(i\right)}}}^2}\right)} + \lambda \left( \sum_{i=0}^{N}\sigma^2_{r_i} - T\right),\\
&= \sum_{i} \!\log_2{\!\left(1+
\frac{\sigma_{1_i}^2(\gamma^\star_{{s}_{\rho{\left(i\right)}}}+\varepsilon_i)}{1+\gamma_{{r}_i}\sigma_{{s}_{\rho{\left(i\right)}}}^2}\right)} + \lambda \left( \sum_{i=0}^{N}\sigma^2_{r_i} - T\right),\\
&= \sum_{i} \!\log_2{\!\left((1+\frac{\varepsilon}{\eta})\left(1+
\frac{\sigma_{1_i}^2\gamma^\star_{{s}_{\rho{\left(i\right)}}}}{1+\gamma_{{r}_i}\sigma_{{s}_{\rho{\left(i\right)}}}^2}\right)\right)} + \lambda \left( \sum_{i=0}^{N}\sigma^2_{r_i} - T\right),\\
&= \sum_{i} \!\log_2{\!(1+\frac{\varepsilon}{\eta})}+\sum_i\log_2{\left(1+
\frac{\sigma_{1_i}^2\gamma^\star_{{s}_{\rho{\left(i\right)}}}}{1+\gamma_{{r}_i}\sigma_{{s}_{\rho{\left(i\right)}}}^2}\right)} + \lambda \left( \sum_{i=0}^{N}\sigma^2_{r_i} - T\right).
\end{align}
Now notice that as $\sum_{i} \!\log_2{\!(1+\frac{\varepsilon}{\eta})}$ is a constant we have $\frac{\partial \sum_{i} \!\log_2{\!(1+\frac{\varepsilon}{\eta})}}{\partial \sigma^2_{\mathrm{r}_i}}=0$ and $\frac{\partial \sum_{i} \!\log_2{\!(1+\frac{\varepsilon}{\eta})}}{\partial \lambda}=0$. As a result, the optimum interference allocation for $\boldsymbol{\gamma}'_\mathrm{r}$
 is the same as that of ${\boldsymbol{\gamma}^\star}_\mathrm{r}$. Similarly to the case of minimization, here we have $\sum_{i}\varepsilon_i= \varepsilon$. Also we have $\varepsilon_i \geq 0$ and there exist at least one $i'$ for which we have $\varepsilon_{i'} > 0$. Finally as $f_i(x) = \log \left(1+\frac{{\sigma^2}_{1_i}x}{1+{{\sigma^\star}^2}_{r_{\rho\left(i\right)}}\gamma_{r_i}}\right)$ is a monotonically increasing function of $x$, we conclude $R^{\text{FD}}_{\text{sr}}(\boldsymbol{\gamma}^\star_{s})<R^{\text{FD}}_{\text{sr}}(\boldsymbol{\gamma'}_{s})$ which contradicts the first assumption of ${\boldsymbol{\gamma}^\star}_{s}$ being the optimal source power allocation, and the proof is complete.

\section{Appendix IV}
First, we show $R^{\text{FD}}_\text{sr}$ is a decreasing function of $T$ and an increasing function of ${P}_{s}$. It is sufficient to show $\frac{d R^{\text{FD}}_\text{sr}}{d P_{{s}}} \geq 0$ and $\frac{d R^{\text{FD}}_\text{sr}}{d T} \leq 0$. We have
\begin{align}
        \frac{d R^{\text{FD}}_{\text{sr}}}{d P_\text{s}} &=\frac{\sum_{i} \frac{\partial R^{\text{FD}}_\text{sr}}{\partial \gamma_{s_{\rho\left(i\right)}}}d \gamma_{s_{\rho\left(i\right)}}}{\sum_{i}^{} \frac{\partial P_\text{s}}{\partial \gamma_{s_{\rho\left(i\right)}}}d \gamma_{s_{\rho\left(i\right)}}} = \frac{\sum_{i} \frac{{\sigma^2}_{1_i}d \gamma_{s_{\rho\left(i\right)}}}{1+{\sigma^2}_{r_{\rho\left(i\right)}}\gamma_{r_i}+{\sigma^2}_{1_i}\gamma_{s_{\rho\left(i\right)}}}}{\sum_{i} d \gamma_{s_{\rho\left(i\right)}}} \geq \frac{\sum_{i}\phi_1d \gamma_{s_{\rho\left(i\right)}}}{\sum_{i}^{} d \gamma_{s_{\rho\left(i\right)}}}=\phi_1>0,\\
    \frac{d R^{\text{FD}}_\text{sr}}{d T} &=\frac{\sum_{i} \frac{\partial R^{\text{FD}}_\text{sr}}{\partial {\sigma^2}_{r_{\rho\left(i\right)}}}d {\sigma^2}_{r_{\rho\left(i\right)}}}{\sum_{i}^{} \frac{\partial T}{\partial \sigma^2_{r_{\rho\left(i\right)}}}d {\sigma^2}_{r_{\rho\left(i\right)}}} \\&= \frac{\sum_{i}^{}\frac{-{\sigma^2}_{1_{i}}\gamma_{s_{\rho\left(i\right)}}{\gamma_r}_i}{\left(1+{\sigma^2}_{r_{\rho\left(i\right)}}\gamma_{r_i}\right)\left(1+{\sigma^2}_{1_{i}}\gamma_{s_{\rho\left(i\right)}}+{\sigma^2}_{r_{i}}\gamma_{r_i}\right)}d {\sigma^2}_{r_{\rho\left(i\right)}}}{\sum_{i}^{} d {\sigma^2}_{r_{\rho\left(i\right)}}} \leq \frac{\sum_{i}^{}-\phi_2d \sigma^2_{r_{\rho\left(i\right)}}}{\sum_{i}^{} d {\sigma^2}_{r_{\rho\left(i\right)}}}=-\phi_2\leq0,
\end{align}
where 
\begin{align}
    \phi_1\stackrel{.}{=} \min_i \Bigg\{ \frac{\sigma^2_{1_i}}{1+\sigma^2_{r_{\rho\left(i\right)}}\gamma_{r_i}+\sigma^2_{1_i}\gamma_{s_{\rho\left(i\right)}}}\Bigg\} 
\end{align} 
and 
\begin{align}
    \phi_2\stackrel{.}{=} \min_i \Bigg\{\frac{-\sigma^2_{1_{i}}\gamma_{s_{\rho\left(i\right)}}\gamma_{r_i}}{\left(1+\sigma^2_{r_{\rho\left(i\right)}}\gamma_{r_i}\right)\left(1+\sigma^2_{1_{i}}\gamma_{s_{\rho\left(i\right)}}+\sigma^2_{r_{\rho\left(i\right)}}\gamma_{r_i}\right)}\Bigg\}
\end{align}
respectively.

Next, we show $g(P_{r})=R^{\text{FD}}_\text{sr}(P_{r})-R^{\text{FD}}_\text{rd}(P_{r})$ is a monotonically decreasing function of ${P}_{r}$. It is sufficient to show $\frac{d R^{\text{FD}}_\text{sr}}{d P_{{r}}} \leq 0$ and $\frac{d R^{\text{FD}}_\text{rd}}{d P_{{r}}} > 0$. We have
\begin{align}
    d {R^{\text{FD}}}_\text{rd} =& \sum_{i}^{} \frac{\partial R^{\text{FD}}_\text{rd}}{\partial \gamma_{r_i}}d \gamma_{r_i}= \sum_{i} \frac{{\sigma^2}_{2_i}}{1+\sigma^2_{2_i}\gamma_{r_i}}d \gamma_{r_i}\\
    d {R^{\text{FD}}}_\text{sr} =& \sum_{i}^{} \frac{\partial R^{\text{FD}}_\text{sr}}{\partial \gamma_{r_i}}d \gamma_{r_i}\nonumber
    =\sum_{i} \frac{-{\sigma^2}_{1_{i}}\gamma_{s_{\rho\left(i\right)}}{\sigma^2}_{r_{\rho\left(i\right)}}}{\left(1+{\sigma^2}_{r_{\rho\left(i\right)}}\gamma_{r_i}\right)\left(1+{\sigma^2}_{1_{i}}\gamma_{s_{\rho\left(i\right)}}+{\sigma^2}_{r_{\rho\left(i\right)}}\gamma_{r_i}\right)}d \gamma_{r_i}\\
    d P_\text{r} =& \sum_{i} \frac{\partial P_\text{r}}{\partial \gamma_{r_i}}d \gamma_{r_i}=\sum_{i} d \gamma_{r_i}.
\end{align}
Now we define
\begin{align}
     &\psi_1 \stackrel{.}{=} \min_i \Bigg\{ \frac{\sigma^2_{2_i}}{1+\sigma^2_{2_i}\gamma_{r_i}}\Bigg\}\\
     &\psi_2 \stackrel{.}{=} \min_i \Bigg\{ \frac{\sigma^2_{1_{i}}\gamma_{s_{\rho\left(i\right)}}\sigma^2_{r_{\rho\left(i\right)}}}{\left(1+\sigma^2_{r_{\rho\left(i\right)}}\gamma_{r_i}\right)\left(1+\sigma^2_{1_{i}}\gamma_{s_{\rho\left(i\right)}}+\sigma^2_{r_{\rho\left(i\right)}}\gamma_{r_i}\right)}\Bigg\}.
\end{align}
It is obvious that $\psi_1>0$ and $\psi_2\geq0$. Now we have
\begin{align}
    \frac{d R^{\text{FD}}_\text{sr}}{d P_{{r}}} = \frac{\sum_{i}^{} \frac{-\sigma^2_{1_{i}}\gamma_{s_{\rho\left(i\right)}}\sigma^2_{r_{\rho\left(i\right)}}}{\left(1+\sigma^2_{r_{\rho\left(i\right)}}\gamma_{r_i}\right)\left(1+\sigma^2_{1_{i}}\gamma_{s_{\rho\left(i\right)}}+\sigma^2_{r_{\rho\left(i\right)}}\gamma_{r_i}\right)}d \gamma_{r_i}}{\sum_{i}^{}d \gamma_{r_i}} \leq \frac{\sum_{i}^{} -\psi_2 d \gamma_{r_i}}{\sum_{i}^{}d \gamma_{r_i}} = -\psi_2 \leq 0,
\end{align}
and 
\begin{align}
    \frac{d R^{\text{FD}}_\text{rd}}{d P_{{r}}} &= \frac{\sum_{i}^{} \frac{\sigma^2_{2_i}}{1+\sigma^2_{2_i}\gamma_{r_i}}d \gamma_{r_i}}{\sum_{i}^{}d \gamma_{r_i}}\geq \frac{\sum_{i}^{} \psi_1 d \gamma_{r_i}}{\sum_{i}^{}d \gamma_{r_i}} = \psi_1 > 0.
\end{align}
Finally, one can conclude
\begin{align}
    \frac{d g}{d P_{\text{r}}} &= \frac{d R^{\text{FD}}_\text{sr}}{d P_{{r}}}-\frac{d R^{\text{FD}}_\text{rd}}{d P_{{r}}}\leq -\psi_2 - \psi_1 <0.
\end{align}
\section{Appendix V}
Here we show that if $\gamma_{r_{i}} \geq \gamma_{r_{i+1}}$ and $\sigma_{1_i}^2\gamma_{{s}_{\rho{\left(i\right)}}} \geq \sigma_{1_{i+1}}^2\gamma_{{s}_{\rho\left(i+1\right)}}$ then $\gamma_{{r}_i}\sigma_{{s}_{\rho{\left(i\right)}}}^2 \geq \gamma_{{r}_{i+1}}\sigma_{{r}_{\rho\left(i+1\right)}}^2$. First we define $f(x,y) = \sqrt{x^2 + axy}-x-b, \ x \geq 0, \ y \geq 0$ in which $a$ and $b$ are positive constants. Now we have,
\begin{align}
    \frac{\partial f}{\partial y} = \frac{ax}{2\sqrt{x^2+axy}} \geq 0.
\end{align}
Also for $\frac{\partial f}{\partial x}$ we have
\begin{align}
    \frac{\partial f}{\partial x} = \frac{2x+ay}{2\sqrt{x^2+axy}} - 1\geq0.
\end{align}
One can check that for positive values $x, \ y$ and $a$, we always have $\frac{2x+ay}{2\sqrt{x^2+axy}} \geq 1$. As a result $f$ is an increasing function of both $x$ and $y$. The rest of the proof is as follows
\begin{align}
    \sigma^2_{r_i}\gamma_{r_i} =& \left[\frac{\sqrt{\left(\sigma^2_{1_i}\gamma_{s_{\rho\left(i\right)}}\right)^2+\frac{4\sigma^2_{1_i}\gamma_{s_{\rho\left(i\right)}}\gamma_{r_i}}{\lambda}}-\sigma^2_{1_i}\gamma_{s_{\rho\left(i\right)}}-2(\sigma^2_{t}\!+\!T_1P_s)}{2}\right]^+ \\\stackrel{(a)}{\geq}&
    \left[\frac{\sqrt{\left(\sigma^2_{1_i}\gamma_{s_{\rho\left(i\right)}}\right)^2+\frac{4\sigma^2_{1_i}\gamma_{s_{\rho\left(i\right)}}\gamma_{r_{i+1}}}{\lambda}}-\sigma^2_{1_i}\gamma_{s_{\rho\left(i\right)}}-2(\sigma^2_{t}\!+\!T_1P_s)}{2}\right]^+ \\ \stackrel{(b)}{\geq}&
    \left[\frac{\sqrt{\left(\sigma^2_{1_{i+1}}\gamma_{s_{\rho\left(i+1\right)}}\right)^2\!+\!\frac{4\sigma^2_{1_{i+1}}\gamma_{s_{\rho\left(i+1\right)}}\gamma_{r_{i+1}}}{\lambda}}\!}{2}\nonumber\right.\\&-\left.\frac{-\!\sigma^2_{1_{i+1}}\gamma_{s_{\rho\left(i+1\right)}}\!-\!2(\sigma^2_{t}\!+\!T_1P_s)}{2}\vphantom{\frac{\sqrt{\left(\sigma^2_{1_{i+1}}\gamma_{s_{\rho\left(i+1\right)}}\right)^2\!+\!\frac{4\sigma^2_{1_{i+1}}\gamma_{s_{\rho\left(i+1\right)}}\gamma_{r_{i+1}}}{\lambda}}\!}{2}}\right]^+\\
    =&\sigma^2_{r_{i+1}}\gamma_{r_{i+1}},
\end{align}
in which $(a)$ holds because $\gamma_{r_{i}} \geq \gamma_{r_{i+1}}$ and $(b)$ holds because $\sigma_{1_i}^2\gamma_{s_{\rho{\left(i\right)}}} \geq \sigma_{1_{i+1}}^2\gamma_{s_{\rho\left(i+1\right)}}$.

\color{black}
\bibliographystyle{IEEEtran} 
\bibliography{reference}
\end{document}

%% file: SystemModel.tex
\begin{tikzpicture}

\coordinate (a) at (1.4,2);
\definecolor{cadmiumgreen}{rgb}{0.0, 0.42, 0.24}
\foreach \y in {1,3,...,16}{
    \foreach \x in {1,3,...,16}{
\fill[yellow, scale = 0.15] ($(a)+(\x,\y)$)  +($(a)+(0, 1)$) -- +($(a)+(1, 1)$) -- +($(a)+(1,0)$) -- +($(a)$) -- cycle;
        \fill[cadmiumgreen, scale = 0.15] ($(a)+(\x-1,\y-1)$)  +($(a)+(0, 1)$) -- +($(a)+(1, 1)$) -- +($(a)+(1,0)$) -- +($(a)$) -- cycle;
        \fill[cadmiumgreen, scale = 0.15] ($(a)+(\x,\y-1)$)  +($(a)+(0, 1)$) -- +($(a)+(1, 1)$) -- +($(a)+(1,0)$) -- +($(a)$) -- cycle;
        \fill[cadmiumgreen, scale = 0.15] ($(a)+(\x+1,\y-1)$)  +($(a)+(0, 1)$) -- +($(a)+(1, 1)$) -- +($(a)+(1,0)$) -- +($(a)$) -- cycle;
        \fill[cadmiumgreen, scale = 0.15] ($(a)+(\x-1,\y)$)  +($(a)+(0, 1)$) -- +($(a)+(1, 1)$) -- +($(a)+(1,0)$) -- +($(a)$) -- cycle;
        \fill[cadmiumgreen, scale = 0.15] ($(a)+(\x+1,\y)$)  +($(a)+(0, 1)$) -- +($(a)+(1, 1)$) -- +($(a)+(1,0)$) -- +($(a)$) -- cycle;
        \fill[cadmiumgreen, scale = 0.15] ($(a)+(\x-1,\y+1)$)  +($(a)+(0, 1)$) -- +($(a)+(1, 1)$) -- +($(a)+(1,0)$) -- +($(a)$) -- cycle;
        \fill[cadmiumgreen, scale = 0.15] ($(a)+(\x,\y+1)$)  +($(a)+(0, 1)$) -- +($(a)+(1, 1)$) -- +($(a)+(1,0)$) -- +($(a)$) -- cycle;
        \fill[cadmiumgreen, scale = 0.15] ($(a)+(\x+1,\y+1)$)  +($(a)+(0, 1)$) -- +($(a)+(1, 1)$) -- +($(a)+(1,0)$) -- +($(a)$) -- cycle;
        }}

\draw (0,0) rectangle (1,2.5);
\TxAntenna{1}{1.8}{0.8};
\draw (1.2,1.6) circle (0.01cm);
\draw (1.2,1.4) circle (0.01cm);
\draw (1.2,1.2) circle (0.01cm);
\TxAntenna{1}{0.2}{0.8};
\node at (1.3,1.8){1};
\node at (1.3,0.2){$N_t$};

\node[rotate=90] at (0.5,1.25){Source};

\draw[fill=green,opacity=0.5] (3,0) rectangle (5,2.5);

\RxAntenna{3}{1.8}{0.8};
\node at (2.7,1.8){1};
\node at (2.7,0.2){$K_r$};
\draw (2.8,1.6) circle (0.01cm);
\draw (2.8,1.4) circle (0.01cm);
\draw (2.8,1.2) circle (0.01cm);
\RxAntenna{3}{0.2}{0.8};

\TxAntenna{5}{1.8}{0.8};
\node at (5.3,1.8){1};
\node at (5.3,0.2){$K_t$};
\draw (5.2,1.6) circle (0.01cm);
\draw (5.2,1.4) circle (0.01cm);
\draw (5.2,1.2) circle (0.01cm);
\TxAntenna{5}{0.2}{0.8};

\node at (4,1.25){Relay};

\draw (7,0) rectangle (8,2.5);
\RxAntenna{7}{1.8}{0.8};
\node at (6.7,1.8){1};
\node at (6.7,0.2){$N_r$};
\draw (6.8,1.6) circle (0.01cm);
\draw (6.8,1.4) circle (0.01cm);
\draw (6.8,1.2) circle (0.01cm);
\RxAntenna{7}{0.2}{0.8};


\node[rotate=90] at (7.5,1.25){Destination};

\draw[fill=yellow,opacity=0.3] (2,1.25) ellipse (0.5cm and 1.25cm);
\node at (2,1.25){$\mathbf{H}_1$};
\draw[fill=yellow,opacity=0.3] (6,1.25) ellipse (0.5cm and 1.25cm);
\node at (6,1.25){$\mathbf{H}_2$};
\draw[rotate around={-35:(2,4)}, fill=yellow,opacity=0.3] (2,4) ellipse (0.25cm and 1.55cm);
\node at (2,4){$\mathbf{H}_{SI}$};
\draw[rotate around={35:(6,4)}, fill=yellow,opacity=0.3] (6,4) ellipse (0.25cm and 1.55cm);
\node at (6,4){$\mathbf{H}_{ID}$};
\draw[rotate=90, fill=yellow,opacity=0.3] (2.9,-4) ellipse (0.25cm and .9cm);
\node at (4,2.9){$\mathbf{H}_{\mathrm{r}}$};
\draw[rotate around={25:(5,3.5)}, fill=yellow,opacity=0.3] (5,3.5) ellipse (0.25cm and .8cm);
\node at (5,3.5){$\mathbf{H}_{RI}$};
\draw[rotate around={-25:(3,3.5)}, fill=yellow,opacity=0.3] (3,3.5) ellipse (0.25cm and .8cm);
\node at (3,3.5){$\mathbf{H}_{IR}$};



\node at (4,6.8){IRS};
\end{tikzpicture}


%% file: threetotwo.tex
 \begin{tikzpicture}[scale=1]

    \def\R{sqrt(3)}
    \coordinate (O) at (0,0,0);
    \fill[ball color=white!10, opacity=1.0] (O) circle (\R); 
    \begin{scope}[ shift={(0,0)}, rotate=0]
      \draw[-latex, opacity=0.5] (O)--(3.5,0,0) node[anchor=east] {$X$};
      \draw[-latex, opacity=0.5] (O)-- (0,2.8,0) node[anchor=south] {$Y$};
      \draw[-latex, opacity=0.5] (O)--(0,0,2.5) node[anchor=south] {$Z$};
      \draw[fill=blue]  coordinate (O) circle (1pt) node[anchor=south  east] {$O$};

      \draw[opacity=0.8]
      (2,-2,2)   coordinate (A) --
      (-2,-2,2)  coordinate (B) --
      (-2,2,2)   coordinate (C) --
      (2,2,2)    coordinate (D) -- cycle
      (2,2,-2)   coordinate (E) --
      (-2,2,-2)  coordinate (F) --
      (-2,-2,-2) coordinate (G) --
      (2,-2,-2)  coordinate (H) -- cycle
      (A)--(H) (B)--(G) (D)--(E) (C)--(F);

      \foreach \l in {A,B,C,D,E,F,G,H}
      \draw[fill=black] (\l) circle (1pt);
    \end{scope}

    \coordinate (O) at (10,0,0);
    \fill[color=blue, opacity=0.5] (O) circle (\R);

      \draw[-latex, opacity=0.5] (10,0)--(10,3) node[anchor=east] {$X'$};
      \draw[-latex, opacity=0.5] (10,0)-- (13,0) node[anchor=south] {$Y'$};

      \draw[fill=blue]  coordinate (O) circle (1pt) node[anchor=south  east] {$O$};

      \draw[opacity=0.8]
      (12,-2,2)   coordinate (AA) --
      (8,-2,2)  coordinate (AB) --
      (8,2,2)   coordinate (AC) --
      (8,2,-2)  coordinate (AD) --
      (12,2,-2)  coordinate (AE) --
      (12,-2,-2)  coordinate (AF) -- cycle;

      \foreach \l in {AA,AB,AC,AD,AE,AF}
      \draw[fill=black] (\l) circle (1pt);
      \draw[-to] (4,0) -- (6,0); 
      \node at (5,.5){$f$};
  \end{tikzpicture}

%% file: georep.tex
\begin{tikzpicture}
\draw[red,dashed] (0,0) circle (7cm);
\node[trapezium, draw,trapezium left angle=60, trapezium right angle=120,shade,shading=axis,shading angle=100,inner ysep=2.5cm, outer sep=0pt,] at (0,0) {};
\draw[rotate=60, fill = applegreen] (0,0) ellipse (3cm and 1.5cm);
\draw[blue, fill = amber] (0,0) circle (1.5cm);
\draw[thick,-] (-7,0) edge[out=0, in=180] (7,0);
\draw[thick,-] (0,-7) edge[out=90, in=270] (0,7);
\draw[thick,-] (0,-7) edge[out=90, in=270] (0,7);
\draw[thick,-] (0,0) edge[out=150, in=-30] (-6.06,3.5);
\node at (.3,-.3){O};
\node at (-1.2,.3){E};
\node at (-1.8,1.4){F};
\node at (-5.7,3.6){G};

\end{tikzpicture}

%% file: histog.tex
%
%
\definecolor{mycolor1}{rgb}{0.00000,0.44700,0.74100}%
\begin{tikzpicture}

\begin{axis}[%
width=4.521in,
height=3.566in,
at={(0.758in,0.481in)},
scale only axis,
unbounded coords=jump,
xmin=0,
xmax=50,
xlabel={Number of iterations},
xlabel near ticks,
xmajorgrids,
ymin=0,
ymax=1,
ylabel={Probability},
ylabel near ticks,
ymajorgrids,
axis background/.style={fill=white},
legend style={legend cell align=left, align=left, draw=white!15!black}
]
\addplot [color=mycolor1]
  table[row sep=crcr]{%
-inf	0\\
2	0\\
2	0.779396499266206\\
3	0.779396499266206\\
3	0.887586355013065\\
4	0.887586355013065\\
4	0.920184701292193\\
5	0.920184701292193\\
5	0.935071768622257\\
6	0.935071768622257\\
6	0.943304578157998\\
7	0.943304578157998\\
7	0.948126140960017\\
8	0.948126140960017\\
8	0.95161613630669\\
9	0.95161613630669\\
9	0.954211261051652\\
10	0.954211261051652\\
10	0.956473493932777\\
11	0.956473493932777\\
11	0.958166589111215\\
12	0.958166589111215\\
12	0.959669971722089\\
13	0.959669971722089\\
13	0.960879836775602\\
14	0.960879836775602\\
14	0.961942943050435\\
15	0.961942943050435\\
15	0.962837813652146\\
16	0.962837813652146\\
16	0.963675412535347\\
17	0.963675412535347\\
17	0.964351934710241\\
18	0.964351934710241\\
18	0.964903175000895\\
19	0.964903175000895\\
19	0.96540788202026\\
20	0.96540788202026\\
20	0.965866055768336\\
21	0.965866055768336\\
21	0.96634928589326\\
22	0.96634928589326\\
22	0.966775244299674\\
23	0.966775244299674\\
23	0.967140351505172\\
24	0.967140351505172\\
24	0.967523356122705\\
25	0.967523356122705\\
25	0.967859827468948\\
26	0.967859827468948\\
26	0.968135447614275\\
27	0.968135447614275\\
27	0.968396749829975\\
28	0.968396749829975\\
28	0.968622257221606\\
29	0.968622257221606\\
29	0.968808390306762\\
30	0.968808390306762\\
30	0.969012420803952\\
31	0.969012420803952\\
31	0.969180656477073\\
32	0.969180656477073\\
32	0.969352471632602\\
33	0.969352471632602\\
33	0.969538604717758\\
34	0.969538604717758\\
34	0.969692522461252\\
35	0.969692522461252\\
35	0.969771271074203\\
36	0.969771271074203\\
36	0.969939506747324\\
37	0.969939506747324\\
37	0.970093424490819\\
38	0.970093424490819\\
38	0.97025092171672\\
39	0.97025092171672\\
39	0.970354726706518\\
40	0.970354726706518\\
40	0.970469270143537\\
41	0.970469270143537\\
41	0.970580234098149\\
42	0.970580234098149\\
42	0.970669721158321\\
43	0.970669721158321\\
43	0.970773526148119\\
44	0.970773526148119\\
44	0.970866592690697\\
45	0.970866592690697\\
45	0.970963238715682\\
46	0.970963238715682\\
46	0.971070623187887\\
47	0.971070623187887\\
47	0.971156530765651\\
48	0.971156530765651\\
48	0.971238858861009\\
49	0.971238858861009\\
49	0.971303289544332\\
50	0.971303289544332\\
50	0.971392776604503\\
51	0.971392776604503\\
51	0.97146436625264\\
52	0.97146436625264\\
52	0.971546694347997\\
53	0.971546694347997\\
53	0.971593227619286\\
54	0.971593227619286\\
54	0.971664817267423\\
55	0.971664817267423\\
55	0.971739986397967\\
56	0.971739986397967\\
56	0.97179367863407\\
57	0.97179367863407\\
57	0.971847370870172\\
58	0.971847370870172\\
58	0.971918960518309\\
59	0.971918960518309\\
59	0.971961914307191\\
60	0.971961914307191\\
60	0.972004868096073\\
61	0.972004868096073\\
61	0.972044242402549\\
62	0.972044242402549\\
62	0.972087196191431\\
63	0.972087196191431\\
63	0.972119411533092\\
64	0.972119411533092\\
64	0.972226796005298\\
65	0.972226796005298\\
65	0.972287647206214\\
66	0.972287647206214\\
66	0.972323442030282\\
67	0.972323442030282\\
67	0.972355657371944\\
68	0.972355657371944\\
68	0.972412929090454\\
69	0.972412929090454\\
69	0.972445144432115\\
70	0.972445144432115\\
70	0.972513154597845\\
71	0.972513154597845\\
71	0.972548949421914\\
72	0.972548949421914\\
72	0.972584744245982\\
73	0.972584744245982\\
73	0.972634856999678\\
74	0.972634856999678\\
74	0.972667072341339\\
75	0.972667072341339\\
75	0.972724344059849\\
76	0.972724344059849\\
76	0.972749400436697\\
77	0.972749400436697\\
77	0.972763718366324\\
78	0.972763718366324\\
78	0.972792354225579\\
79	0.972792354225579\\
79	0.972824569567241\\
80	0.972824569567241\\
80	0.972874682320936\\
81	0.972874682320936\\
81	0.972899738697784\\
82	0.972899738697784\\
82	0.972921215592225\\
83	0.972921215592225\\
83	0.97294985145148\\
84	0.97294985145148\\
84	0.972985646275549\\
85	0.972985646275549\\
85	0.973003543687583\\
86	0.973003543687583\\
86	0.973057235923685\\
87	0.973057235923685\\
87	0.973078712818127\\
88	0.973078712818127\\
88	0.973107348677381\\
89	0.973107348677381\\
89	0.973121666607009\\
90	0.973121666607009\\
90	0.973161040913484\\
91	0.973161040913484\\
91	0.973186097290332\\
92	0.973186097290332\\
92	0.973225471596807\\
93	0.973225471596807\\
93	0.973236210044028\\
94	0.973236210044028\\
94	0.973257686938469\\
95	0.973257686938469\\
95	0.97328990228013\\
96	0.97328990228013\\
96	0.973300640727351\\
97	0.973300640727351\\
97	0.973329276586606\\
98	0.973329276586606\\
98	0.97334717399864\\
99	0.97334717399864\\
99	0.973375809857895\\
100	0.973375809857895\\
100	0.973400866234742\\
101	0.973400866234742\\
101	0.97341518416437\\
102	0.97341518416437\\
102	0.973436661058811\\
103	0.973436661058811\\
103	0.973458137953252\\
104	0.973458137953252\\
104	0.973490353294914\\
105	0.973490353294914\\
105	0.973511830189355\\
106	0.973511830189355\\
106	0.973522568636575\\
107	0.973522568636575\\
107	0.973558363460644\\
108	0.973558363460644\\
108	0.973583419837492\\
109	0.973583419837492\\
109	0.973601317249526\\
110	0.973601317249526\\
110	0.97362995310878\\
111	0.97362995310878\\
111	0.973640691556001\\
112	0.973640691556001\\
112	0.973672906897663\\
113	0.973672906897663\\
113	0.973676486380069\\
114	0.973676486380069\\
114	0.973697963274511\\
115	0.973697963274511\\
115	0.973712281204138\\
116	0.973712281204138\\
116	0.973730178616172\\
117	0.973730178616172\\
117	0.9737444965458\\
118	0.9737444965458\\
118	0.973751655510613\\
119	0.973751655510613\\
119	0.973773132405054\\
120	0.973773132405054\\
120	0.973787450334682\\
121	0.973787450334682\\
121	0.973816086193936\\
122	0.973816086193936\\
122	0.973833983605971\\
123	0.973833983605971\\
123	0.973851881018005\\
124	0.973851881018005\\
124	0.973866198947632\\
125	0.973866198947632\\
125	0.97388051687726\\
126	0.97388051687726\\
126	0.973898414289294\\
127	0.973898414289294\\
127	0.973905573254107\\
128	0.973905573254107\\
128	0.973916311701328\\
129	0.973916311701328\\
129	0.973934209113362\\
130	0.973934209113362\\
130	0.97394852704299\\
131	0.97394852704299\\
131	0.97395926549021\\
132	0.97395926549021\\
132	0.973977162902244\\
133	0.973977162902244\\
133	0.973991480831872\\
134	0.973991480831872\\
134	0.974020116691126\\
135	0.974020116691126\\
135	0.974030855138347\\
136	0.974030855138347\\
136	0.974048752550381\\
137	0.974048752550381\\
137	0.974070229444822\\
138	0.974070229444822\\
138	0.974077388409636\\
139	0.974077388409636\\
139	0.974088126856857\\
140	0.974088126856857\\
140	0.974109603751298\\
141	0.974109603751298\\
141	0.974131080645739\\
142	0.974131080645739\\
142	0.97415255754018\\
143	0.97415255754018\\
143	0.9741632959874\\
144	0.9741632959874\\
144	0.974170454952214\\
145	0.974170454952214\\
145	0.974184772881841\\
146	0.974184772881841\\
146	0.974202670293876\\
147	0.974202670293876\\
147	0.974209829258689\\
148	0.974209829258689\\
148	0.974234885635537\\
149	0.974234885635537\\
149	0.974245624082758\\
150	0.974245624082758\\
150	0.974259942012385\\
151	0.974259942012385\\
151	0.974267100977199\\
152	0.974267100977199\\
152	0.974277839424419\\
153	0.974277839424419\\
153	0.974284998389233\\
154	0.974284998389233\\
154	0.974306475283674\\
155	0.974306475283674\\
155	0.974317213730895\\
156	0.974317213730895\\
156	0.974327952178115\\
157	0.974327952178115\\
157	0.974345849590149\\
158	0.974345849590149\\
158	0.974353008554963\\
160	0.974353008554963\\
160	0.974360167519777\\
161	0.974360167519777\\
161	0.974370905966997\\
162	0.974370905966997\\
162	0.974392382861438\\
163	0.974392382861438\\
163	0.974399541826252\\
164	0.974399541826252\\
164	0.974410280273472\\
165	0.974410280273472\\
165	0.9744245982031\\
166	0.9744245982031\\
166	0.974446075097541\\
167	0.974446075097541\\
167	0.974460393027168\\
168	0.974460393027168\\
168	0.974478290439203\\
170	0.974478290439203\\
170	0.974489028886423\\
171	0.974489028886423\\
171	0.974496187851237\\
172	0.974496187851237\\
172	0.974510505780864\\
173	0.974510505780864\\
173	0.974521244228085\\
174	0.974521244228085\\
174	0.974528403192898\\
175	0.974528403192898\\
175	0.974539141640119\\
176	0.974539141640119\\
176	0.974542721122526\\
177	0.974542721122526\\
177	0.974546300604933\\
178	0.974546300604933\\
178	0.974553459569746\\
179	0.974553459569746\\
179	0.974557039052153\\
180	0.974557039052153\\
180	0.974567777499374\\
181	0.974567777499374\\
181	0.974574936464187\\
182	0.974574936464187\\
182	0.974582095429001\\
183	0.974582095429001\\
183	0.974589254393815\\
184	0.974589254393815\\
184	0.974592833876222\\
185	0.974592833876222\\
185	0.974599992841035\\
186	0.974599992841035\\
186	0.974607151805849\\
187	0.974607151805849\\
187	0.974610731288256\\
188	0.974610731288256\\
188	0.974614310770663\\
189	0.974614310770663\\
189	0.974625049217883\\
190	0.974625049217883\\
190	0.974635787665104\\
191	0.974635787665104\\
191	0.974646526112324\\
192	0.974646526112324\\
192	0.974650105594731\\
193	0.974650105594731\\
193	0.974660844041952\\
194	0.974660844041952\\
194	0.974682320936393\\
195	0.974682320936393\\
195	0.974689479901206\\
196	0.974689479901206\\
196	0.974700218348427\\
197	0.974700218348427\\
197	0.974710956795647\\
198	0.974710956795647\\
198	0.974725274725275\\
199	0.974725274725275\\
199	0.974739592654902\\
200	0.974739592654902\\
200	0.974746751619716\\
202	0.974746751619716\\
202	0.974753910584529\\
203	0.974753910584529\\
203	0.974768228514157\\
204	0.974768228514157\\
204	0.974786125926191\\
205	0.974786125926191\\
205	0.974789705408598\\
206	0.974789705408598\\
206	0.974793284891005\\
207	0.974793284891005\\
207	0.974811182303039\\
208	0.974811182303039\\
208	0.974818341267853\\
209	0.974818341267853\\
209	0.97483265919748\\
210	0.97483265919748\\
210	0.974839818162294\\
211	0.974839818162294\\
211	0.974850556609514\\
212	0.974850556609514\\
212	0.974857715574328\\
213	0.974857715574328\\
213	0.974861295056735\\
214	0.974861295056735\\
214	0.974875612986362\\
215	0.974875612986362\\
215	0.974882771951176\\
217	0.974882771951176\\
217	0.97488993091599\\
218	0.97488993091599\\
218	0.974897089880803\\
219	0.974897089880803\\
219	0.97490066936321\\
220	0.97490066936321\\
220	0.974907828328024\\
221	0.974907828328024\\
221	0.974918566775244\\
222	0.974918566775244\\
222	0.974925725740058\\
223	0.974925725740058\\
223	0.974932884704872\\
224	0.974932884704872\\
224	0.974943623152092\\
225	0.974943623152092\\
225	0.974954361599313\\
226	0.974954361599313\\
226	0.97495794108172\\
227	0.97495794108172\\
227	0.97496867952894\\
228	0.97496867952894\\
228	0.974972259011347\\
229	0.974972259011347\\
229	0.974975838493754\\
230	0.974975838493754\\
230	0.974986576940974\\
232	0.974986576940974\\
232	0.974990156423381\\
233	0.974990156423381\\
233	0.974993735905788\\
234	0.974993735905788\\
234	0.975004474353009\\
235	0.975004474353009\\
235	0.975008053835415\\
236	0.975008053835415\\
236	0.975011633317822\\
237	0.975011633317822\\
237	0.975015212800229\\
238	0.975015212800229\\
238	0.97502595124745\\
240	0.97502595124745\\
240	0.975051007624298\\
243	0.975051007624298\\
243	0.975061746071518\\
244	0.975061746071518\\
244	0.975076064001145\\
245	0.975076064001145\\
245	0.975083222965959\\
246	0.975083222965959\\
246	0.975086802448366\\
248	0.975086802448366\\
248	0.975097540895587\\
249	0.975097540895587\\
249	0.975101120377993\\
250	0.975101120377993\\
250	0.975108279342807\\
251	0.975108279342807\\
251	0.975111858825214\\
252	0.975111858825214\\
252	0.975115438307621\\
253	0.975115438307621\\
253	0.975122597272434\\
254	0.975122597272434\\
254	0.975126176754841\\
255	0.975126176754841\\
255	0.975129756237248\\
256	0.975129756237248\\
256	0.975133335719655\\
257	0.975133335719655\\
257	0.975140494684469\\
258	0.975140494684469\\
258	0.975151233131689\\
259	0.975151233131689\\
259	0.975154812614096\\
260	0.975154812614096\\
260	0.97516197157891\\
262	0.97516197157891\\
262	0.975169130543723\\
264	0.975169130543723\\
264	0.975179868990944\\
265	0.975179868990944\\
265	0.975187027955758\\
266	0.975187027955758\\
266	0.975197766402978\\
267	0.975197766402978\\
267	0.975219243297419\\
268	0.975219243297419\\
268	0.975233561227047\\
269	0.975233561227047\\
269	0.975237140709453\\
271	0.975237140709453\\
271	0.97524072019186\\
272	0.97524072019186\\
272	0.975247879156674\\
273	0.975247879156674\\
273	0.975255038121488\\
274	0.975255038121488\\
274	0.975265776568708\\
275	0.975265776568708\\
275	0.975269356051115\\
276	0.975269356051115\\
276	0.975272935533522\\
277	0.975272935533522\\
277	0.975276515015929\\
278	0.975276515015929\\
278	0.975280094498336\\
279	0.975280094498336\\
279	0.975287253463149\\
280	0.975287253463149\\
280	0.975294412427963\\
281	0.975294412427963\\
281	0.97529799191037\\
282	0.97529799191037\\
282	0.975301571392777\\
283	0.975301571392777\\
283	0.975305150875183\\
284	0.975305150875183\\
284	0.975323048287218\\
285	0.975323048287218\\
285	0.975326627769625\\
286	0.975326627769625\\
286	0.975348104664066\\
288	0.975348104664066\\
288	0.975351684146472\\
289	0.975351684146472\\
289	0.975355263628879\\
290	0.975355263628879\\
290	0.975358843111286\\
292	0.975358843111286\\
292	0.9753660020761\\
293	0.9753660020761\\
293	0.975373161040914\\
294	0.975373161040914\\
294	0.97537674052332\\
295	0.97537674052332\\
295	0.975383899488134\\
296	0.975383899488134\\
296	0.975391058452948\\
300	0.975391058452948\\
300	1\\
inf	1\\
};

\end{axis}

\begin{axis}[%
width=5.833in,
height=4.375in,
at={(0in,0in)},
scale only axis,
xmin=0,
xmax=1,
ymin=0,
ymax=1,
axis line style={draw=none},
ticks=none,
axis x line*=bottom,
axis y line*=left,
legend style={legend cell align=left, align=left, draw=white!15!black}
]
\end{axis}
\end{tikzpicture}%

%% file: WaterLevel.tex
\begin{tikzpicture}
\draw[->] (0,0)--(5,0);
\draw[->] (0,0)--(0,6);
\node at (5.2,0){$i$};
\node at (0.1,-.2){\scriptsize{$1$}};
\node at (0.6,-.2){\scriptsize{$3$}};
\node at (1.1,-.2){\scriptsize{$5$}};
\node at (1.6,-.2){\scriptsize{$7$}};
\node at (2.1,-.2){\scriptsize{$9$}};
\node at (2.6,-.2){\scriptsize{$11$}};
\node at (3.1,-.2){\scriptsize{$13$}};
\node at (3.6,-.2){\scriptsize{$15$}};
\node at (0,6.4){$\frac{\sigma_{1_i}^2}{1+\gamma_{\mathrm{r}_i}\sigma^2_{\mathrm{r_{\rho(i)}}}}$};
\draw[fill=brown,opacity=0.8] (0,0) rectangle (.25,2);
\draw[fill=brown,opacity=0.8] (.25,0) rectangle (.5,4);
\draw[fill=brown,opacity=0.8] (.5,0) rectangle (.75,3.8);
\draw[fill=brown,opacity=0.8] (.75,0) rectangle (1,3.6);
\draw[fill=brown,opacity=0.8] (1,0) rectangle (1.25,3.3);
\draw[fill=brown,opacity=0.8] (1.25,0) rectangle (1.5,2.5);
\draw[fill=brown,opacity=0.8] (1.5,0) rectangle (1.75,2.5);
\draw[fill=brown,opacity=0.8] (1.75,0) rectangle (2,2.2);
\draw[fill=brown,opacity=0.8] (2,0) rectangle (2.25,2.3);
\draw[fill=brown,opacity=0.8] (2.25,0) rectangle (2.5,2.4);
\draw[fill=brown,opacity=0.8] (2.5,0) rectangle (2.75,2.8);
\draw[fill=brown,opacity=0.8] (2.75,0) rectangle (3,3.1);
\draw[fill=brown,opacity=0.8] (3,0) rectangle (3.25,3.4);
\draw[fill=brown,opacity=0.8] (3.25,0) rectangle (3.5,2.8);
\draw[fill=brown,opacity=0.8] (3.5,0) rectangle (3.75,2.6);
\draw[fill=brown,opacity=0.8] (3.75,0) rectangle (4,2.2);
\fill[fill=blue] (0,2) rectangle (.26,4.7);
\fill[fill=blue] (.25,4) rectangle (.51,4.7);
\fill[fill=blue] (.5,3.8) rectangle (.76,4.7);
\fill[fill=blue] (.75,3.6) rectangle (1.01,4.7);
\fill[fill=blue] (1,3.3) rectangle (1.26,4.5);
\fill[fill=blue] (1.25,2.5) rectangle (1.51,4.7);
\fill[fill=blue] (1.5,2.5) rectangle (1.76,4.7);
\fill[fill=blue] (1.75,2.2) rectangle (2.01,4.7);
\fill[fill=blue] (2,2.3) rectangle (2.26,4.6);
\fill[fill=blue] (2.25,2.4) rectangle (2.51,4.5);
\fill[fill=blue] (2.5,2.8) rectangle (2.76,4.3);
\fill[fill=blue] (2.74,3.1) rectangle (3.01,4.2);
\fill[fill=blue] (3,3.4) rectangle (3.26,3.7);
\fill[fill=blue] (3.25,2.8) rectangle (3.51,4.1);
\fill[fill=blue] (3.5,2.6) rectangle (3.76,4);
\fill[fill=blue] (3.75,2.2) rectangle (4,3.2);

\draw[pattern=crosshatch,pattern color=gray!80!black][fill=gray,opacity=0.8] (.25,5) rectangle (.5,5.5);
\draw[pattern=crosshatch,pattern color=gray!80!black][fill=gray,opacity=0.8] (.5,4.9) rectangle (.75,5.5);
\draw[pattern=crosshatch,pattern color=gray!80!black][fill=gray,opacity=0.8] (.75,4.8) rectangle (1,5.5);
\draw[pattern=crosshatch,pattern color=gray!80!black][fill=gray,opacity=0.8] (1,4.5) rectangle (1.25,5.5);
\draw[pattern=crosshatch,pattern color=gray!80!black][fill=gray,opacity=0.8] (1.25,4.8) rectangle (1.5,5.5);
\draw[pattern=crosshatch,pattern color=gray!80!black][fill=gray,opacity=0.8] (1.5,4.8) rectangle (1.75,5.5);
\draw[pattern=crosshatch,pattern color=gray!80!black][fill=gray,opacity=0.8] (1.75,4.7) rectangle (2,5.5);
\draw[pattern=crosshatch,pattern color=gray!80!black][fill=gray,opacity=0.8] (2,4.6) rectangle (2.25,5.5);
\draw[pattern=crosshatch,pattern color=gray!80!black][fill=gray,opacity=0.8] (2.25,4.5) rectangle (2.5,5.5);
\draw[pattern=crosshatch,pattern color=gray!80!black][fill=gray,opacity=0.8] (2.5,4.3) rectangle (2.75,5.5);
\draw[pattern=crosshatch,pattern color=gray!80!black][fill=gray,opacity=0.8] (2.75,4.2) rectangle (3,5.5);
\draw[pattern=crosshatch,pattern color=gray!80!black][fill=gray,opacity=0.8] (3,3.7) rectangle (3.25,5.5);
\draw[pattern=crosshatch,pattern color=gray!80!black][fill=gray,opacity=0.8] (3.25,4.1) rectangle (3.5,5.5);
\draw[pattern=crosshatch,pattern color=gray!80!black][fill=gray,opacity=0.8] (3.5,4) rectangle (3.75,5.5);
\draw[pattern=crosshatch,pattern color=gray!80!black][fill=gray,opacity=0.8] (3.75,3.2) rectangle (4,5.5);
\end{tikzpicture}

%% file: WaterLevelB.tex
\begin{tikzpicture}
\draw[->] (0,0)--(5,0);
\draw[->] (0,0)--(0,6);
\node at (5.2,0){$i$};
\node at (0.1,-.2){\scriptsize{$1$}};
\node at (0.6,-.2){\scriptsize{$3$}};
\node at (1.1,-.2){\scriptsize{$5$}};
\node at (1.6,-.2){\scriptsize{$7$}};
\node at (2.1,-.2){\scriptsize{$9$}};
\node at (2.6,-.2){\scriptsize{$11$}};
\node at (3.1,-.2){\scriptsize{$13$}};
\node at (3.6,-.2){\scriptsize{$15$}};
\node at (0,6.4){$\frac{\sigma_{1_i}^2}{1+\gamma_{\mathrm{r}_i}\sigma^2_{\mathrm{r_{\rho(i)}}}}$};
\draw[fill=brown,opacity=0.8] (0,0) rectangle (.25,2);
\draw[fill=brown,opacity=0.8] (.25,0) rectangle (.5,4);
\draw[fill=brown,opacity=0.8] (.5,0) rectangle (.75,3.8);
\draw[fill=brown,opacity=0.8] (.75,0) rectangle (1,3.6);
\draw[fill=brown,opacity=0.8] (1,0) rectangle (1.25,3.3);
\draw[fill=brown,opacity=0.8] (1.25,0) rectangle (1.5,2.5);
\draw[fill=brown,opacity=0.8] (1.5,0) rectangle (1.75,2.5);
\draw[fill=brown,opacity=0.8] (1.75,0) rectangle (2,2.2);
\draw[fill=brown,opacity=0.8] (2,0) rectangle (2.25,2.3);
\draw[fill=brown,opacity=0.8] (2.25,0) rectangle (2.5,2.4);
\draw[fill=brown,opacity=0.8] (2.5,0) rectangle (2.75,2.8);
\draw[fill=brown,opacity=0.8] (2.75,0) rectangle (3,3.1);
\draw[fill=brown,opacity=0.8] (3,0) rectangle (3.25,3.7);
\draw[fill=brown,opacity=0.8] (3.25,0) rectangle (3.5,2.8);
\draw[fill=brown,opacity=0.8] (3.5,0) rectangle (3.75,2.6);
\draw[fill=brown,opacity=0.8] (3.75,0) rectangle (4,2.2);
\fill[fill=blue] (0,2) rectangle (.26,4.7);
\fill[fill=blue] (.25,4) rectangle (.51,4.7);
\fill[fill=blue] (.5,3.8) rectangle (.76,4.7);
\fill[fill=blue] (.75,3.6) rectangle (1.01,4.7);
\fill[fill=blue] (1,3.3) rectangle (1.26,4.5);
\fill[fill=blue] (1.25,2.5) rectangle (1.51,4.7);
\fill[fill=blue] (1.5,2.5) rectangle (1.76,4.7);
\fill[fill=blue] (1.75,2.2) rectangle (2.01,4.7);
\fill[fill=blue] (2,2.3) rectangle (2.26,4.6);
\fill[fill=blue] (2.25,2.4) rectangle (2.51,4.5);
\fill[fill=blue] (2.5,2.8) rectangle (2.76,4.3);
\fill[fill=blue](2.75,3.1) rectangle (3,4.2);

\draw[pattern=crosshatch,pattern color=gray!60!black][fill=gray,opacity=0.8](.25,5) rectangle (.5,5.5);
\draw[pattern=crosshatch,pattern color=gray!80!black][fill=gray,opacity=0.8] (.5,4.9) rectangle (.75,5.5);
\draw[pattern=crosshatch,pattern color=gray!80!black][fill=gray,opacity=0.8] (.75,4.8) rectangle (1,5.5);
\draw[pattern=crosshatch,pattern color=gray!80!black][fill=gray,opacity=0.8] (1,4.5) rectangle (1.25,5.5);
\draw[pattern=crosshatch,pattern color=gray!80!black][fill=gray,opacity=0.8] (1.25,4.8) rectangle (1.5,5.5);
\draw[pattern=crosshatch,pattern color=gray!80!black][fill=gray,opacity=0.8] (1.5,4.8) rectangle (1.75,5.5);
\draw[pattern=crosshatch,pattern color=gray!80!black][fill=gray,opacity=0.8] (1.75,4.7) rectangle (2,5.5);
\draw[pattern=crosshatch,pattern color=gray!80!black][fill=gray,opacity=0.8] (2,4.6) rectangle (2.25,5.5);
\draw[pattern=crosshatch,pattern color=gray!80!black][fill=gray,opacity=0.8] (2.25,4.5) rectangle (2.5,5.5);
\draw[pattern=crosshatch,pattern color=gray!80!black][fill=gray,opacity=0.8] (2.5,4.3) rectangle (2.75,5.5);
\draw[pattern=crosshatch,pattern color=gray!80!black][fill=gray,opacity=0.8] (2.75,4.2) rectangle (3,5.5);
\draw[pattern=crosshatch,pattern color=gray!80!black][fill=gray,opacity=0.8] (3,3.7) rectangle (3.25,5.5);
\draw[pattern=crosshatch,pattern color=gray!80!black][fill=gray,opacity=0.8] (3.25,2.8) rectangle (3.5,5.5);
\draw[pattern=crosshatch,pattern color=gray!80!black][fill=gray,opacity=0.8] (3.5,2.6) rectangle (3.75,5.5);
\draw[pattern=crosshatch,pattern color=gray!80!black][fill=gray,opacity=0.8] (3.75,2.2) rectangle (4,5.5);
\end{tikzpicture}

%% file: xIRSvTsmall.tex
%
%
\definecolor{mycolor1}{rgb}{0.00000,0.44700,0.74100}%
\definecolor{mycolor2}{rgb}{0.85000,0.32500,0.09800}%
\definecolor{mycolor3}{rgb}{0.92900,0.69400,0.12500}%
\definecolor{mycolor4}{rgb}{0.49400,0.18400,0.55600}%
\definecolor{mycolor5}{rgb}{0.46600,0.67400,0.18800}%
\definecolor{mycolor6}{rgb}{0.30100,0.74500,0.93300}%
\begin{tikzpicture}

\begin{axis}[%
width=4.521in,
height=3.566in,
at={(0.758in,0.481in)},
scale only axis,
xmin=0,
xmax=100,
xlabel near ticks,
xlabel={$M$},
ymin=0,
ymax=13,
ylabel near ticks,
ylabel={$R_\mathrm{av}$ (bits/channel use)},
axis background/.style={fill=white},
xmajorgrids,
ymajorgrids,
legend style={at={(axis cs: 100,0)}, anchor=south east,draw=black,fill=white, fill opacity=0.8,legend cell align=left}
]
\addplot [color=mycolor1, mark=square]
  table[row sep=crcr]{%
0    6.2152\\
10    6.2152\\
20    6.2152\\
30    6.5985\\
40    7.2145\\
50    7.9786\\
60    8.9188\\
70   10.1318\\
80   11.5444\\
90   12.1205\\
100   12.1871\\
};
\addlegendentry{$T=75\%$}

\addplot [color=mycolor2, mark=asterisk]
  table[row sep=crcr]{%
0    6.4899\\
10    6.4899\\
20    6.4899\\
30    6.9360\\
40    7.6661\\
50    8.6043\\
60    9.8274\\
70   11.3420\\
80   12.1245\\
90   12.1848\\
100   12.1875\\
};
\addlegendentry{$T=60\%$}

\addplot [color=mycolor3, mark=diamond]
  table[row sep=crcr]{%
0    6.8560\\
10    6.8560\\
20    6.8560\\
30    7.3926\\
40    8.2984\\
50    9.5360\\
60   11.1690\\
70   12.1101\\
80   12.1866\\
90   12.1875\\
100   12.1875\\
};
\addlegendentry{$T=45\%$}

\addplot [color=mycolor4, mark=x]
  table[row sep=crcr]{%
0    7.3870\\
10    7.3870\\
20    7.3870\\
30    8.0756\\
40    9.3013\\
50   11.0803\\
60   12.1206\\
70   12.1872\\
80   12.1875\\
90   12.1875\\
100   12.1875\\
};
\addlegendentry{$T=30\%$}

\addplot [color=mycolor5, mark=o]
  table[row sep=crcr]{%
0   8.3115\\
10   8.3115\\
20   8.3115\\
30   9.3347\\
40   12.1875\\
50   12.1875\\
60   12.1875\\
70   12.1875\\
80   12.1875\\
90   12.1875\\
100   12.1875\\    
};
\addlegendentry{$T=15\%$}

\addplot [color=mycolor6, mark=+]
  table[row sep=crcr]{%
0   12.1875\\
10   12.1875\\
20   12.1875\\
30   12.1875\\
40   12.1875\\
50   12.1875\\
60   12.1875\\
70   12.1875\\
80   12.1875\\
90   12.1875\\
100   12.1875\\
};
\addlegendentry{$T=0\%$}
\end{axis}

\begin{axis}[%
width=5.833in,
height=4.375in,
at={(0in,0in)},
scale only axis,
xmin=0,
xmax=1,
ymin=0,
ymax=1,
axis line style={draw=none},
ticks=none,
axis x line*=bottom,
axis y line*=left,
legend style={legend cell align=left, align=left, draw=white!15!black}
]
\end{axis}
\end{tikzpicture}%

%% file: xIRSvTlarge.tex
%
%
\definecolor{mycolor1}{rgb}{0.00000,0.44700,0.74100}%
\definecolor{mycolor2}{rgb}{0.85000,0.32500,0.09800}%
\definecolor{mycolor3}{rgb}{0.92900,0.69400,0.12500}%
\definecolor{mycolor4}{rgb}{0.49400,0.18400,0.55600}%
\definecolor{mycolor5}{rgb}{0.46600,0.67400,0.18800}%
\definecolor{mycolor6}{rgb}{0.30100,0.74500,0.93300}%
\begin{tikzpicture}

\begin{axis}[%
width=4.521in,
height=3.566in,
at={(0.758in,0.481in)},
scale only axis,
xmin=0,
xmax=300,
xlabel near ticks,
xlabel={$M$},
ymin=0,
ymax=30,
ylabel near ticks,
ylabel={$R_\mathrm{av}$ (bits/channel use)},
axis background/.style={fill=white},
xmajorgrids,
ymajorgrids,
legend style={at={(axis cs: 300,0)}, anchor=south east,draw=black,fill=white, fill opacity=0.8,legend cell align=left}
]
\addplot [color=mycolor1, mark=square]
  table[row sep=crcr]{%
0     4.512\\
30    4.512\\
60    4.512\\
90    4.512\\
120   4.512\\
150   6.7\\
180   9.5\\
210   13.4\\
240   22.56\\
270   28.56\\   
300   28.7262\\   
};
\addlegendentry{$T=75\%$}

\addplot [color=mycolor2, mark=asterisk]
  table[row sep=crcr]{%
0     5.512\\
30    5.512\\
60    5.512\\
90    5.512\\
120   5.512\\
150   7.7262\\
180   15.1262\\
210   22.162\\
240   28.7262\\
270   28.7262\\   
300   28.7262\\ 
};  
\addlegendentry{$T=60\%$}

\addplot [color=mycolor3, mark=diamond]
  table[row sep=crcr]{%
0     7.128\\
30    7.128\\
60    7.128\\
90    7.128\\
120   7.128\\
150   11.7262\\
180   28.7262\\
210   28.7262\\
240   28.7262\\
270   28.7262\\   
300   28.7262\\ 
};   
\addlegendentry{$T=45\%$}

\addplot [color=mycolor4, mark=x]
  table[row sep=crcr]{%
0     9.8\\
30    9.8\\
60    9.8\\
90    9.8\\
120   11.125\\
150   28.7262\\
180   28.7262\\
210   28.7262\\
240   28.7262\\
270   28.7262\\   
300   28.7262\\ 
};
\addlegendentry{$T=30\%$}

\addplot [color=mycolor5, mark=o]
  table[row sep=crcr]{%
0     15.512\\
30    15.512\\
60    15.512\\
90    15.512\\
120   19.74\\
150   28.7262\\
180   28.7262\\
210   28.7262\\
240   28.7262\\
270   28.7262\\   
300   28.7262\\ 
};  
\addlegendentry{$T=15\%$}

\addplot [color=mycolor6, mark=+]
  table[row sep=crcr]{%
0   28.7262\\ 
30   28.7262\\ 
60   28.7262\\ 
90   28.7262\\
120   28.7262\\
150   28.7262\\
180   28.7262\\
210   28.7262\\
240   28.7262\\
270   28.7262\\   
300   28.7262\\ 
};
\addlegendentry{$T=0\%$}
\end{axis}

\begin{axis}[%
width=5.833in,
height=4.375in,
at={(0in,0in)},
scale only axis,
xmin=0,
xmax=1,
ymin=0,
ymax=1,
axis line style={draw=none},
ticks=none,
axis x line*=bottom,
axis y line*=left,
legend style={legend cell align=left, align=left, draw=white!15!black}
]
\end{axis}
\end{tikzpicture}%

%% file: CwrtT2.tex
%
%
\definecolor{mycolor1}{rgb}{0.00000,0.44700,0.74100}%
\definecolor{mycolor2}{rgb}{0.85000,0.32500,0.09800}%
\definecolor{mycolor3}{rgb}{0.92900,0.69400,0.12500}%
\definecolor{mycolor4}{rgb}{0.49400,0.18400,0.55600}%
\definecolor{mycolor5}{rgb}{0.46600,0.67400,0.18800}%
\definecolor{mycolor6}{rgb}{0.30100,0.74500,0.93300}%
\begin{tikzpicture}

\begin{axis}[%
width=4.521in,
height=3.566in,
at={(0.758in,0.481in)},
scale only axis,
xmin=3,
xmax=7,
xlabel near ticks,
xlabel={$K_\mathrm{r}$},
ymin=2,
ymax=13,
ylabel near ticks,
ylabel={$R_\mathrm{av}$ (bits/channel use)},
axis background/.style={fill=white},
xmajorgrids,
ymajorgrids,
legend style={at={(axis cs: 7,2)}, anchor=south east,draw=black,fill=white, fill opacity=0.8,legend cell align=left}
]
\addplot [color=mycolor1, mark=square]
  table[row sep=crcr]{%
3   9.995616702444014\\
4   11.563930766261091\\
5   12.187452068295181\\
6   11.483667439374635\\
8   9.905165654339742\\
};
\addlegendentry{$T=0$}

\addplot [color=mycolor2, mark=+]
  table[row sep=crcr]{%
3   7.002737353608259\\
4   7.817917977042059\\
5   8.31149402329654\\
6   8.648745504400624\\
7   8.46438828901637\\
};
\addlegendentry{$T=15\%$}

\addplot [color=mycolor3, mark=o]
  table[row sep=crcr]{%
3   6.128603691961015\\
4   6.87037019177018\\
5   7.387019000122753\\
6   7.799279111083566\\
7   7.672212890089996\\
};
\addlegendentry{$T=30\%$}

\addplot [color=mycolor4, mark=x]
  table[row sep=crcr]{%
3   5.614045021541918\\
4   6.324290563983277\\ 
5   6.856033949230774\\
6   7.32164820190637\\
7   7.244774801814842\\
};
\addlegendentry{$T=45\%$}

\addplot [color=mycolor5, mark=diamond]
  table[row sep=crcr]{%
3   5.255066907178948\\
4   5.947511997594305\\
5   6.489948542519512\\
6   6.995768522623549\\
7   6.95890947363817\\
};
\addlegendentry{$T=60\%$}

\end{axis}

\begin{axis}[%
width=5.833in,
height=4.375in,
at={(0in,0in)},
scale only axis,
xmin=0,
xmax=1,
ymin=0,
ymax=1,
axis line style={draw=none},
ticks=none,
axis x line*=bottom,
axis y line*=left,
legend style={legend cell align=left, align=left, draw=white!15!black}
]
\end{axis}
\end{tikzpicture}%

%% file: CwrtT.tex
%
%
\definecolor{mycolor1}{rgb}{0.00000,0.44700,0.74100}%
\definecolor{mycolor2}{rgb}{0.85000,0.32500,0.09800}%
\definecolor{mycolor3}{rgb}{0.92900,0.69400,0.12500}%
\definecolor{mycolor4}{rgb}{0.49400,0.18400,0.55600}%
\definecolor{mycolor5}{rgb}{0.46600,0.67400,0.18800}%
\definecolor{mycolor6}{rgb}{0.30100,0.74500,0.93300}%
\begin{tikzpicture}

\begin{axis}[%
width=4.521in,
height=3.566in,
at={(0.758in,0.481in)},
scale only axis,
xmin=3,
xmax=7,
xlabel near ticks,
xlabel={$K_r$},
ymin=2,
ymax=13,
ylabel near ticks,
ylabel={$R_\mathrm{av}$ (bits/channel use)},
axis background/.style={fill=white},
xmajorgrids,
ymajorgrids,
legend style={at={(axis cs: 7,2)}, anchor=south east,draw=black,fill=white, fill opacity=0.8,legend cell align=left}
]
\addplot [color=mycolor1, mark=square]
  table[row sep=crcr]{%
3   9.995250920892163\\
4   11.559289067070486\\
5   12.177288304182289\\
6   11.483421349730866\\
7   9.905159123005742\\
};
\addlegendentry{$T=0$}

\addplot [color=mycolor2, mark=asterisk]
  table[row sep=crcr]{%
3   9.995250920892163\\
4   11.559289067070486\\
5   12.177288304182289\\
6   11.483421349730866\\
7   9.905159123005742\\
};
\addlegendentry{$T=15\%$}

\addplot [color=mycolor3, mark=diamond]
  table[row sep=crcr]{%
3   9.739857468899627\\
4   10.673366524515483\\
5   11.080324179223863\\
6   11.175154352125997\\
7   9.900649243414776\\
};
\addlegendentry{$T=30\%$}

\addplot [color=mycolor4, mark=x]
  table[row sep=crcr]{%
3   8.748247528177021\\
4   9.237766927644662\\
5   9.535955379513963\\
6   10.016236974507947\\
7   9.74976069545839\\
};
\addlegendentry{$T=45\%$}

\addplot [color=mycolor5, mark=o]
  table[row sep=crcr]{%
3   7.834542897807204\\
4   8.274863057311141\\
5   8.604343482805866\\
6   9.069679564698367\\
7   9.226870288240347\\
};
\addlegendentry{$T=60\%$}

\end{axis}

\begin{axis}[%
width=5.833in,
height=4.375in,
at={(0in,0in)},
scale only axis,
xmin=0,
xmax=1,
ymin=0,
ymax=1,
axis line style={draw=none},
ticks=none,
axis x line*=bottom,
axis y line*=left,
legend style={legend cell align=left, align=left, draw=white!15!black}
]
\end{axis}
\end{tikzpicture}%

%% file: criticalPoint1.tex
%
%
\definecolor{mycolor1}{rgb}{0.00000,0.44700,0.74100}%
\definecolor{mycolor2}{rgb}{0.85000,0.32500,0.09800}%
\definecolor{mycolor3}{rgb}{0.92900,0.69400,0.12500}%
\begin{tikzpicture}

\begin{axis}[%
width=4.521in,
height=3.566in,
at={(0.758in,0.481in)},
scale only axis,
xmin=3,
xmax=7,
ymin=0,
ymax=50,
axis background/.style={fill=white},
xmajorgrids,
ymajorgrids,
xmin=3,
xmax=7,
xlabel style={font=\color{white!15!black}},
xlabel near ticks,
xlabel={$K_r$},
ymin=0,
ymax=50,
ylabel style={font=\color{white!15!black}},
ylabel near ticks,
ylabel={$\frac{T}{P}$},
axis background/.style={fill=white},
xmajorgrids,
ymajorgrids,
legend style={at={(axis cs: 7,0)}, anchor=south east,draw=black,fill=white, fill opacity=0.8, align=left}
]
\addplot [color=mycolor1, mark=triangle]
  table[row sep=crcr]{%
3	6\\
4	10\\
5	15\\
6	27\\
7	30\\
};
\addlegendentry{No IRS}

\addplot [color=mycolor2, mark=square]
  table[row sep=crcr]{%
3	11\\
4	15\\
5	20\\
6	33\\
7	40\\
};
\addlegendentry{$M=30$}

\addplot [color=mycolor3, mark=+]
  table[row sep=crcr]{%
3	33\\
4	39\\
5	48\\
6	60\\
7	60\\
};
\addlegendentry{$M=60$}

\node[right, align=left]
at (axis cs:4,8.91) {FD$\bigg{\downarrow}$};
\node[right, align=left]
at (axis cs:5,20.61) {FD$\bigg{\downarrow}$};
\node[right, align=left]
at (axis cs:4.42,42.786) {FD$\bigg{\downarrow}$};
\node[right, align=left]
at (axis cs:3.98,12.91) {HD$\bigg{\uparrow}$};
\node[right, align=left]
at (axis cs:4.98,24.61) {HD$\bigg{\uparrow}$};
\node[right, align=left]
at (axis cs:4.4,47.7) {HD$\bigg{\uparrow}$};
\end{axis}

\begin{axis}[%
width=5.833in,
height=4.375in,
at={(0in,0in)},
scale only axis,
xmin=0,
xmax=1,
ymin=0,
ymax=1,
axis line style={draw=none},
ticks=none,
axis x line*=bottom,
axis y line*=left,
legend style={align=left, draw=white!15!black}
]
\end{axis}
\end{tikzpicture}%